\documentclass[fleqn,usenatbib]{mnras}

\usepackage[OT2,T1]{fontenc}
\usepackage{ae,aecompl}

\usepackage{times}
\usepackage{epsfig}
\usepackage[]{graphicx} 
\usepackage{pdfpages}
\usepackage{verbatim} 
\usepackage[fleqn]{amsmath}
\usepackage{amssymb}
\usepackage{calc}
\usepackage[normal]{threeparttable}
\usepackage{longtable}
\usepackage{pdflscape}
\newcommand{\comments}[1]{} 
\usepackage{placeins} 
\usepackage{float}
\usepackage{soul}
\graphicspath{{./plots/}}
\title[Cold gas feeding an interacting young radio galaxy]{PKS\,B1740$\mathbf{-}$517: An ALMA view of the cold gas feeding a distant interacting young radio galaxy}

\author[J.~R. Allison et al.]{
J.~R. Allison,$^{1,2}$\thanks{E-mail: james.allison@physics.ox.ac.uk}
E.~K. Mahony,$^{3}$
V.~A. Moss,$^{4,5}$
E.~M. Sadler,$^{2,3,5}$
M.~T. Whiting,$^{3}$
\newauthor
R.~F. Allison,$^{6}$
J. Bland-Hawthorn,$^{2,5}$
S.~J. Curran,$^{7}$
B.~H.~C. Emonts,$^{8}$
C.~D.~P. Lagos,$^{2,9}$
\newauthor
R. Morganti,$^{4,10}$
G. Tremblay,$^{11}$
M. Zwaan,$^{12}$
C.~S. Anderson,$^{13}$
J.~D. Bunton,$^{3}$
\newauthor
and M.~A. Voronkov$^{3}$
\\
$^{1}$Sub-Dept. of Astrophysics, Department of Physics, University of Oxford, Denys Wilkinson Building, Keble Rd., Oxford, OX1 3RH, UK\\
$^{2}$ARC Centre of Excellence for All-sky Astrophysics in 3 Dimensions (ASTRO 3D)\\
$^{3}$CSIRO Astronomy and Space Science, PO Box 76, Epping NSW 1710, Australia\\
$^{4}$ASTRON, The Netherlands Institute for Radio Astronomy, Postbus 2, NL-7900 AA Dwingeloo, the Netherlands\\
$^{5}$Sydney Institute for Astronomy, School of Physics A28, University of Sydney, NSW 2006, Australia\\
$^{6}$Heilbronn Institute for Mathematical Research, Bristol, UK\\
$^{7}$School of Chemical and Physical Sciences, Victoria University of Wellington, PO Box 600, Wellington 6140, New Zealand\\
$^{8}$National Radio Astronomy Observatory, 520 Edgemont Road, Charlottesville, VA 22903, USA\\
$^{9}$International Centre for Radio Astronomy Research (ICRAR), M468, University of Western Australia, 35 Stirling Hwy, Crawley, WA 6009, Australia.\\
$^{10}$Kapteyn Astronomical Institute, University of Groningen, Postbus 800, NL-9700 AV Groningen, the Netherlands\\
$^{11}$Harvard-Smithsonian Center for Astrophysics, 60 Garden St., Cambridge, MA 02138, USA\\
$^{12}$European Southern Observatory, Karl-Schwarzschild-Str. 2, D85748 Garching, Germany\\
$^{13}$CSIRO Astronomy and Space Science, 26 Dick Perry Avenue, Kensington WA 6151, Australia\\
}
\date{Accepted XXX. Received YYY; in original form ZZZ}

\pubyear{2017}

\begin{document}
\label{firstpage}
\pagerange{\pageref{firstpage}--\pageref{lastpage}}
\maketitle
\begin{abstract}
{Cold neutral gas is a key ingredient for growing the stellar and
central black hole mass in galaxies throughout cosmic history. We have used the Atacama Large Millimetre Array (ALMA) to detect a rare example of redshifted $^{12}$CO\,(2-1) absorption in PKS\,B1740$-$517, a young ($t \sim 1.6 \times 10^{3}$\,yr) and luminous ($L_{\rm 5\,GHz} \approx 6.6 \times 10^{43}$\,erg\,s$^{-1}$) radio galaxy at $z = 0.44$ that is undergoing a tidal interaction with at least one lower-mass companion. The coincident \mbox{H\,{\sc i}} 21-cm and molecular absorption have very similar line profiles and reveal a reservoir of cold gas ($M_{\rm gas} \sim 10^{7} - 10^{8}$\,$M_{\odot}$), likely distributed in a disc or ring within a few kiloparsecs of the nucleus. A separate \mbox{H\,{\sc i}} component is kinematically distinct and has a very narrow line width ($\Delta{v_{\rm FWHM}} \lesssim 5$\,km\,s$^{-1}$), consistent with a single diffuse cloud of cold ($T_{\rm k} \sim 100$\,K) atomic gas. The $^{12}$CO\,(2-1) absorption is not associated with this component, which suggests that the cloud is either much smaller than 100\,pc along our sight-line and/or located in low-metallicity gas that was possibly tidally stripped from the companion. We argue that the gas reservoir in PKS\,B1740$-$517 may have accreted onto the host galaxy $\sim$\,50\,Myr before the young radio AGN was triggered, but has only recently reached the nucleus. This is consistent with the paradigm that powerful luminous radio galaxies are triggered by minor mergers and interactions with low-mass satellites and represent a brief, possibly recurrent, active phase in the life cycle of massive early type galaxies.}
\end{abstract}




\section{Introduction}\label{section:introduction}

It is now well established that supermassive black hole (SMBH) mass in galaxies is strongly correlated with the properties of their stellar bulge, namely velocity dispersion, luminosity and mass (see \citealt{Kormendy:2013} for a review). This correlation implies that the evolution of massive galaxies is physically connected to the growth of their SMBH. However, given the disparate scales of the sphere of influence of the SMBH ($\sim 1$\,pc) and the stellar bulge of a massive galaxy ($\sim 1$\,kpc), the exact physical mechanisms building such a relationship are not yet clear. Models invoke both negative (\citealt{Silk:1998,Fabian:1999,King:2003}) and positive feedback from the SMBH (e.g. \citealt{Ishibashi:2012,Silk:2013}), during a particularly active phase known as quasar mode feedback, or regulation of SMBH growth by the stellar bulge itself (e.g. \citealt{Umemura:2001}). Alternatively, such a tight correlation between bulge size and SMBH may simply arise indirectly as a natural consequence of several major galaxy mergers averaged over time (e.g. \citealt{Jahnke:2011}). 

It is clear that a ready supply of cold gas (kinetic temperature $T_{\rm k} \lesssim 10^{4}$\,K) is a common factor driving both stellar mass and SMBH growth in massive galaxies. At low to intermediate redshifts the most radiatively efficient active galactic nuclei (AGN), with high accretion rates ($L_{\rm bol} \gtrsim 0.01\,L_{\rm Edd}$),\footnote{$L_{\rm bol}$ is the total bolometric luminosity of the AGN and $L_{\rm Edd}$ is the Eddington luminosity limit, defined as $L_{\rm Edd} = \frac{4\pi{G}{m_{\rm p}}{c}}{\sigma_{T}}{M_{\rm SMBH}}$; where $G$ is the gravitational constant, $m_{\rm p}$ is the proton mass, $c$ is the speed of light in vacuum, $\sigma_{T}$ is the electron Thomson scattering cross section, and $M_{\rm SMBH}$ is the SMBH mass.} are typically hosted by star-forming galaxies with a central young stellar population and ample cold gas reservoirs (e.g. \citealt{Kauffmann:2003, Kauffmann:2007, Kauffmann:2009, LaMassa:2013}). This class of luminous AGN have a rapidly evolving volume density which mirrors that of the global star formation rate (e.g. \citealt{Ueda:2003, Shankar:2009}). In contrast, AGN that are radiatively inefficient (i.e. $L_{\rm bol} \lesssim 0.01\,L_{\rm Edd}$), but emit synchroton radiation at radio wavelengths, are typically hosted by higher mass passive galaxies that are globally poor in neutral gas (e.g. \citealt{Best:2012}; see \citealt{Heckman:2014} for a review). Regular mechanical feedback from their radio jets (i.e. radio mode feedback) is thought to regulate the cooling of gas from the halo onto these massive galaxies and is an important factor in quenching further growth (e.g. {\citealt{Nulsen:2000, Croton:2006}). At higher redshifts ($z > 1.5$) there is evidence for a reversal in the stellar masses of these two populations, consistent with downsizing of SMBH growth over cosmic time (\citealt{Delvecchio:2017}).

Although this two-flavour model elegantly explains the observed duality of the AGN population, and the role of SMBHs in galaxy evolution, the detailed mechanics of how the gas reaches the nucleus are not yet well determined from observation. In the case of radiatively efficient AGN (i.e. Seyferts, high excitation radio galaxies, and quasars), the cold gas is thought to originate from an existing cold reservoir and loses angular momentum through secular processes and merger events (e.g. \citealt{Hopkins:2006, Hopkins:2008}). For radiatively inefficient AGN (i.e. LINERS and low excitation radio galaxies) the mechanism by which gas reaches the centre of the galaxy is less clear. The simplest model that reproduces the observed correlation between accretion rate and radio jet power (e.g. \citealt{Allen:2006, Hardcastle:2007}, see also \citealt{Russell:2013}) is that the galactic halo supplies the AGN with hot gas via Bondi accretion (\citealt{Bondi:1952}). However, this simplified model assumes spherical symmetry of the accreting material and negligible angular momentum. A more plausible scenario predicted by recent hydrodynamic simulations is that cool gas clouds condense out of the hot halo as a result of turbulence, which in turn is regulated by heating from the radio jets, and lose their angular momentum through collisions close to the Bondi radius (e.g. \citealt{Gaspari:2013, Li:2014a}). Direct evidence for this chaotic cold accretion is seen both from filamentary gas structures in well-resolved nearby cool-core clusters (e.g. \citealt{Salome:2006, Lim:2008,McNamara:2014,Russell:2016}) and inward radial motion of individual atomic and molecular gas clouds seen in absorption towards radio AGN (e.g. \citealt{David:2014, Maccagni:2014,Tremblay:2016,Maccagni:2018}). 

The most luminous and strongly accreting active galactic nuclei are thought to be triggered by interactions between galaxies, which efficiently remove angular momentum from the cold gas and funnel it towards the nucleus (e.g. \citealt{Sanders:1988, Hopkins:2008, Ellison:2011}), briefly, but significantly, elevating the activity above that generated by stochastic accretion of the gas. While major mergers between gas-rich massive galaxies are thought to trigger quasars and drive the formation of massive spheroids throughout the history of the Universe (e.g. \citealt{Treister:2012}), minor mergers and interactions (with a mass ratio greater than 3) can also trigger efficient AGN accretion in otherwise passive galaxies. Nearby powerful radio galaxies that are not in the centres of massive cool-core clusters are typically associated with an ongoing minor merger or interaction, indicating that the gas is efficiently accreted through tidal stripping (\citealt{RamosAlmeida:2011,RamosAlmeida:2012}). Direct evidence of ongoing star formation and radio AGN activity triggered by a minor interaction has been seen in the young ($t \sim 10^{5}$\,yr) compact steep spectrum (CSS) radio galaxy MRC\,B1221$-$423 (\citealt{Johnston:2005,Johnston:2010,Anderson:2013}).

Clearly measuring the kinematics and abundance of cold neutral gas in the host galaxies of AGN is crucial to understanding their prevalence and evolution throughout cosmological history. However, due to current limitations on the sensitivity and observable bands of radio telescopes, most of our knowledge of the cold gas associated with AGN comes from studying nearby objects. Absorption lines, which are effectively detectable at any luminosity distance for a given population of background sources, afford us an opportunity to extend our knowledge of the cold gas in active galaxies beyond $z \sim 0.1$ to cosmological distances. Particularly neutral atomic hydrogen (\mbox{H\,{\sc i}}), which is abundant in the interstellar medium (ISM) and readily detectable via the 21-cm and Lyman\,$\alpha$ lines, is an effective tracer of neutral gas in galaxies to high redshift (see \citealt{Kanekar:2004,Morganti:2015} and \citealt{Wolfe:2005} for reviews). 

In the case of detecting the cold gas feeding radio-loud AGN at cosmological distances (i.e. $z \gg 0.1$) there has been mixed success (e.g. \citealt{Uson:1991, Carilli:1998, Moore:1999, Vermeulen:2003, Ishwara-Chandra:2003, Curran:2011a, Curran:2013a, Allison:2015a, Yan:2016, Aditya:2017, Aditya:2018}), due primarily to the low detection rates associated with luminous AGN (see e.g. \citealt{Curran:2010}) and the targeted nature of these searches driven by limitations in usable radio bands and survey speeds of existing radio telescopes. However, new pathfinder telescopes to the planned Square Kilometre Array (SKA) provide an exciting opportunity to survey the cold gas in active galaxies using the 21-cm absorption line. Recently, \cite{Allison:2015a} (hereafter A15) and \cite{Moss:2017} successfully demonstrated this with the Australian Square Kilometre Array Pathfinder (ASKAP; \citealt{Johnston:2007}). The combination of a wide instantaneous bandwidth (300\,MHz), and a band that at frequencies below 1\,GHz is typically clear of radio frequency interference (see e.g. \citealt{Allison:2017}), enables 21-cm line surveys to be carried out without the need for existing spectroscopically-determined redshifts. Future wide-field 21-cm line surveys with telescopes such as ASKAP will search for the cold gas in thousands of radio galaxies (\citealt{Maccagni:2017, Morganti:2018}). 

In this paper we present results from observations with the Atacama Large Millimetre Array (ALMA) of PKS\,B1740$-$517, a compact (${\Delta{d}_{\rm src} \sim 300\,\mathrm{pc}}$) GHz peaked spectrum (GPS) radio source associated with a luminous AGN at a redshift of $z = 0.44$. A15 discovered \mbox{H\,{\sc i}} absorption in the host galaxy of this source using the ASKAP Boolardy Engineering Test Array (BETA; \citealt{Hotan:2014,McConnell:2016}). Here we have used ALMA to search for $^{12}$CO\,(2-1) emission and absorption, tracing the cold molecular component of the neutral gas, and to provide a stronger constraint on the high frequency continuum and hence spectral age of the source. We also present further ASKAP observations of the \mbox{H\,{\sc i}} absorption line, allowing us to better constrain the kinematics of the line-of-sight atomic gas. We structure the paper as follows. In \autoref{section:PKSB1740-517} we describe the properties of PKS\,B1740$-$517 and its immediate environment. We present a brief description of our observations and data analysis in \autoref{section:observations_data}. We present our results in \autoref{section:results} and in \autoref{section:discussion} we discuss the implications for understanding this radio galaxy and future searches for molecular absorption at cosmological distances. We summarise our conclusions in \autoref{section:conclusions}. In all distance calculations dependent on cosmological parameters  we adopt a flat $\Lambda$ cold dark matter ($\Lambda$CDM) cosmology with $H_{0} = 70$\,km\,s$^{-1}$\,Mpc$^{-1}$, $\Omega_{\rm M} = 0.3$ and $\Omega_{\Lambda} = 0.7$.

\section{PKS B1740-517: A distant young radio galaxy}\label{section:PKSB1740-517}

We summarise the known properties of PKS\,B1740$-$517 and its host galaxy in \autoref{table:properties}. The radio source is bright ($S_{\rm 843\,MHz} = 8.2$\,Jy; \citealt{Mauch:2003}), and resolved by very long baseline interferometry (VLBI; \citealt{King:1994}) into two compact ($\lesssim 10$\,mas) components separated by 52\,mas. The radio spectral energy distribution (SED) is peaked at about 1\,GHz indicating that it is an intrinsically young source and/or strongly confined by a dense ambient medium.

The redshift of $z = 0.44$ (corresponding to a comoving distance of approximately 1700\,Mpc) was first determined using the 21-cm line in absorption and later confirmed through measurement of the optical spectrum (A15). The optical spectrum is typical of an emission-line radio galaxy and exhibits radiatively efficient accretion on to the AGN ($\mathrm{EW}_{\rm [OIII]}$ = $23.73 \pm 0.95$\,\AA, $\mbox{[O\,{\sc iii}]}/\mathrm{H}\,{\beta} = 5.60 \pm 1.43$; A15). The \mbox{[O\,{\sc iii}]} $\lambda 5007$ line has two peaks that are separated by approximately 500\,km\,s$^{-1}$ in radial velocity. These velocity components are spatially resolved by GMOS long-slit observations (A15) into two distinct components separated by 1.3\,kpc, indicating either rotating circumnuclear, or outflowing, ionised gas, on scales extending beyond the central region of the AGN.

\begin{table} 
   \caption{Selected properties of PKS\,B1740-517 and its host galaxy.}\label{table:properties}
 \begin{threeparttable}
\begin{tabular}{lcll}
    \hline
    \multicolumn{1}{l}{Property} & \multicolumn{1}{c}{Value} &
\multicolumn{1}{l}{Unit} & \multicolumn{1}{l}{Reference} \\   
    \hline
	$\rmn{RA}(\rmn{J}2000)$ & $17^{\rmn{h}} 44^{\rmn{m}} 25\fs450$ & & \citet{Fey:2009} \\
    $\rmn{Dec.}(\rmn{J}2000)$ & $-51\degr 44\arcmin 43\farcs79$ & & \citet{Fey:2009} \\
    $\Delta{\theta}_{\rm src}$ & 52 & mas & \citet{King:1994} \\
	$S_{\rm 843\,MHz}$ & 8.2 & Jy & \citet{Mauch:2003} \\ 
    W4\,$(22\,\mu{\rm m})$ & 7.9 & mag & \citet{Wright:2010} \\
    W3\,$(12\,\mu{\rm m})$ & 10.5 & mag & \citet{Wright:2010} \\
    W2\,$(4.6\,\mu{\rm m})$ & 13.3 & mag & \citet{Wright:2010} \\
	W1\,$(3.4\,\mu{\rm m})$ & 14.4 & mag & \citet{Wright:2010} \\
    $K_{\rm s}$ & $15.3$ & mag & \citet{Skrutskie:2006} \\
    $R$ & $20.8$ & mag & \citet{Burgess:2006} \\
	$b_{J}$ & $>21.5$ & mag & \citet{Burgess:2006} \\
    $z$ & 0.4413 & & A15 \\
    $P_{\rm 1.4\,GHz}$ & $3.0 \times 10^{27}$ & W\,Hz\,$^{-1}$ & This work\\
    $L_{\rm 5\,GHz}$ & $6.6 \times 10^{43}$ & erg\,s$^{-1}$ & This work\\
    $L_{\rm 12\,\mu{m}}$ & 5.8 $\times 10^{45}$ & erg\,s$^{-1}$ & This work\\
	$L_{\rm 3.4\,\mu{m}}$ & 3.7 $\times 10^{44}$ & erg\,s$^{-1}$ & This work\\
    $L_{\rm [O\,\sc II]\,\lambda{3727}}$\tnote{$a$} & $1.0 \times 10^{41}$ & erg\,s$^{-1}$ & A15 \\
    $L_{\rm [O\,\sc III]\,\lambda{5007}}$\tnote{$a$} & $3.4 \times 10^{41}$ & erg\,s$^{-1}$ & A15 \\
    $L_{\rm 2 - 10\,keV}$ & $6.2 \times 10^{44}$  & erg\,s$^{-1}$ & A15 \\ 
    $N_{\rm H,X}$ & $1.2 \times 10^{22}$ & cm$^{-2}$ & A15 \\
    $M_{\ast}$ & $\sim 5 \times 10^{10}$ & $\mathrm{M}_{\odot}$ & This work \\
    \hline
  \end{tabular}
   \begin{tablenotes}
   \item[$a$] corrected for Galactic extinction ($A_{\rm v} \approx 0.66$\,mag;  \citealt{Schlegel:1998}). 
   \end{tablenotes}
 \end{threeparttable}
\end{table}

The AGN is X-ray luminous ($L_{\rm 2 - 10\,keV} \approx 6.2 \times 10^{44}$\,erg\,s$^{-1}$) when compared with  other young compact radio galaxies (e.g. \citealt{Tengstrand:2009, Siemiginowska:2016}). Fitting a simple absorption and power law model to the X-ray spectrum reveals an intrinsic absorbing medium with hydrogen column density $N_{\rm H, X} \sim 10^{22}$\,cm$^{-2}$ (A15). However, the photon spectral index is particularly hard ($\Gamma \approx 0.8$), which may indicate either secondary emission from scattering and a Compton-thick medium ($N_{\rm H, X} > 10^{24}$\,cm$^{-2}$), or inverse-Compton scattering in the radio lobes (e.g. \citealt{Ostorero:2010}). 

At near and mid-infrared wavelengths, photometry from the Two Micron All-Sky Survey (2MASS; \citealt{Skrutskie:2006}) and the Wide-Field Infrared Survey Explorer (WISE; \citealt{Wright:2010}) indicates a massive host galaxy with $M_{\ast} \sim 5 \times 10^{10}$\,M$_{\odot}$, assuming a stellar mass-to-light ratio of $M_{\ast}/L \sim 0.6$ (\citealt{Meidt:2014}). The rest frame WISE colours ([3.4]$-$[4.6] = 0.67 and [4.6]$-$[12] = 3.4) are typical for a Seyfert-like AGN. 

In \autoref{figure:PKS1740-517_gemini_alma_image} we show a three-colour optical image of the immediate environment, constructed from $g^{\prime}r^{\prime}i^{\prime}$-band observations with the 8-m Gemini-South telescope by A15. At least two candidate companion galaxies can be seen within a projected distance of approximately 11\,kpc, which are connected to PKS\,B1740$-$527 by a bridge of emission that indicates a possible ongoing tidal interaction. We also show the optical spectrum of one of these candidate companion galaxies, which based on the [O\,{\sc ii}]\,$\lambda 3727$ and blended Mg emission lines has a redshift $z = 0.44080$ that is consistent with that of PKS\,B1740$-$517. There is therefore evidence that this radio galaxy is located within an interacting group, which may have triggered the most recent activity.

\begin{figure*}
\centering
\includegraphics[width=0.35\textwidth]{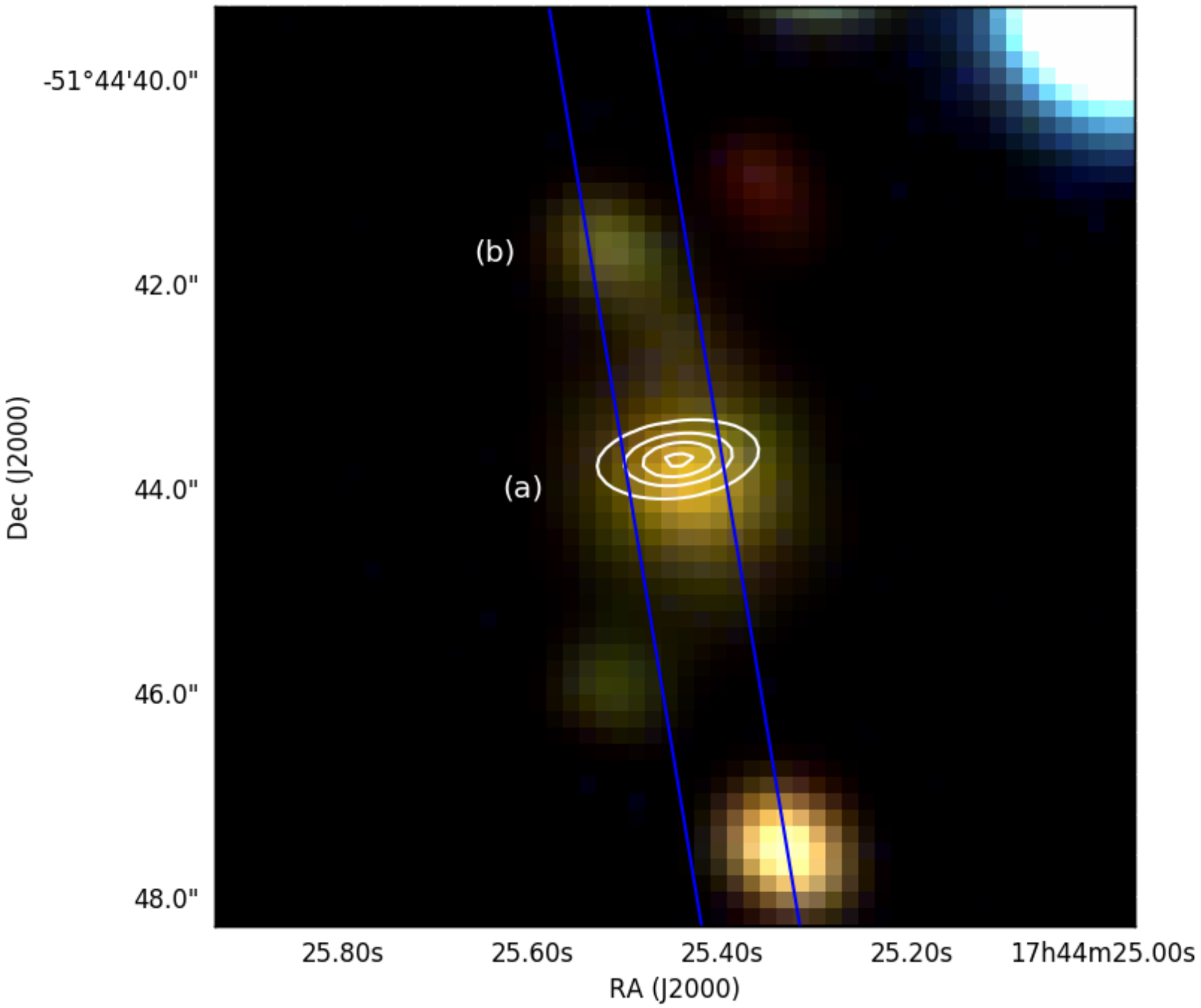}
\includegraphics[width=0.6\textwidth]{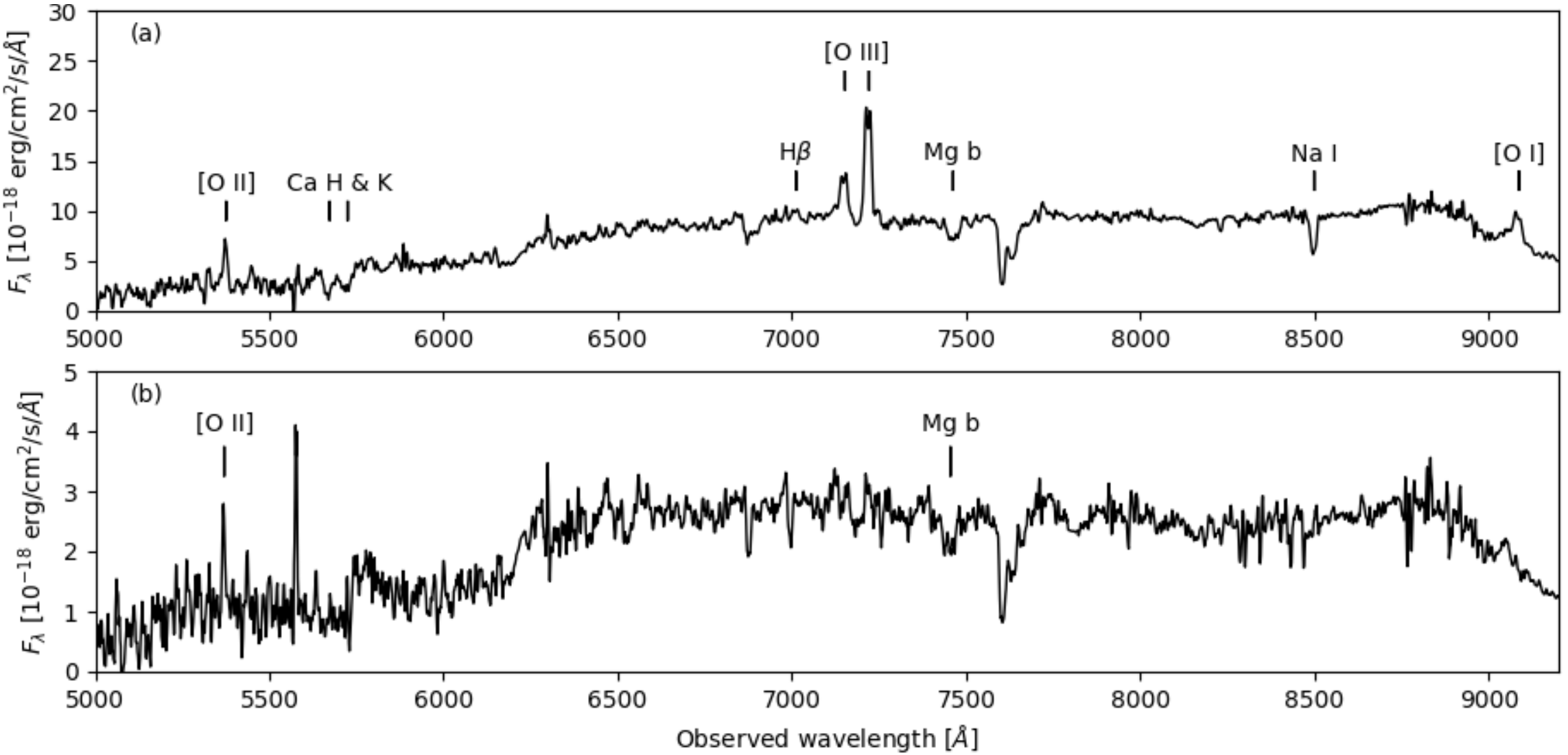}
\caption{Left: A three-colour optical image of PKS\,B1740$-$517 constructed from $g^{\prime}r^{\prime}i^{\prime}$-band observations with the 8-m Gemini-South telescope (A15). At least one candidate companion galaxy (b) is evident, and a possible tidal interaction. To the south east, off the slit, we also see evidence of a possible second companion. The diagonal blue lines indicate the position and width of the slit. Overlaid are ALMA 152\,GHz continuum contours at 20\,per\,cent intervals of the peak continuum ($S_{152\,\rm{GHz}} = 74.3$\,mJy). Right: Gemini spectra extracted at the position of PKS\,B1740$-$517 (a) and candidate companion galaxy (b). Identified spectral lines are as labeled.}\label{figure:PKS1740-517_gemini_alma_image}
\end{figure*}

\section{Observations and Data}\label{section:observations_data}

\subsection{The Atacama Large Millimetre Array}\label{section:alma_observations}

Millimetre wavelength observations of PKS\,B1740$-$517 were carried using ALMA on 2016 July 20 to detect either line emission or absorption from the $J = 1 \rightarrow 2$ rotational transition of Carbon Monoxide ($^{12}$CO). Our observations were prepared and submitted using Cycle-3 Phase-II version of the ALMA Observing Tool\footnote{\url{https://almascience.nrao.edu/documents-and-tools/cycle3/alma-ot-refmanual}}. PKS\,B1740$-$517 was observed for a total of 23\,mins, in three scans each bracketed by 30\,sec observations of a nearby phase calibrator (J1753$-$5015). A further 5\,min observation of J1924-2914 was used for calibration of the flux scale and spectral bandpass. 

We centred one baseband on the expected frequency of 159.9\,GHz for the redshifted $^{12}$CO\,(2-1) line, with $3840 \times 0.488$\,MHz channels ($\Delta{v} \approx 0.9$\,km\,s$^{-1}$) spanning rest-frame radial velocities in the range $v \approx \pm 1760$\,km\,s$^{-1}$. After Hanning smoothing the effective resolution is twice the channel spacing, $\Delta{v} \approx 1.8$\,km\,s$^{-1}$. We used the other three available basebands to provide a sensitive continuum flux density measurement of PKS\,B1740$-$517 at 150\,GHz. These were centred at sky frequencies of 146, 148, and 158\,GHz, each with 128 channels spanning a bandwidth of 1875\,MHz. At 159.9\,GHz, the primary beam full width at half power of the 12-m antennas is 33\,arcsec, and the array configuration, spanning baseline lengths between approximately 15.1\,m and 1.1\,km, is sensitive to angular scales between about 0.35 and 26\,arcsec with an elongation by a factor of approximately 2 in the East-West direction.

Data importing, flagging and calibration were carried out using version 4.5.3 of the {\sc CASA}\footnote{\url{http://casa.nrao.edu}} package (\citealt{McMullin:2007}) and the standard ALMA science pipeline tools\footnote{\url{https://almascience.nao.ac.jp/documents-and-tools/cycle-2/alma-pipeline-reference-manual}}. Using visibility data from the continuum bands we generated a two-term Taylor multi-frequency synthesis image with Uniform weighting, giving a resolution of $0.33 \times 0.76$\,arcsec at 152\,GHz. The continuum emission from PKS\,B1740$-$517 is unresolved by our ALMA observations, and has a total flux density of $72.5\pm0.3$\,mJy at 152\,GHz. However, we note that future observations with ALMA that use the maximum 16\,km antenna baseline ($\Delta{\theta_{\rm psf}} \approx 26$\,mas at 150\,GHz) would be well matched to the known source separation ($\Delta{\theta_{\rm src}} = 52$\,mas at 2.3\,GHz; \citealt{King:1994}).

Imaging of the $^{12}$CO\,(2-1) spectral line band was carried out using two different approaches for detecting absorption and emission, respectively. For detecting $^{12}$CO\,(2-1) absorption towards the continuum source we imaged the visibilities at the full 2\,km\,s$^{-1}$ velocity resolution. Since the continuum source is not resolved with uniform weighting we used natural weighting (with a resolution of $0.5 
\times 1.1$\,arcsec) to maximise detected signal-to-noise, and extracted the spectrum from the cube at the position of peak continuum flux density. We fitted a 1st-order polynomial to the continuum, which was then used to normalise the spectrum to units of percentage absorption. We show the spectrum in \autoref{figure:PKS1740-517_absorption_spectrum}, overlaid with the \mbox{H\,{\sc i}} absorption detected with ASKAP. Although $^{12}$CO\,(2-1) absorption is clearly detected above the RMS noise level of 3.1\,per\,cent per 2\,km\,s$^{-1}$, there is a noticeable absence of molecular gas detected at the velocity of the narrower deeper \mbox{H\,{\sc i}} component. We also do not detect any $^{12}$CO\,(2-1) emission in this spectrum towards the continuum source above the RMS noise level of 1.98\,mJy\,beam$^{-1}$ per 2\,km\,s$^{-1}$. 

For detecting $^{12}$CO\,(2-1) emission within the host galaxy we first imaged and subtracted the continuum directly from the visibilities using {\sc UVSUB}, followed by a first-order polynomial using {\sc UVCONTSUB3} to remove any residual signal. We then formed a data cube in 10\,km\,s$^{-1}$ velocity bins, again using natural weighting to optimise detected signal-to-noise. We extracted the flux density spectrum by summing pixels per spatial resolution element within 1.75\,arcsec (a projected distance of 10\,kpc) of PKS\,B1740$-$517. This spectrum (\autoref{figure:PKS1740-517_emission_spectrum}) shows no evidence of $^{12}$CO\,(2-1) emission above the RMS noise level of 2.10\,mJy per 10\,km\,s$^{-1}$. 

\subsection{The Australian SKA Pathfinder}

\begin{table*} 
   \caption{Summary of our observations using the ASKAP BETA 
     and ASKAP$-$12 telescopes (\citealt{Hotan:2014, McConnell:2016}). Each observation is assigned a unique scheduling block
     identification (SBID) number for that telescope. $t_{\rm int}$ denotes the on-source observing time. $\sigma_{\rm chan}$ is the rms noise per 18.5\,kHz channel at the \mbox{H\,{\sc i}} line, as a fraction of the continuum flux density. $S_{\rm cont}$ is the total continuum flux density measured using the full bandwidth.}\label{table:askap_observations}
 \begin{threeparttable}
  \begin{tabular}{lllllccccc}
    \hline
    \multicolumn{1}{l}{SBID}  &
\multicolumn{1}{l}{Date} & \multicolumn{1}{c}{Frequency\,band} & \multicolumn{1}{l}{ASKAP antennas\tnote{$a$}} &
\multicolumn{1}{c}{$t_\mathrm{int}$} & \multicolumn{1}{c}{$\sigma_{\rm chan}$} & \multicolumn{1}{c}{$S_{\rm cont}$} \\
\multicolumn{1}{c}{} & \multicolumn{1}{c}{} & \multicolumn{1}{c}{[MHz]} & \multicolumn{1}{c}{} & \multicolumn{1}{c}{[h]} & \multicolumn{1}{c}{[per\,cent]} & \multicolumn{1}{c}{[Jy]} \\   
    \hline
    ASKAP-BETA & & & & \\
    93\tnote{$b$} & 2014 June 24 & 711.5 $-$ 1015.5 & 1, 8, 9, 15 & 11.5 & 0.24 & 8.0 \\ 
    315\tnote{$b$} & 2014 August 03 & 711.5 $-$ 1015.5 & 1, 3, 6, 8, 15 & 3.0 & 0.59 & 7.9 \\ 
    514\tnote{$b$} & 2014 September 01 & 711.5 $-$ 1015.5 & 1, 6, 8, 15 & 11.9 & 0.20 & 7.4 \\ 
    1808 & 2015 May 14 & 711.5 $-$ 1015.5 & 1, 3, 6, 8, 9, 15 & 12.0 & 0.17 & 7.4 \\ 
    1994 & 2015 June 25 & 711.5 $-$ 1015.5 & 1, 6, 8, 15 & 7.5 & 0.80 & 8.5 \\             
    2009 & 2015 June 27 & 711.5 $-$ 1015.5 & 1, 6, 8, 15 & 11.0 & 0.47 & 6.6 \\             
    3073 & 2015 November 09 & 711.5 $-$ 1015.5 & 1, 3, 6, 8, 15 & 6.0 & 0.32 & 6.8 \\
    3085 & 2015 November 11 & 711.5 $-$ 1015.5 & 1, 3, 6, 8, 15 & 12.0 & 0.22 & 6.9 \\
    3086 & 2015 November 12 & 711.5 $-$ 1015.5 & 1, 3, 6, 8, 15 & 12.0 & 0.22 & 6.8 \\
    3088 & 2015 November 13 & 711.5 $-$ 1015.5 & 1, 3, 6, 8, 15 & 12.0 & 0.22 & 6.9 \\
    & & & & & & \\
    ASKAP-12 & & & & & & & \\
    3304 & 2017 January 31 & 799.4 $-$ 1039.4 & 1, 2, 4, 5, 6, 10, 12, 13, & 2.0 & 0.16 & 8.3 \\
    & & & 14, 19, 24, 26, 27, 28 & & & \\
    \hline
  \end{tabular}
   \begin{tablenotes}
   \item[$a$] {See \citet{Hotan:2014} and Lee-Waddell et al. in preparation for details of the ASKAP-BETA and ASKAP-12 antenna positions.} 
   \item[$b$] {Originally published by \citet{Allison:2015a}.}
   \end{tablenotes}
 \end{threeparttable}
\end{table*}

Further 21-cm observations of PKS\,B1740$-$517 were undertaken as part of ASKAP commissioning and early science verification. We used two developmental stages of this telescope -- the six-antenna Boolardy Engineering Test Array (ASKAP-BETA) and ASKAP-12 (\citealt{Hotan:2014, McConnell:2016}) -- which differ principally in the design of their Phased Array Feeds (PAFs; \citealt{Hay:2008}) and backend. ASKAP-12 is equipped with new Mark II PAFs, which have improved sensitivity at 1400\,MHz (\citealt{Chippendale:2015}) and will be used for the completed ASKAP telescope. We used the same observing strategy as described by A15, whereby we pointed the telescope at PKS\,B1740$-$517 and maximum S/N PAF beams were electronically formed in an on-sky pattern that included a central beam on the source. For each observation we included a short 15\,min scan of PKS\,B1934$-$638 to calibrate the flux scale (using the model of \citealt{Reynolds:1994}) and to obtain initial solutions for the antenna gains. Further self-calibration, continuum-subtraction and imaging of the target field were carried out using the \texttt{MIRIAD}\footnote{\url{http://www.atnf.csiro.au/computing/software/miriad/}} package (\citealt{Sault:1995}) as described by A15, but using a higher 1\,MHz resolution to form the coarse-channel data (see \citealt{Allison:2017}). 

Increased backend capacity means that ASKAP-12 forms PAF beams at 1\,MHz intervals (c.f. 4 and 5\,MHz intervals for ASKAP-BETA). The element weights used to form PAF beams are applied uniformly in these intervals before channelization at full 18.5\,kHz resolution. This process creates discontinuous jumps in amplitude and phase as a function frequency, which typically dominate the bandpass on these scales. In the case of ASKAP-BETA, these intervals in frequency equate to radial velocities $\Delta{v} > 1000$\,km\,s$^{-1}$ at $z_{\rm HI} = 0.44$ and so on these scales we can simply fit and subtract any residual continuum signal in the target spectrum using UVLIN (see A15). However,  for the smaller beam forming intervals of ASKAP-12, this approach would lead to subtraction of real spectral features with linewidths $\Delta{v} > 300$\,km\,s$^{-1}$ at $z_{\rm HI} = 0.44$. In the case of PKS\,B1740$-$517 we did not detect any significant features in the ASKAP BETA spectra with these linewidths. Furthermore, since the ASKAP-12 data contribute only an additional 11\,per\,cent to the S/N in the average spectrum, this affect is unlikely to influence detection of broad features that already exist in the BETA data.

We summarise our observations in \autoref{table:askap_observations}. The source is not spatially resolved by either ASKAP-BETA or ASKAP-12. The averaged spectrum (shown in \autoref{figure:PKS1740-517_absorption_spectrum}) was constructed to have optimal S/N by weighting each observation by its inverse variance, with a resulting RMS fractional continuum absorption 
of $0.07$\,per\,cent per 5.6\,km\,s$^{-1}$ at the \mbox{H\,{\sc i}} line. Two distinct velocity components are seen in \mbox{H\,{\sc  i}} absorption. The third component, a possible broad $\Delta{v} \approx 300$\,km\,s$^{-1}$ outflow reported by A15, remains tentative, and corroborating evidence from either deeper observations or other gas transitions are needed to resolve its authenticity.

\section{Results}\label{section:results}

\subsection{Constraints on the line-of-sight cold gas}

Using ALMA we have detected $^{12}$CO\,(2-1) absorption in the host galaxy of PKS\,B1740$-$517. In \autoref{figure:PKS1740-517_absorption_spectrum} we show the ALMA spectrum overlaid with the \mbox{H{\sc i}} absorption detected with ASKAP. Since information about the gas kinematics encoded in the spectrum is limited by the spatial extent of the continuum source, it is difficult to determine the three dimensional motion of the gas from the unresolved absorption spectrum alone. However, the absorption lines allow us to probe the neutral gas at small scales towards to the AGN and we can use the known continuum source morphology, the line-of-sight gas kinematics, and simple priors for the gas distribution to infer likely models. In the following we discuss our interpretation of the features seen in the ALMA and ASKAP spectra, and how the former has allowed us to further constrain our understanding of the cold gas in the host galaxy.

\subsubsection{The broad absorption - a circumnuclear reservoir}

The \mbox{H\,{\sc i}} absorption in PKS\,B1740$-$517 has two distinct spectral components that are separated by a radial velocity of $\Delta{v} \approx 115$\,km\,s$^{−1}$. The broader component has a FWHM of approximately 60\,km\,s$^{-1}$ and a redshift that is consistent with the centroid of the double-peaked [O\,{\sc iii}]\,$\lambda{5007}$ emission line, indicating that the gas motion is symmetric about the galactic nucleus. The coincident $^{12}$CO\,(2-1) line has a similar width and is therefore unlikely to arise in a single molecular cloud; the velocity dispersion is significantly larger than observed in any Galactic molecular cloud and the inferred linear size would be larger than a few kiloparsecs (\citealt{Larson:1981, Solomon:1987, Heyer:2009}). It is more likely that the CO absorption is arising from several blended spectral features from individual parsec-scale molecular clouds embedded in a larger gaseous structure with a velocity gradient. 

\begin{figure}
\centering
\includegraphics[width=0.5\textwidth]{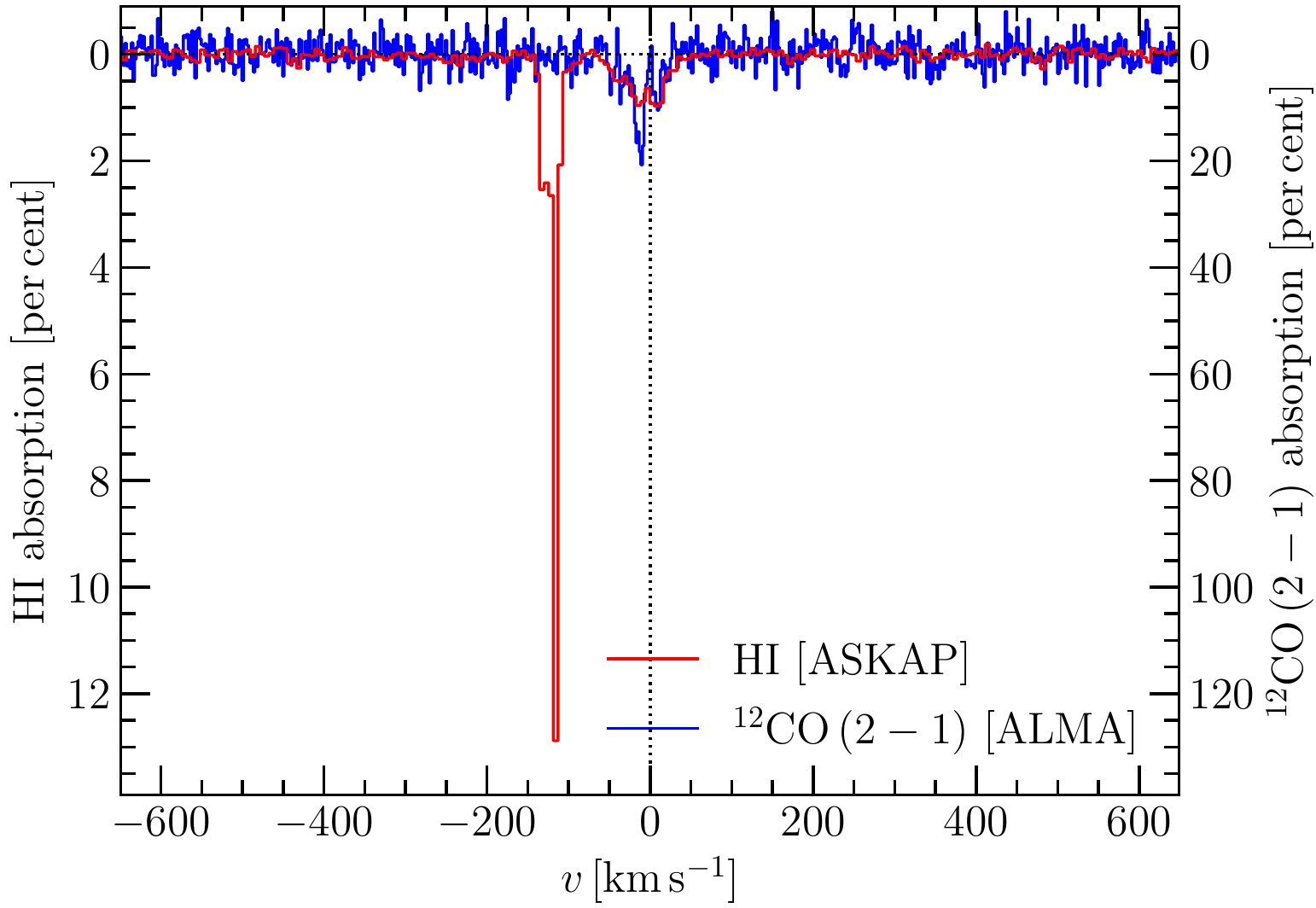}
\caption{ALMA spectrum of PKS\,B1740$-$517 showing $^{12}$CO\,(2-1) absorption towards the radio-loud AGN. Superposed is the average ASKAP 21-cm spectrum showing \mbox{H\,{\sc i}} absorption. The data are given as a fraction of the total continuum flux density measured at that transition. The ALMA and ASKAP data have been binned to velocity resolutions of 2.0 and 5.6\,km\,s$^{-1}$, respectively. The velocity axis is given in the rest frame defined by the broader \mbox{H\,{\sc i}} and CO absorption (at $z = 0.44185$), which is consistent with the AGN rest frame defined by the centroid of the double-peaked [O\,{\sc iii}]\,$\lambda{5007}$ emission line (see A15).}\label{figure:PKS1740-517_absorption_spectrum}
\end{figure}

In \autoref{figure:PKS1740-517_absorption_zoom} we show in the detail the absorption spectrum, which reveals sub-structure in both the \mbox{H\,{\sc i}} and $^{12}$CO\,(2-1) lines centred at $v \approx 0$\,km\,s$^{-1}$. The molecular absorption possibly has more sub-structure, albeit at higher resolution and lower signal-to-noise; several individual components are apparent with widths $\Delta{v} \sim 10$\,km\,s$^{-1}$. If the molecular gas is optically thick to $^{12}$CO\,(2-1) then the seemingly low absorbed fraction of the total continuum (i.e. less than 100\,per\,cent) is likely the result of patchy obscuration of the background source as a function of radial velocity (see e.g. \citealt{Wiklind:1997}). This would be consistent with CO absorption through a line-of-sight distribution of discrete cold molecular cores, while the \mbox{H\,{\sc i}} absorption arises from more diffuse cold-phase atomic structures as observed in the ISM of the Milky Way and Local Group galaxies (see e.g. \citealt{Heiles:2003b, Gibson:2005, Braun:2009, Braun:2012}). 

\begin{figure}
\centering
\includegraphics[width=0.5\textwidth]{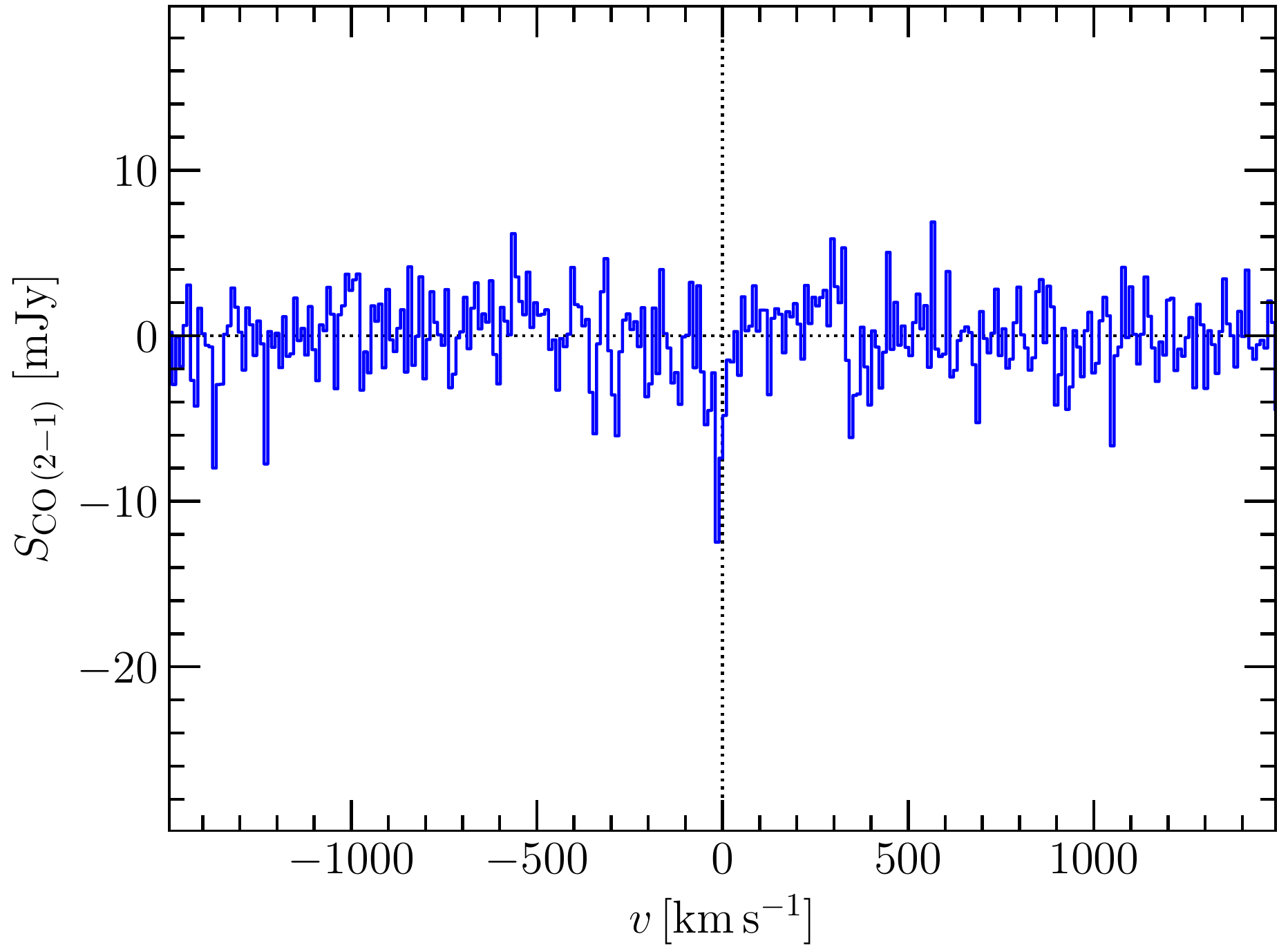}
\caption{ALMA spectrum showing no evidence of CO\,(2-1) emission within a projected distance of 10\,kpc of PKS\,B1740$-$517. The data have been binned to a velocity resolution of 10\,km\,s$^{-1}$. The velocity axis is as defined in \autoref{figure:PKS1740-517_absorption_spectrum}.} \label{figure:PKS1740-517_emission_spectrum}
\end{figure}

From very long baseline interferometry at 2.3\,GHz (\citealt{King:1994}) we know that the source components are separated in projection by approximately $300$\,pc and each has a size less than approximately $50$\,pc, which are thus likely to be fully subtended by any coherent structure in the foreground interstellar medium (ISM). Assuming uniformly distributed optically thick CO clouds ($\tau > 1$) of radius $r_{\rm mol}$ and filling factor $f_{\rm mol}$ within a roughly spherical volume of characteristic radius $R$, the expected areal covering fraction (allowing for overlapping clouds) is then given by 
\begin{equation}\label{equation:covering_fraction}
\langle{c_{\rm mol}}\rangle \approx 1 - \left[1 -\frac{3}{4}\left(\frac{r_{\rm mol}}{R}\right)^{2}\right]^{n_{\rm mol}},
\end{equation}
where 
\begin{equation}
n_{\rm mol} = f_{\rm mol}\,\left(\frac{R}{r_{\rm mol}}\right)^{3}.
\end{equation}
For the Milky Way ISM, $f_{\rm mol} \lesssim 1$\,per\,cent and $r_{\rm mol} \sim 10$\,pc, gives $\langle{c_{\rm mol}}\rangle \lesssim 86$\,per\,cent for $R \sim 2$\,kpc (typical of the radial extent of molecular discs in massive early type galaxies; e.g. \citealt{Davis:2013}). Assuming that the molecular gas has a dispersion per opaque cloud of $\sim$\,4\,km\,s$^{-1}$ leads to a covering fraction of less than $\sim 10$\,per\,cent over a total line width of $\sim 50$\,km\,s$^{-1}$, which is consistent with the absorbed fraction of total continuum flux density seen in the ALMA spectrum.

The \mbox{H\,{\sc i}} absorption, which is assumed to trace the more diffuse cold gas, appears to have two prominent peaks, separated by a radial velocity of $\Delta{v} \approx 20$\,km\,s$^{-1}$. Such symmetric double-peaked structure could arise from a gaseous disc, or ring, which is illuminated by continuum emission from the compact-double radio source. Given that the flux densities of the source components are in the ratio $\sim$\,4:1 (\citealt{King:1994}) it may seem surprising that the optical depths of the two peaks are almost equal. However several effects, including the detailed radial velocity distribution of the HI gas in front of each source (including turbulence) and differences in the physical conditions of the gas at different positions, could easily account for this.

Using the separation of the velocity peaks ($\Delta{v}$) and source components ($\Delta{d}_{\rm src}$) we can then obtain an approximate estimate of the line-of-sight extent of the neutral gas ($\Delta{d}_{\rm gas}$). For gas moving in a circular orbit with velocity $v_{\rm circ}$ with respect to the galactic centre, the radial velocity separation observed in the absorption spectrum is given by (see \autoref{figure:schematic})
\begin{equation}
\Delta{v} \approx v_{\rm circ}\cos{(i)}\cos{(\phi)}\cos{(\theta)},
\end{equation}
where
\begin{equation}
\cos{(\theta)} = \frac{\Delta{d}_{\rm src}}{\Delta{d}_{\rm gas}},
\end{equation}
and $i$ and $\phi$ are angles of the plane of rotation with respect to our sight line and the projected source-AGN axis, respectively. We can then infer the line-of-sight extent of the absorbing gas by rearranging this expression, as follows
\begin{equation}
\Delta{d}_{\rm gas} \approx \frac{v_{\rm circ}}{\Delta{v}}\cos{(i)}\cos{(\phi)}\,\Delta{d}_{\rm src}.
\end{equation}
Following this simple geometrical argument, large radial velocities imply that the absorbing gas is close to the radio source, while gas that is much further away will have an observed radial velocity close to zero. Deviations from a regular circular orbit will result in an error in the inferred radial distance of the absorbing gas. Given that $\Delta{d_{\rm src}} \sim 300$\,pc, $\Delta{v} \sim 20$\,km\,s$^{-1}$, $0 \leq i \leq 90$\degr and $0 \leq \phi \leq 90$\degr, and assuming $v_{\rm circ} \sim 200$\,km\,s$^{-1}$, we estimate that the illuminated absorbing gas lies radially within $\Delta{d}_{\rm gas} \sim 2$\,kpc of the AGN. A15 found similar double-peaked structure in the [O\,{\sc iii}]\,$\lambda$5007 emission line, separated by $\Delta{v} \sim 500$\,km\,s$^{-1}$ and resolved into two spatially distinct components separated by 1.3\,kpc, indicating a possible ionised counterpart to the neutral gas seen in absorption. The neutral and ionised gas detected towards PKS\,B1740$-$517 are thus consistent with a disc or ring within a few kiloparsecs of the AGN.

\begin{figure}
\centering
\includegraphics[width=0.5\textwidth]{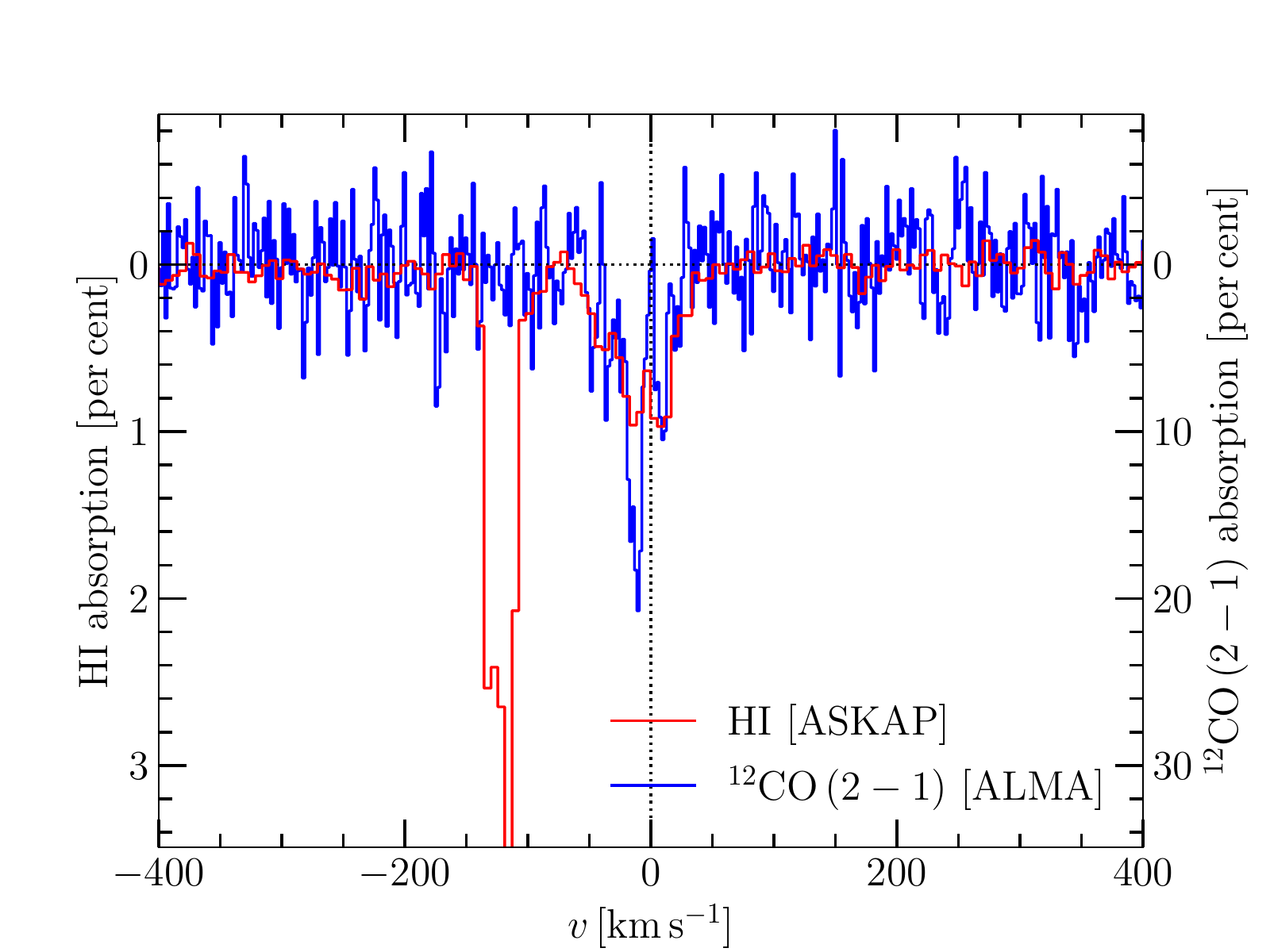}
\caption{An enlarged version of the absorption spectrum shown in \autoref{figure:PKS1740-517_absorption_spectrum}, highlighting the velocity structure at $v \sim 0$\,km\,s$^{-1}$ that may indicate illumination of a gaseous ring or disc.}\label{figure:PKS1740-517_absorption_zoom}
\end{figure}

\begin{figure}
\centering
\includegraphics[width=0.4\textwidth]{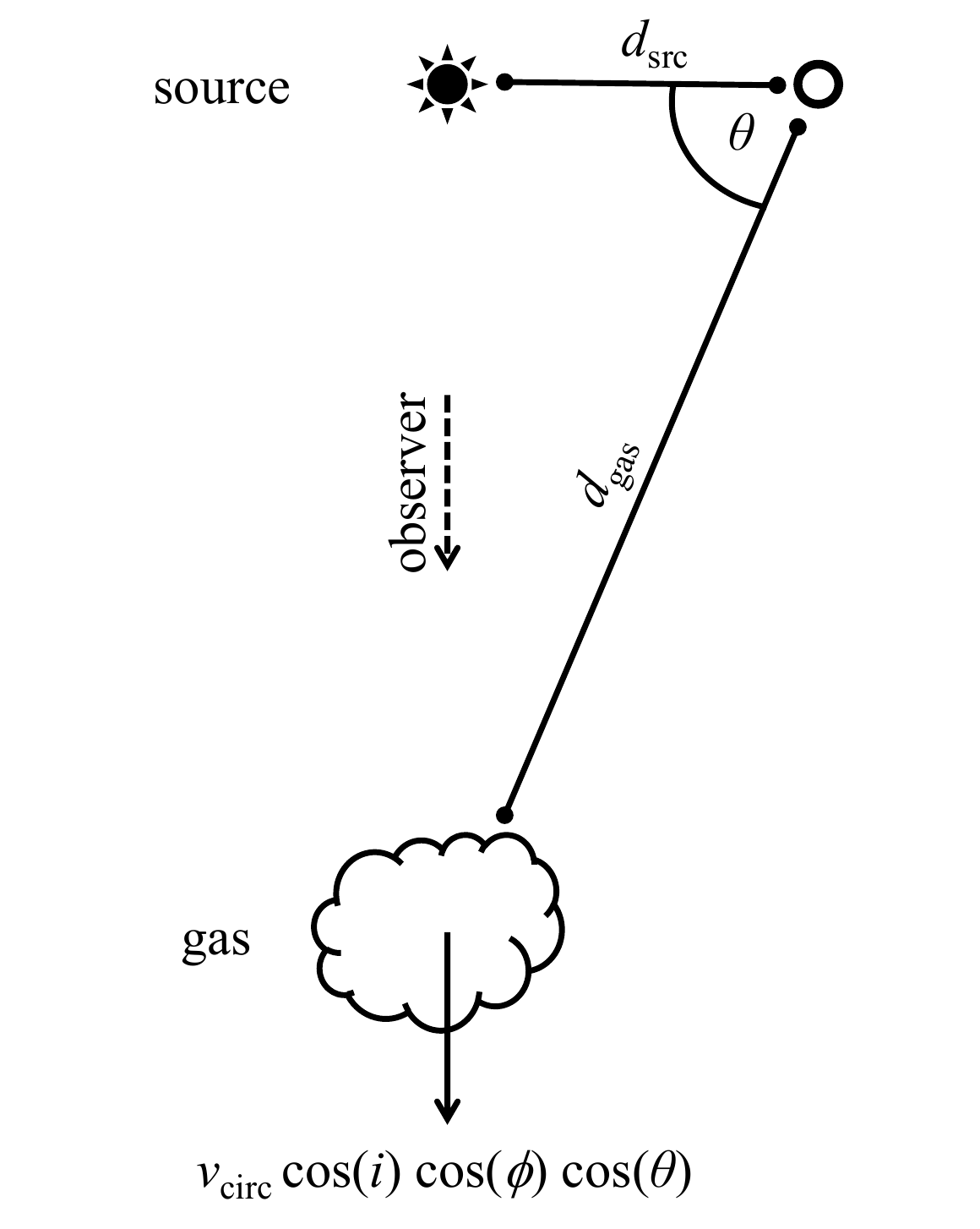}
\caption{A schematic diagram (not to scale) showing the relative positions of the AGN (origin), radio source component (star) and absorbing gas (cloud). The gas is assumed to be moving in a circular orbit with velocity $v_{\rm circ}$. The plane of rotation is centred on the origin and orientated with inclination angle $i$ to the line-of-sight and position angle $\phi$ with respect to the radio source axis. The component of velocity in the direction of the observer is indicated.}\label{figure:schematic}
\end{figure}

We can estimate the column density of neutral atomic and molecular gas from the integrated absorption line strength in the ASKAP and ALMA spectra. In general the column density of a gaseous species X populating an upper level $u$ of a transition $u \rightarrow l$, with optical depth $\tau_{{\rm X}(ul)}$ integrated over rest frame velocity $v$, is given by 
\begin{equation}
N_{{\rm X}(u)} = \frac{8\pi}{c^{3}}\frac{1}{A_{ul}}\frac{\nu_{ul}^{3}}{e^{h\nu_{ul}/k_{\rm B}T_{\rm ul}}-1}\int{\tau_{{\rm X}(ul)}\,\mathrm{d}{v}},
\end{equation}
where $\nu_{ul}$ is the rest frequency, $A_{ul}$ is the Einstein coefficient for spontaneous emission and $T_{\rm ul}$ is the excitation temperature. The total column density for species X is calculated by multiplying this by the partition function and dividing by the relative population of the upper level $u$, giving 
\begin{equation}
N_{\rm X} = \frac{8\pi}{c^{3}}\frac{\nu_{ul}^{3}}{g_{u}}\frac{f(T_{\rm ul})}{A_{ul}}\int{\tau_{{\rm X}(ul)}\,\mathrm{d}{v}},
\end{equation}
where 
\begin{equation}
f(T_{\rm ul}) = \frac{Q(T_{\rm ul})\,e^{E_{l}/k_{\rm B}T_{\rm ul}}}{1-e^{-h\nu_{ul}/k_{\rm B}T_{\rm ul}}},	
\end{equation}
$g_{u}$ is the statistical weight of $u$, $E_{l}$ is the lower energy level, and $Q(T_{\rm ul})$ is the partition function assuming a single excitation temperature (e.g. \citealt{Wiklind:1995}). 

For the 21-cm line of \mbox{H\,{\sc i}}, $h\nu_{ul} \ll k_{\rm B} T_{\rm ul}$ and $A_{ul} = 2.85 \times 10^{-15}$\,s$^{-1}$, giving
\begin{equation}
N_{\rm HI} \approx 1.823\times10^{20}\left[{\frac{T_{\rm spin}}{100\,\mathrm{K}}}\right]\,\left[{\frac{\int{\tau_{\rm HI}\,\mathrm{d}{v}}}{1\,\mathrm{km}\,\mathrm{s}^{-1}}}\right]\,\mathrm{cm^{-2}},
\end{equation}
where $T_{\rm spin}$ is the excitation temperature for the spin-flip transition. Since our observations with ASKAP and ALMA do not resolve the spatial structure of PKS\,B1740$-$517, we can only estimate the average optical depth over the source cross-section, giving $\int{\tau_{\rm HI}\,\mathrm{d}{v}} \approx 0.549 \pm 0.019$\,km\,s$^{-1}$. If the \mbox{H\,{\sc i}} does not subtend all of the source then this can underestimate the true column of atomic gas along our line-of-sight. However, we  note that the source components have an angular size less than 10\,mas, which corresponds to a physical scale of approximately 57\,pc at $z = 0.44$. This is similar to the scales typically measured for coherent absorbing \mbox{H\,{\sc i}} structures resolved in studies of the Local Group (e.g. \citealt{Braun:2012} and references therein) and distant galaxies (e.g. \citealt{Lane:2000,Gupta:2012,Srianand:2013,Borthakur:2014,Biggs:2016,Dutta:2016,Gupta:2018}), implying that the covering factor may well be close to unity.

In the absence of independent measurements of the \mbox{H\,{\sc i}} column density from Lyman-$\alpha$ absorption or 21-cm line emission, a canonical spin temperature of $T_{\rm spin} \sim 100$\,K is often adopted based on the typical value for \mbox{H\,{\sc i}} in the Milky Way (e.g. \citealt{Heiles:2003b}). However, given the proximity of the atomic gas to PKS\,B1740$-$517, a luminous AGN that is a strong source of continuum radiation both at 21-cm and Lyman-$\alpha$ wavelengths, the spin temperature could be significantly larger than this value (\citealt{Bahcall:1969}). For example, \cite{Holt:2006} used an independent estimate of the neutral gas column from extinction to derive a spin temperature lower limit of 3000\,K in absorbing gas in the host galaxy of the obscured quasar PKS\,1549$-$79. For PKS\,B1740$-$517, A15 compared the total hydrogen column density from the X-ray spectrum and the total integrated \mbox{H\,{\sc i}} 21-cm optical depth from the ASKAP spectrum to estimate an upper limit for the spin temperature of $T_{\rm spin} \lesssim 2430 \pm 1050$\,K. We therefore consider spin temperatures in the range $100$ to $2000$\,K, giving an average \mbox{H\,{\sc i}} column density in front of the source in the range $N_{\rm HI} \sim 0.1 - 2 \times 10^{21}$\,cm$^{-2}$. Assuming that the atomic gas has a scale height $h \sim 150$\,pc (so that the disc obscures both source components) and a radial extent $\sim 2$\,kpc from the AGN, we obtain an enclosed mass for \mbox{H\,{\sc i}} in the range $M_{\rm HI} \sim 0.3 - 6 \times 10^{7}$\,M$_{\odot}$.

For the $J = 2\rightarrow{1}$ rotational transition of $^{12}$CO, $A_{ul} = 7.36 \times 10^{-7}$\,s$^{-1}$ (e.g. \citealt{Chandra:1996}) and we can use the rigid-rotor approximation, so that $h\,\nu_{ul}/k_{\rm B} \approx 2(J_l+1)B = 4B$ and $E_{l}/k_{\rm B} \approx J_l(J_l+1)B = 2B$, where the rotational constant $B \approx 2.766$\,K. In the Milky Way ISM the typical temperature of the CO gas is $\sim 16$\,K (\citealt{Heyer:2009}). However, close to a radio-loud active galactic nucleus the CO could be excited to higher temperatures through shocks and turbulence in the gas (e.g. \citealt{Papadopoulos:2008, Dasyra:2012}). Under the assumption that $T_{\rm ul} \gg B$, the partition function can be approximated by $Q(T_{\rm ul}) \approx T_{\rm ul}/B$. Assuming an excitation temperature $T_{\rm ul} \gtrsim 10$\,K, we obtain a $^{12}$CO column density of 
\begin{equation}
N_{\rm CO} \gtrsim 1.5 \times 10^{16} \, \left[{\frac{\int{\tau_{\rm CO}\,\mathrm{d}{v}}}{5\,\mathrm{km}\,\mathrm{s}^{-1}}}\right]\,\mathrm{cm^{-2}}.
\end{equation}
As with the \mbox{H\,{\sc i}} absorption, the $^{12}$CO\,(2-1) optical depth integrated over the line shown in \autoref{figure:PKS1740-517_absorption_spectrum} is an average over the total source cross-section. However, unlike the \mbox{H\,{\sc i}}, the molecular gas is distributed in clumpy dense clouds, which significantly reduces the covering fraction against the continuum source. Given this uncertainty about the covering fraction we obtain a lower limit for the integrated optical depth of $\int{\tau_{\rm CO}\,\mathrm{d}{v}} \gtrsim 4.78 \pm 0.47 $\,km\,s$^{-1}$. We estimate a lower limit to the average CO column density in front of the source of $1.5 \times 10^{16}$\,cm$^{-2}$ and, for a Milky Way $^{12}$CO-to-H$_{2}$ abundance ratio in dense clouds of $Z_{\rm CO} \sim 1 \times 10^{-4}$ (e.g. \citealt{Sofia:2004}), a corresponding cold H$_{2}$ column density lower limit of $N_{\rm H_{2}} \gtrsim 1.5 \times 10^{20}$\,cm$^{-2}$. We note that the $^{12}$CO-to-H$_{2}$ abundance ratio can be two orders of magnitude lower in the diffuse interstellar medium than in dense clouds, in which case the total H$_{\rm 2}$ column density in front of the radio AGN could be much higher than this lower limit (\citealt{Burgh:2007}).

We can compare this lower limit with the non-detection of $^{12}$CO\,(2-1) emission on the same sight-line to see if they are consistent. Assuming optically thick CO emission from the molecular clouds, and $L_{\rm CO(2-1)} = L_{\rm CO(1-0)}$, the H$_{\rm 2}$ column density is given by 
\begin{equation}
N_{\rm H_2} \approx 1.22 \times 10^{3} \,X_{\rm CO}\,(\Delta\theta_{\rm maj}\,\Delta\theta_{\rm min})^{-1}\,\nu_{\rm obs}^{-2}\,\int{S_{\rm CO}\,\mathrm{d}v}\,\mathrm{cm^{-2}},
\end{equation}
where $X_{\rm CO}$ is the CO-to-H$_{2}$ conversion factor, in units of $\mathrm{cm^{-2}\,[K\,km\,s^{-1}]^{-1}}$, $\Delta\theta_{\rm maj}$ and $\Delta\theta_{\rm min}$ are the major and minor axes of the spatial region from which the spectrum was measured, in units of arcsec, $\nu_{\rm obs}$ is the observed frequency in GHz, and $\int{S_{\rm CO}\,\mathrm{d}v}$ is the line integrated flux density in units of mJy\,km\,s$^{-1}$. For the ALMA spectrum shown in \autoref{figure:PKS1740-517_absorption_spectrum} and \autoref{figure:PKS1740-517_absorption_zoom}, the spatial resolution is $0.5 \times 1.1$\,arcsec and the observed frequency is $159.9$\,GHz. We assume a line width of $400$\,km\,s$^{-1}$ for the undetected line (based on typical rotation velocities of molecular discs in massive galaxies; \citealt{Helfer:2003, Davis:2013}) and measure an RMS of 1.98\,mJy per 2\,km\,s$^{-1}$, giving a 5-$\sigma$ detection limit for the line integrated flux density of $279$\,mJy\,km\,s$^{-1}$. However, the detected absorption will act to disguise any possible emission, effectively lowering our detection sensitivity to emission over that region of the spectrum. We take the somewhat conservative approach and add the missing integrated flux density in the absorbed region of the spectrum, equal to 300\,mJy\,km\,s$^{-1}$, giving an upper limit for undetected emission of $\int{S_{\rm CO}\,\mathrm{d}v} < 579$\,mJy\,km\,s$^{-1}$. 

The $X_{\rm CO}$ conversion factor depends on the physical properties of the molecular gas, including density, temperature and metallicity  (see \citealt{Bolatto:2013} for a review).  At solar metallicities in the inner disc of the Milky Way, $X_{\rm CO} \approx 2 \times 10^{20} \mathrm{cm^{-2}\,[K\,km\,s^{-1}]^{-1}}$, but can take significantly lower values in the low density turbulent gas of ultraluminous infrared galaxies ($X_{\rm CO} \approx 0.4 \times 10^{20} \mathrm{cm^{-2}\,[K\,km\,s^{-1}]^{-1}}$), or higher values in low metallicity environments. A sub-solar CO-to-H$_{2}$ conversion factor could be expected for molecular gas near to the radio source, where heightened temperatures and turbulence could lead to optically thin CO emission (as for example the jet-driven molecular outflows seen in IC\,5063; {\citealt{Oosterloo:2017}). Without further information we use $X_{\rm CO}$ for molecular gas in the Milky Way, which gives $N_{\rm H_2} \lesssim 1 \times 10^{22}$\,cm$^{-2}$, consistent with the lower limit estimated from the absorption line. Using the same disc model as the atomic gas, we use our two limits on the $H_{2}$ column density to calculate an enclosed mass for cold molecular gas in the range $M_{\rm H_2} \sim 0.9 - 60 \times 10^{7}$\,M$_{\odot}$. 

We note that both the \mbox{H\,{\sc i}} and \mbox{H$_{2}$} masses inferred here are consistent with the lower end of the mass distributions for nearby massive early-type galaxies with detected neutral gas (e.g. \citealt{Young:2011,Serra:2012}). However, we stress that our estimates are only for line-of-sight gas detected in absorption towards the radio source, and may underestimate the total masses in the host galaxy. In \autoref{figure:PKS1740-517_emission_spectrum} we show that no $^{12}$\,CO(2-1) emission is detected within a projected distance of 10\,kpc of PKS\,B1740$-$517 to the same level as seen towards the nucleus. It it is possible that most of the cold gas detected in absorption was accreted from its companion through tidal interaction, in which case a future gas-rich minor merger may increase the gas mass significantly (e.g. \citealt{Davis:2015}). 

The \mbox{H\,{\sc i}} and \mbox{H$_{2}$} column densities equate to a total gas surface mass density in the range $\Sigma_{\rm gas} \sim 4.4 - 221$\,M$_{\odot}$\,pc$^{-2}$ (with a  37\,per\,cent correction for helium and metals). The Kennicutt-Schmidt law (\citealt{Kennicutt:1998}) then gives a star formation rate surface density in the range $\Sigma_{\rm SFR} \sim 0.002 - 0.5$\,M$_{\odot}$\,yr$^{-1}$\,kpc$^{-2}$ in the direction towards the AGN. We separately estimate the total star formation rate from the O\,[{\sc ii}]\,$\lambda$3727 emission line (see \autoref{figure:PKS1740-517_gemini_alma_image}) using the following relationship given by \cite{Kewley:2004},
\begin{equation}\label{equation:OII_to_SFR}
\left[\frac{\mathrm{SFR}}{\mathrm{M_{\odot}\,yr^{-1}}}\right] = (6.58 \pm 1.65) \left[\frac{L_{\rm [OII]}}{10^{42}\,\mathrm{erg\,s^{-1}}}\right].
\end{equation}
This is an upper limit since there is likely to be significant contribution to the emission line from AGN-driven ionisation. The O\,[{\sc ii}]\,$\lambda$3727 luminosity, corrected for Galactic extinction, is $L_{\rm [OII]} \approx 1.01 \times 10^{41}$\,erg\,s$^{-1}$, giving a total SFR $\lesssim 0.66$\,M$_{\odot}$\,yr$^{-1}$. The spectrum was extracted over a region of 0.7\,arcsec (equal to 4\,kpc at $z = 0.44$), giving an average SFR surface density upper limit of $\Sigma_{\rm SFR} \lesssim 0.05$\,M$_{\odot}$\,yr$^{-1}$\,kpc$^{-2}$. We therefore do not find evidence for any significant ongoing star formation (i.e. starburst behaviour) in PKS\,B1740$-$517. This may not be surprising given both the timescales involved in triggering the AGN following an interaction and the suppression by several orders magnitude of star formation efficiency seen in minor mergers involving massive early type galaxies at lower redshifts (e.g. \citealt{VanDeVoort:2018}).

\begin{figure*}
\centering
\includegraphics[width=1.0\textwidth]{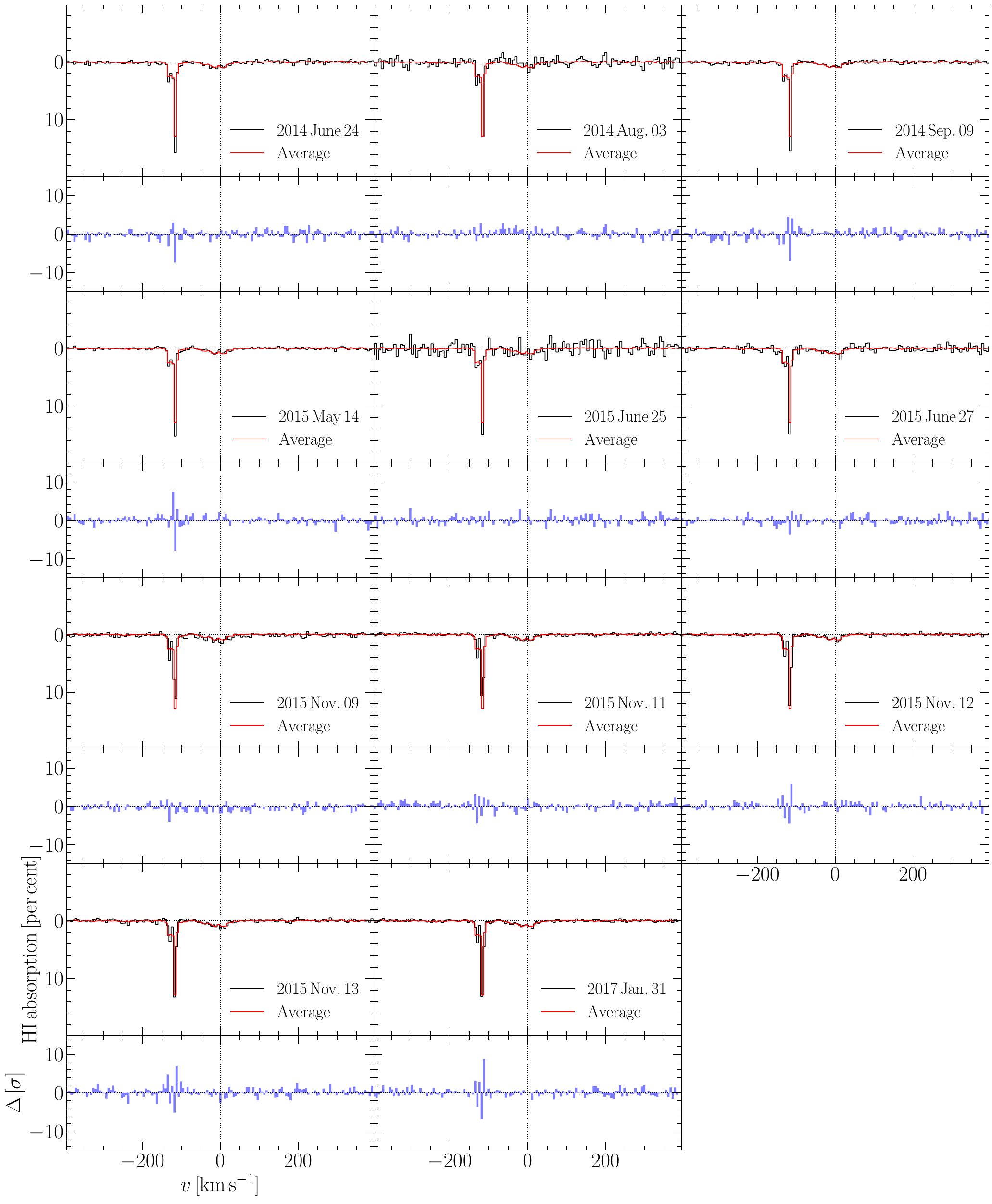}
\caption{Spectra showing \mbox{H\,{\sc i}} absorption for each of the 11 epochs observed with ASKAP. The red line in each panel is the variance-weighted arithmetic mean of the data from all observations. Note that the affect of averaging over the line, which moves with respect to a fixed frequency grid, is to smooth out structures on the scale of a single channel spacing. The difference spectra (blue filled) are shown in the panels below each epoch and are given in units of the RMS noise per channel. We attribute the symmetrically-distributed noise artefacts at the position of the deep narrow component to re-gridding the under-sampled component to a common channel.}\label{figure:PKS1740-517_askap_comparison}
\end{figure*}

\subsubsection{The narrow HI absorption}

We consider next the physical origin of the deeper narrow \mbox{H\,{\sc i}} absorption, which rather strikingly has no associated CO absorption in the ALMA spectrum. Under our previously stated assumptions about the AGN rest frame, this component of the atomic gas is blue shifted by 115\,km\,s$^{-1}$ with respect to the nucleus and hence has radial motion towards the observer. However, the line width ($\Delta{v_{\rm FHWM}} < 5$\,km\,s$^{-1}$) is inconsistent with jet-driven gas outflows that are typically associated with young radio sources (e.g. \citealt{Holt:2008}, \citealt{Morganti:2005b}). Hydrodynamic simulations of interactions between radio jets and the interstellar medium (e.g. \citealt{Wagner:2012, Mukherjee:2018}) predict dispersion velocities between $\sim$100 and 1000\,km\,s$^{-1}$, resulting from turbulence generated as the jet passes through disc. It is unlikely that the gas could be accelerated by the jet to more than 100\,km\,s$^{-1}$ and retain such a low dispersion.  

We show in \autoref{figure:PKS1740-517_askap_comparison} the \mbox{H\,{\sc i}} spectrum for each epoch observed with ASKAP. Although the narrow line is under-sampled it does move systematically with respect to the observed frequency grid as a result of the orbital motion of the Earth, thus indicating that it has a finite width on scales comparable to the resolution of the telescope. For example, between 2015 November 09  and November 11 the line peak is seen to cross  channels, thereby inverting the profile shape. In order to measure the line width we repeat the method described by A15, jointly modeling the narrower \mbox{H\,{\sc i}} component as the superposition of two Gaussians and taking into account the spectral resolution of the telescope by convolving with the fine filter bank response (\citealt{Tuthill:2012}). Assuming no astrophysical variation in the absorption line over the timescale of a few years (no variation in the equivalent width is evident), we obtain a FHWM line width for the deep narrow component of $\Delta{v_{\rm FHWM}} = 4.77_{-0.07}^{+0.07}$\,km\,s$^{-1}$. Since this equates to a kinetic temperature upper limit of $T_{\rm k} \lesssim 500$\,K, we are likely seeing absorption through a single diffuse structure of cold neutral medium (CNM; $T_{\rm k} \sim 100$\,K, e.g \citealt{Wolfire:2003,Heiles:2003b}). Surprisingly, this feature is devoid of the $^{12}$CO\,(2-1) absorption detected elsewhere towards PKS\,B1740$-$517. 

Before discussing a physical interpretation for the absence of CO absorption associated with the narrow \mbox{H\,{\sc i}} line, we should first consider that the subtended continuum emission may be too faint at mm-wavelengths to sufficiently detect any CO that is present. Assuming that the flux density ratio of the two source components at 2.3\,GHz can be extrapolated to all frequencies, this interpretation would imply that the narrow \mbox{H\,{\sc i}} line is located in front of the weaker component, giving a large peak optical depth of $\tau_{\rm HI} \gtrsim 0.9$. Assuming a spin temperature in the range 100 and 500\,K (set by the upper limit on the kinetic temperature), the average column density of the atomic gas is then between $N_{\rm HI} \sim 1$ and $5 \times 10^{21}$\,cm$^{-2}$, consistent with a cold CNM structure. If the detected $^{12}$CO\,(2-1) absorption is seen only in front of the stronger component then given the 3.9:1 ratio of the source components at 2.3\,GHz (\citealt{King:1994}) we predict a peak absorbed fraction of approximately 16\,per\,cent. If the molecular gas has a similar distribution and physical state in front of both source components, then any absorption associated with the weaker component is expected to have a peak in the ALMA spectrum (\autoref{figure:PKS1740-517_absorption_zoom}) of approximately 4\,per\,cent. The RMS per 2\,km\,s$^{-1}$ is 3\,per\,cent of the total continuum flux density, which for a single molecular cloud of velocity dispersion $\Delta{v} \sim 4$\,km\,s$^{-1}$ (\citealt{Heyer:2015}) would give a 3-$\sigma$ detection limit on the peak of approximately 6\,per\,cent. The ALMA data are therefore consistent with not detecting $^{12}$CO\,(2-1) absorption, if the sight-line giving rise to the narrow \mbox{H\,{\sc i}} line is seen only toward the weaker source component. 

\subsubsection{An accreting cold atomic gas cloud?}

It is also possible that our sight-line misses the dense molecular gas entirely, particularly if the \mbox{H\,{\sc i}} structure giving rise to the narrow line is much smaller than the supposed large-scale disc associated with the broader absorption line. If we assume that the molecular clouds are well mixed with the cold atomic gas, then we can use \autoref{equation:covering_fraction} to estimate the expected relative covering fraction for different sizes of the absorbing region. We further assume that since molecular clouds likely form in the CNM they will have on average a larger volume filling fraction than in the large-scale warm ISM, so that $f_{\rm mol}/f_{\rm CNM} \lesssim 30$\,per\,cent. For an absorbing region in the atomic gas extending $R = 100$\,pc along our line-of-sight, we estimate an upper limit molecular covering fraction of $\sim 90$\,per\,cent, which decreases to $\sim 34$\,per\,cent for $R = r = 10$\,pc. This would suggest that the gas giving rise to the narrow \mbox{H\,{\sc i}} line is distributed on scales smaller than 100\,pc. The atomic cloud/structure clearly has very different kinematics to the aforementioned disc, and may be on a non-circular orbit that will eventually accrete onto the AGN itself. Evidence is found for similar non-circular kinematics of individual clouds detected in absorption in nearby radio AGN (e.g. \citealt{Maccagni:2014, Tremblay:2016, Maccagni:2018}).

\subsubsection{A low-metallicity cloud in the tidal stream?}

Alternatively, this narrow component of the \mbox{H\,{\sc i}} line may arise in atomic gas that has physically distinct properties to that of the broader \mbox{H\,{\sc i}} and CO absorption. In their 21-cm absorption survey of quasar-galaxy pairs using the Green Bank Telescope and Jansky Very Large Array, \cite{Borthakur:2011, Borthakur:2014} report a similarly deep ($\tau \sim 1.37$) and narrow ($\Delta{v} \sim 1.1$\,km\,s$^{-1}$) \mbox{H\,{\sc  i}} absorption line in UCG\,7408, a low-metallicity ($\mathrm{[O/H]} \approx -0.4$) dwarf galaxy towards the background quasar J122106.854$+$454852.16. The line width is consistent with cold-phase atomic gas with a kinetic temperature upper limit of $T_{\rm k} \lesssim $\,26\,K, much lower than is typically measured for CNM clouds in the Milky Way ($T_{\rm k} \sim 60$\,K; e.g. \citealt{Heiles:2003b, Murray:2018}). This is also seen in well-studied low-metallicity environments in the local Universe; for example in both the Large (e.g. \citealt{Dickey:1994,Marx-Zimmer:2000}) and Small Magellanic Clouds (e.g. \citealt{Dickey:2000}) where CNM temperatures are typically $T_{\rm k} \lesssim 40$\,K. It is therefore possible that the absence of $^{12}$CO\,(2-1) absorption associated with the narrow \mbox{H\,\sc{i}} absorption in PKS\,B1740$-$517 is the result of the gas having lower metallicity than that seen in the broader absorption. Lower gas-phase metallicity produces fewer dust grains to catalyse \mbox{H$_{2}$} formation and shield the molecular gas from the background UV radiation field, thereby requiring a higher column density of cold atomic gas to self-shield the molecular gas. Our estimate of the \mbox{H\,{\sc i}} column density of CNM that gives rise to the narrow absorption line would require a metallicity that is approximately an order of magnitude lower than solar for H$_{\rm 2}$ not to form (\citealt{Krumholz:2009}). This is a lower limit since the relative abundance of CO to H$_{2}$ is known to decrease strongly as a function of metallicity (e.g. \citealt{wolfire:2010, Bolatto:2013}).

Could this lower metallicity gas be external to the host galaxy of PKS\,B1740$-$517? In \autoref{figure:PKS1740-517_gemini_alma_image} we see possible evidence of an ongoing tidal interaction between the radio galaxy and at least one candidate companion galaxy, including a possible tidal stream. The radial velocity of one companion in the AGN rest frame is $\Delta{v} \approx -220$\,km\,s$^{-1}$, consistent with cosmological N-body simulations of an interacting central galaxy and low-mass satellite with a combined stellar mass of $M_{\ast} \sim 5 \times 10^{10}$\,M$_{\odot}$ (\citealt{Moreno:2013}). The similarity in direction and magnitude of the radial velocities of the companion and the narrow \mbox{H\,{\sc i}} line suggests that we may be seeing absorption through a cold atomic structure within the gas that forms a stream between the two galaxies. It is therefore possible that the absence of CO absorption associated with the narrow \mbox{H\,{\sc i}} absorption in PKS\,B1740$-$517 is evidence for low metallicity gas that is being tidally stripped from a companion satellite galaxy. 

We note that a dense \mbox{H\,{\sc i}} structure in the outskirts of the host galaxy (see e.g. \citealt{Bland-Hawthorn:2017}) could also lack CO gas as a natural consequence of the metallicity gradient. However, the observed radial velocity of this structure with respect to the compact radio source would mean that it is distinctly non-circular at these large radii (see \autoref{figure:schematic}), indicating a tidal interaction either way. In future, more sensitive optical spectroscopy of the tidal feature seen in the Gemini image would enable stronger constraints to be made on the kinematics of the ionised gas both in the candidate companion and tidal bridge, enabling stronger tests of these scenarios. Furthermore, observations of the cold gas at higher spatial resolution using an extended configuration of ALMA, and very long baseline interferometry at 21-cm wavelengths, would enable better determination of the geometry of the absorbing gas with respect to the radio source.

\subsection{Evidence for a young radio AGN}\label{section:spectral_age}

Data from our ALMA observations at 150\,GHz and the ALMA calibrator catalogue at 97.4\,GHz (\citealt{VanKempen:2014}) allow us to determine the spectral energy distribution (SED) of the radio continuum well above the peak frequency, from which we can estimate the spectral age of the radio galaxy. Furthermore, we can use the high frequency end of the radio SED to confirm whether the source has a compact double (i.e consistent steep-spectrum behaviour from the lobes and hot spots) or core-jet morphology (a flat spectrum component). 

\begin{figure}
\centering
\includegraphics[width=0.5\textwidth]{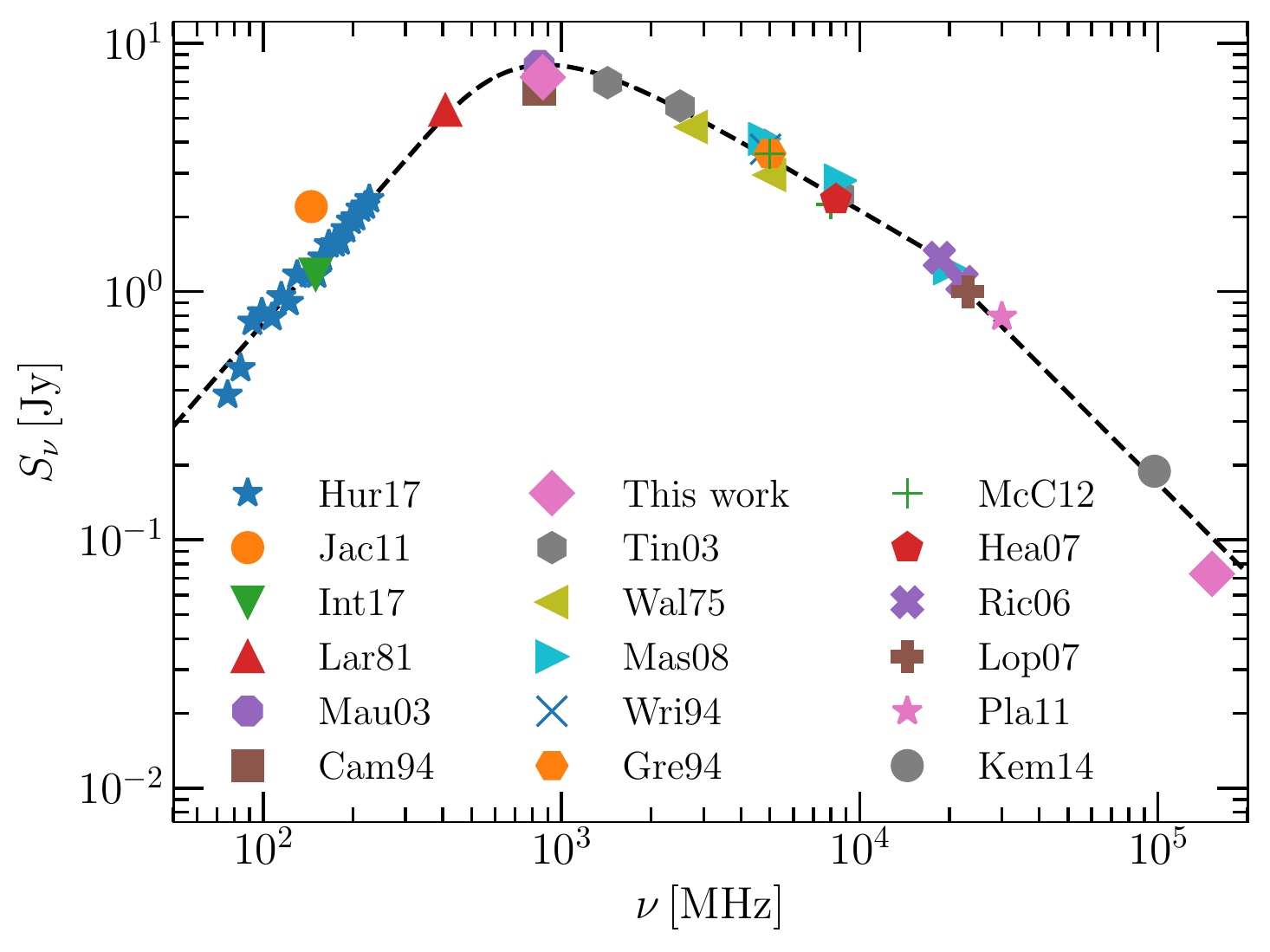}
\caption{The integrated SED of PKS\,B1740$-$517 at radio wavelengths, compiled using our data and those in the literature. The frequency axis is given in the observer
  rest frame. The dashed line denotes a best-fitting model that includes optically thick and thin power-law spectra, and a continuous-injection break at frequencies higher than 16\,GHz.  References for the data: Hur17 -- \citet{Hurley-Walker:2017};  Jac11 -- \citet{Jacobs:2011}; Int17 -- \citet{Intema:2017}; Lar81 -- \citet{Large:1981}; Cam94 --
  \citet{Campbell-Wilson:1994}; Mau03 -- \citet{Mauch:2003}; Tin03 --
  \citet{Tingay:2003}; Wal75 -- \citet{Wall:1975}; Mas08 --
  \citet{Massardi:2008}; Wri94 -- \citet{Wright:1994}; McC12 --
  \citet{McConnell:2012}; \citet{Gregory:1994} -- Gre94; Hea07 --
  \citet{Healey:2007}; Ric06 -- \citet{Ricci:2006}; Lop07 --
  \citet{Lopez-Caniego:2007}; Pla11 --
  \citet{Planck:2011}; Kem14 -- \citet{VanKempen:2014}}\label{figure:PKS1740-517_radio_sed}
\end{figure}

In \autoref{figure:PKS1740-517_radio_sed} we show the integrated SED at radio wavelengths, which is typical of a compact GPS source ($\Delta{d} \lesssim 1$\,kpc) exhibiting optically thick and thin synchrotron emission above and below $\nu \sim 1$\,GHz, respectively (see \citealt{Odea:1998}). The ALMA data show a clear break from a single synchrotron power-law at $\nu \gtrsim 20$\,GHz, from which we conclude that both components identified by \cite{King:1994} are probably associated with the hot spots and lobes of a compact double source.  
We model the SED using the following parameterisation of \cite{Moffet:1975}, adapted to include a high-frequency spectral break that accounts for synchrotron aging with continuous injection of new electrons (\citealt{Kardashev:1962,Murgia:2003}),
\begin{equation}
	S(\nu) = 
\begin{cases}
    \frac{S_{\rm p}}{1 - \mathrm{e}^{-1}}\left(\frac{\nu}{\nu_{\rm p}}\right)^{-\alpha_{\rm tk}}\left[1 - \mathrm{e}^{-\left(\frac{\nu}{\nu_{\rm p}}\right)^{-\alpha_{\rm tk}-\alpha_{\rm tn}}}\right], & \text{if} ~ \nu \leq \nu_{\rm br} \\[12pt]
    S(\nu_{\rm br})\,\left(\frac{\nu}{\nu_{\rm br}}\right)^{\alpha_{\rm tn}-0.5}, & \text{otherwise}
\end{cases}
\end{equation}
where $v_{\rm p}$ and $v_{\rm br}$ are the peak and break frequencies, $\alpha_{\rm tk}$ and $\alpha_{\rm tn}$ are the optically thick and thin spectral indices, and $S_{\rm p}$ is the peak flux density. 

Least squares fitting gives $v_{\rm p} = 786$\,MHz, $v_{\rm br} = 16$\,GHz, $\alpha_{\rm tk} = 1.4$, $\alpha_{\rm tn} = -0.71$, and $S_{\rm p} = 8.1$\,Jy. The optically thick spectral index is consistent with the distribution seen in samples of other peaked-spectrum radio sources (e.g. \citealt{Callingham:2017, Snellen:1998, Odea:1998, deVries:1997}). These are typically shallower than the theoretical value predicted for homogeneous self-absorption ($\alpha_{\rm tk} = 2.5$; e.g. \citealt{Kellerman:1981}) and differences are attributed to either clumpy self-absorption or free-free absorption from a foreground inhomogeneous medium. At frequencies above the spectral peak, the optically thin spectral index is consistent with the canonical injection index of $\alpha_{\rm in} = -0.7$.

We estimate the age of the source using the following relationship (e.g. \citealt{Murgia:2003}),
\begin{equation}
t_{\rm syn} = 5.03\times10^{4}\,B^{-1.5}\,[(1+z)\,\nu_{\rm br}]^{-0.5}\,\mathrm{yr},
\end{equation}
where $B$ is the magnetic field flux density in mG and $\nu_{\rm br}$ is the break frequency in GHz. Assuming  equipartition of the cosmic ray and magnetic field energy densities, equal energy densities of electrons and protons, and source component sizes of $\Delta{\theta} \sim 10$\,mas, we estimate a magnetic flux density of $B \sim 3.4$\,mG. This gives an approximate spectral age for the source of $t_{\rm syn} \sim 1600$\,yr. We note this estimate of the source age involves several assumptions, including magnetic field uniformity, equipartition, and electron mixing (see e.g. \citealt{Tribble:1993, Jones:1999, Blundell:2000}), and should therefore be considered a useful order of magnitude measurement of the source age. The linear separation of the source components ($\Delta{d_{\rm src}} \sim 300$\,pc) implies an expansion velocity of $v_{\rm exp} \sim 0.3$\,c, which is consistent with the distribution of directly measured velocities in the literature (see e.g. \citealt{deVries:2009} and references therein). Since the unresolved emission could be dominated by the younger hot spots, we caution that the source may be older than our estimate based on the integrated SED. However, in the case of intrinsically compact radio sources such as PKS1740$-$517, estimates of the age from spectral modeling are typically consistent to within an order of magnitude with estimates from hot spot advancement (see e.g. \citealt{Murgia:2003}). Therefore the uncertainty in source age is unlikely to affect our conclusions about the relative timescales for the galaxy interaction and AGN triggering. 

The flux density data shown in \autoref{figure:PKS1740-517_radio_sed} span more than 30\,yr of observations, and are therefore potentially subject to both intrinsic source variability and external propagation processes, such as Galactic interstellar scintillation or relative motion of an inhomogeneous medium at the source. PKS\,B1740$-$517 is an intrinsically compact double radio source with no evident Doppler beaming. Such ``galaxy-type'' peaked-spectrum sources (as opposed to ``quasar-type''; \citealt{Snellen:1998}) typically exhibit low levels of variability less than $10$\,per\,cent on year timescales (see e.g. \citealt{Odea:1998, Jauncey:2003, Liu:2009, Orienti:2010}). Intrinsic variability at this level could be due to adiabatic expansion of the self-absorbed young source as it advances through the interstellar medium of the host galaxy (\citealt{Stanghellini:1997, Orienti:2010}, see also \citealt{Tingay:2015}). In the case of PKS\,B1740$-$517, \cite{Gaensler:2000} carried out a time series analysis of observations at 843\,MHz spanning almost 10\,yr with the Molongolo Observatory Synthesis Telescope, detecting variability at the level of $\sim 6.3$\,per\,cent over a characteristic time scale $\sim 6.8$\,yr. At higher frequencies above the turnover, \cite{King:1994} and \cite{Jauncey:2003} report a systematic decrease in flux density at 2.3\,GHz of $\sim 1$\,Jy ($16$\,per\,cent) over almost 2\,yr of observations with the Mount Pleasant 26-m telescope. At 8.4\,GHz they did not find evidence of changes in the flux density at a similar level. 

In \autoref{table:askap_observations} we give the total continuum flux density of PKS\,B1740$-$517 measured from each observation with ASKAP. The standard deviation as a fraction of the median is 8.4\,per\,cent. Although this is consistent with the previously measured low level variability in this source, it is possible that these variations are the result of relative changes in antenna gain between observations of the calibrator PKS\,B1934$-$638 and PKS\,B1740$-$517. We test for this by measuring the standard deviation in flux density for other point sources within the same field. We choose sources that have a flux density greater than 30 times their local image RMS, so that we can detect any systematic behaviour in flux density at the level of $\sim$\,10\,per\,cent. We only select those sources that are located within the full width at half power of the PAF beam, thus avoiding any error contribution associated with sources at the edge of an imperfectly formed PAF beam. We obtain a median standard deviation of $\langle{\sigma}\rangle= 10.2 \pm 1.4$\,per\,cent and a median correlation coefficient with the PKS\,B1740$-$517 flux density of $\langle{\rho}\rangle = 0.5 \pm 0.1$, indicating that systematic errors in the flux density are a significant component of the variance. This is consistent with the error estimated by \cite{Bhandari:2018} for their ASKAP pilot survey for transients and variables. In conclusion, the emission from PKS\,B1740$-$517 may vary at the level of $\lesssim$\,10\,per\,cent over timescales of a few years, but this is not sufficient to affect our conclusions about its spectral behaviour, morphology or age. 

\section{Discussion}\label{section:discussion}

\subsection{PKS\,B1740$-$517 - triggered by an interaction?}

PKS\,B1740$-$517 is a distant young radio galaxy that seems to be undergoing at least one tidal interaction with a lower-mass satellite galaxy. We see evidence for a reservoir of cold atomic and molecular gas distributed in an inner disc or ring and a distinct diffuse atomic gas cloud that appears to have different kinematics, and possibly physical properties, to that of the settled gas. This gas could have accreted onto the host galaxy during tidal interaction with a gas-rich lower-mass companion. The difference in physical properties could be the result of the location of the absorbing gas, where star formation and gas mixing towards the centre of the host galaxy could act to enrich the accreted gas.

The luminosity of the [O\,{\sc ii}]\,$\lambda${3727} line in the companion spectrum is $L_{\rm [OII]} \approx 1.9 \times 10^{40}$\,erg\,s$^{-1}$, which using \autoref{equation:OII_to_SFR} (\citealt{Kewley:2004}) gives a star formation rate of $\mathrm{SFR} \approx 0.13$\,M$_{\odot}$\,yr$^{-1}$. This implies that the companion galaxy has sufficient gas mass ($M_{\rm gas} \sim 10^{8}$\,M$_{\odot}$) to supply that detected in the host galaxy of PKS\,B1740$-$517. The dynamical timescale for the interaction (based on the projected separation and relative radial velocity) is significantly longer than the age of the AGN ($t_{\rm dyn} \sim 5 \times 10^{7}$\,yr), implying that the gas accreted onto the host galaxy well before the most recent episode of activity and may have been prevented from fuelling the AGN by a burst of star formation that has only recently been quenched.

This scenario is strikingly similar to that seen in the radio galaxy MRC\,B1221$-$423, a young ($t \sim 10^{5}$\,yr) compact steep spectrum (CSS) source that appears to be undergoing a minor merger with a companion galaxy (\citealt{Johnston:2005}). Since this source is at a lower redshift ($z = 0.1706$; \citealt{Simpson:1993}), and is older and therefore more extended than PKS\,B1740$-$517, radio imaging reveals a highly distorted morphology including a 180$\degr$ bent jet that is possibly due to interaction with a highly dense medium (\citealt{Johnston:2010}). The neutral gas in MRC\,B1221$-$423 (seen in \mbox{H\,{\sc i}} absorption against the radio continuum) appears to be infalling towards the nucleus at a velocity of $\sim 250$\,km\,s$^{-1}$. In their IFU study of this radio galaxy, \cite{Anderson:2013} find several stellar populations of different ages, associated with dynamically unstable gas during successive periapses of the companion orbit. The radio source was triggered about 100\,Myr after the most recent closest approach, indicating that this is the timescale for gas to be able to sink to the nucleus. We suggest that a similar scenario may have occurred in the host galaxy PKS\,B1740$-$517, whereby we are seeing the activity from the most recent periapsis and accretion of gas. A future resolved study of the ionised emission from PKS\,B1740$-$517 and its companion using the NIRSpec IFU on the James Webb Space Telescope (\citealt{Purll:2017}) would confirm this interpretation. 

This sequence and timescale of events is consistent with that seen for minor mergers in massive early type galaxies at lower redshifts; in a study of 25 nearby dust-lane-selected early-types, \cite{Shabala:2017} find that those with radio-loud AGN tend to have older star burst ages and lower star formation rates, pointing to delayed onset of black hole accretion following the merger. More generally, emission-line AGN are found to host on average stellar populations that are a few hundred Myr older than their more vigorously star forming counterparts (e.g. \citealt{Schawinski:2007, Wild:2010}}). Furthermore, the prevalence of highly accreting AGN identified in galaxies pairs and mergers is also found to peak amongst those that are either post-merger or in the closest pairs, occurring later in the interaction than the onset of star formation (\citealt{Ellison:2013}). PKS\,B1740$-$517 is a rare example of the early stage in a brief active phase of a massive early type galaxy that is triggered by interactions and minor mergers with smaller companions. If this is the most recent event then it is possible that older AGN activity may have been triggered by previous orbits of the companion. We propose that radio emission from such recurrent activity could be detected using new low-frequency radio arrays at high surface brightness sensitivity, for the example the Murchison Widefield Array (MWA; \citealt{Tingay:2013, Bowman:2013}) and in the future SKA-Low. 

\subsection{Molecular absorption in distant radio galaxies}

Finding molecular absorption lines at cosmological distances is an important probe of the physical  conditions and chemistry of cold gas in the ISM of galaxies at earlier epochs (e.g. \citealt{Henkel:2005, Bottinelli:2009, Muller:2014}). In addition to their importance as a tool for determining the role of the neutral ISM in galaxy evolution, absorption lines also provide a useful independent means of determining key parameters in cosmology and fundamental physics. If several species are detected under conditions where the excitation is determined by the cosmic microwave background (CMB) radiation, then their relative strength can be used to independently determine the CMB temperature at that redshift (\citealt{Wiklind:1997, Muller:2013}). Furthermore, comparison of the redshifts of several lines within the same absorber can be used to examine possible cosmological evolution of the electron to proton mass ratio, the fine structure constant and the proton gyromagnetic ratio (e.g. \citealt{Wiklind:1997, Murphy:2001, Darling:2003, Murphy:2008, Kanekar:2011b, Curran:2011d, Ellingsen:2012, Kanekar:2012, Kanekar:2015b, Kanekar:2018}, see also the review by \citealt{Curran:2004a}), thereby testing fundamental physics. 

Surveys for molecular absorption lines in the spectra of bright radio sources should be observationally efficient at finding cold molecular gas in galaxies at cosmologically significant distances. This is principally due to absorption being observable at any luminosity distance, provided there exists a sufficiently bright background source of emission. However, in practice molecular absorption has proven to be relatively difficult to detect at redshifts much beyond $z \sim 0.1$. Including this work, only six molecular absorbers have been detected at cosmological distances. Only one of these was discovered in a genuinely blind search for molecular absorption (PKS\,B1830$-$211; \citealt{Wiklind:1996a}), and subsequent searches (e.g. \citealt{Kanekar:2014}) have not yielded detections. The key limiting factor is thought to be impact parameter, because the molecular component of the ISM is more centrally concentrated than atomic gas in galaxies. This interpretation is certainly borne out in the detected systems; three are intrinsic to the host galaxy of the radio source: PKS\,B1740$-$517 at z = 0.441 (this work), PKS\,B1413$+$135 at $z = 0.247$ (\citealt{Wiklind:1994, Wiklind:1997}), and B1504$+$377 at $z = 0.685$ (\citealt{Wiklind:1996b}). The three intervening absorbers were all found to have sufficiently low impact parameter that they gravitationally lens the background quasar: B0218$+$357 at $z = 0.685$ (\citealt{Wiklind:1995}), PKS\,B1830$-$211 at $z  = 0.886$ (\citealt{Wiklind:1996a, Wiklind:1998, Muller:2014}) and PKS\,B0132$-$097 at $z = 0.765$ (\citealt{Kanekar:2005,Wiklind:2018}). 

Careful selection of targets is therefore key to detecting more molecular absorbers at cosmological distances. The host galaxies of compact steep spectrum and GPS radio sources, with existing detections of cold gas through \mbox{H\,{\sc i}} 21-cm absorption, are one possible avenue of investigation. Of the three known intrinsic molecular absorbers, PKS\,B1740$-$517 and PKS\,B1413$+$135 have an intrinsically compact source morphology. These sources are intrinsically compact and therefore should not extend beyond the stellar disc of their host (\citealt{Odea:1998}), increasing the probability of detecting line-of-sight molecular gas in absorption. Of course these radio galaxies are not representative of the normal galaxy population and therefore any conclusions drawn about the molecular gas content should be considered in the context of their hosts. Several surveys are planned to carry out wide-field blind searches for \mbox{H\,{\sc i}} 21-cm absorption with pathfinder telescopes to the Square Kilometre Array (SKA). A summary of these in the context of intrinsic absorption is provided by \cite{Morganti:2018} (see also \citealt{Maccagni:2017}), and include the First Large Absorption Survey in HI with ASKAP (e.g. \citealt{Johnston:2007, Allison:2016}), the MeerKAT Absorption Line Survey (\citealt{Gupta:2016}), the MeerKAT International GHz Tiered Extragalactic Exploration (MIGHTEE) Survey (\citealt{Jarvis:2017}), and the Search for \mbox{H\,{\sc i}} Absorption with AperTIF (e.g. \citealt{Oosterloo:2009}). Together, these surveys are expected to detect \mbox{H\,{\sc i}} absorption in the host galaxies of several thousand radio sources, spanning redshifts from the nearby Universe to $z \sim 1.44$. Although this seems promising, confined and/or young radio sources typically have steep spectral indices ($\alpha \sim -1$), meaning that their flux density at 200\,GHz can be two orders of magnitude lower than at 1\,GHz. In order to detect molecular absorption with an observed optical depth $\tau \sim 0.1$ within a reasonable observing time with ALMA we require the source to be brighter than $S_{\rm 200\,GHz} \sim 10$\,mJy at 200\,GHz, which corresponds to $S_{\rm 1\,GHz} \sim 1$\,Jy at 1\,GHz. The proportion of detected \mbox{H\,{\sc i}} absorbers with a radio continuum brighter than 1\,Jy is expected to be $\sim 1$\,per\,cent (e.g. \citealt{Condon:1998, Mauch:2003}) of which $\sim 10 - 30$\,per\,cent will be intrinsically compact and young radio galaxies (e.g. \citealt{Odea:1998}). Although this means that only a few detections with these surveys will provide useful targets for molecular absorption line detection, this will greatly improve the existing sample of high redshift systems currently known.

\section{Conclusions}\label{section:conclusions}

Using ALMA we have detected a rare example of $^{12}$CO\,(2-1) absorption at $z = 0.44$ in the host galaxy of the luminous GPS radio source PKS\,B1740$-$517. Comparison of the CO and \mbox{H\,{\sc i}} absorption line profiles suggests that the CO is distributed in a more clumpy medium of dense clouds while the \mbox{H\,{\sc i}} seems to be more consistent with larger diffuse structures. We argue that the broader atomic and molecular absorption lines arise from a central reservoir of cold gas ($M_{\rm gas} \sim 10^{7} - 10^{8}$\,$M_{\odot}$), likely distributed as a disc or ring within a few kiloparsecs of the active nucleus. A separate kinematically-distinct deep ($\tau_{\rm peak} \gtrsim 0.2$) and narrow ($\Delta{v_{\rm FWHM}} \lesssim 5$\,km\,s$^{-1}$) \mbox{H\,{\sc i}} absorption line is consistent with a single diffuse cold atomic structure ($T_{\rm k} \sim 100$\,K). The absence of molecular gas suggests that either this structure is much smaller than 100\,pc along our line-of-sight and/or that it arises from lower metallicity gas, perhaps from a companion galaxy. 

Examination of optical imaging and spectroscopy from previous observations by A15 with the Gemini-South telescope reveal at least one candidate lower-mass companion galaxy that appears to be undergoing a tidal interaction. The dynamical timescale for this interaction is approximately 50\,Myr, implying that the onset of the radio AGN, which we estimate to be only $\sim 1600$\,yr old, is delayed with respect to the bulk accretion of gas from the companion. This is consistent with previous studies of the relative timescales of mergers, star formation and AGN triggering in massive galaxies. PKS\,B1740$-$517 is a luminous and young radio galaxy that provides a rare opportunity to study the cold gas content of massive early type galaxies at cosmological distances. Future surveys for \mbox{H\,{\sc i}} 21-cm absorption will provide further targets for follow up observations with ALMA, but will require careful selection based on the morphology and spectral behaviour of the source.

\section*{Acknowledgements} 

We thank the anonymous referee for their careful comments that helped improve the clarity of this paper. JRA acknowledges support from a Christ Church Career Development Fellowship. Parts of this research were conducted by the Australian Research Council Centre of Excellence for All-sky Astrophysics in 3D (ASTRO 3D) through project number CE170100013. The National Radio Astronomy Observatory is a facility of the National Science Foundation operated under cooperative agreement by Associated Universities, Inc.

This paper makes use of the following ALMA data: ADS/JAO.ALMA\#2015.1.00808.S. ALMA is a partnership of ESO (representing its member states), NSF (USA) and NINS (Japan), together with NRC (Canada), NSC and ASIAA (Taiwan), and KASI (Republic of Korea), in cooperation with the Republic of Chile. The Joint ALMA Observatory is operated by ESO, AUI/NRAO and NAOJ.

The Australian SKA Pathfinder is part of the Australia Telescope National Facility which is managed by CSIRO. Operation of ASKAP is
funded by the Australian Government with support from the National
Collaborative Research Infrastructure Strategy. Establishment of the
Murchison Radio-astronomy Observatory was funded by the Australian
Government and the Government of Western Australia. ASKAP uses
advanced supercomputing resources at the Pawsey Supercomputing
Centre. We acknowledge the Wajarri Yamatji people as the traditional
owners of the Observatory site. 

We have made use of {\sc Astropy}, a
community-developed core {\sc Python} package for astronomy
(\citealt{Astropy:2013}); the NASA/IPAC Extragalactic Database (NED) and Infrared Science Archive (IRSA), which are operated by the Jet Propulsion Laboratory, California
Institute of Technology, under contract with the National Aeronautics
and Space Administration; NASA's Astrophysics Data System
Bibliographic Services; and the VizieR catalogue access tool operated
at CDS, Strasbourg, France.




\bibliographystyle{mnras}
\bibliography{james}

\begin{thebibliography}{}
\makeatletter
\relax
\def\mn@urlcharsother{\let\do\@makeother \do\$\do\&\do\#\do\^\do\_\do\%\do\~}
\def\mn@doi{\begingroup\mn@urlcharsother \@ifnextchar [ {\mn@doi@}
  {\mn@doi@[]}}
\def\mn@doi@[#1]#2{\def\@tempa{#1}\ifx\@tempa\@empty \href
  {http://dx.doi.org/#2} {doi:#2}\else \href {http://dx.doi.org/#2} {#1}\fi
  \endgroup}
\def\mn@eprint#1#2{\mn@eprint@#1:#2::\@nil}
\def\mn@eprint@arXiv#1{\href {http://arxiv.org/abs/#1} {{\tt arXiv:#1}}}
\def\mn@eprint@dblp#1{\href {http://dblp.uni-trier.de/rec/bibtex/#1.xml}
  {dblp:#1}}
\def\mn@eprint@#1:#2:#3:#4\@nil{\def\@tempa {#1}\def\@tempb {#2}\def\@tempc
  {#3}\ifx \@tempc \@empty \let \@tempc \@tempb \let \@tempb \@tempa \fi \ifx
  \@tempb \@empty \def\@tempb {arXiv}\fi \@ifundefined
  {mn@eprint@\@tempb}{\@tempb:\@tempc}{\expandafter \expandafter \csname
  mn@eprint@\@tempb\endcsname \expandafter{\@tempc}}}

\bibitem[\protect\citeauthoryear{{Aditya} \& {Kanekar}}{{Aditya} \&
  {Kanekar}}{2018}]{Aditya:2018}
{Aditya} J.~N.~H.~S.,  {Kanekar} N.,  2018, \mn@doi [MNRAS]
  {10.1093/mnras/stx2325}, \href
  {http://adsabs.harvard.edu/abs/2018MNRAS.473...59A} {473, 59}

\bibitem[\protect\citeauthoryear{{Aditya}, {Kanekar}, {Prochaska}, {Day},
  {Lynam}  \& {Cruz}}{{Aditya} et~al.}{2017}]{Aditya:2017}
{Aditya} J.~N.~H.~S.,  {Kanekar} N.,  {Prochaska} J.~X.,  {Day} B.,  {Lynam}
  P.,   {Cruz} J.,  2017, \mn@doi [MNRAS] {10.1093/mnras/stw3105}, \href
  {http://adsabs.harvard.edu/abs/2017MNRAS.465.5011A} {465, 5011}

\bibitem[\protect\citeauthoryear{{Allen}, {Dunn}, {Fabian}, {Taylor}  \&
  {Reynolds}}{{Allen} et~al.}{2006}]{Allen:2006}
{Allen} S.~W.,  {Dunn} R.~J.~H.,  {Fabian} A.~C.,  {Taylor} G.~B.,   {Reynolds}
  C.~S.,  2006, \mn@doi [MNRAS] {10.1111/j.1365-2966.2006.10778.x}, \href
  {http://adsabs.harvard.edu/abs/2006MNRAS.372...21A} {372, 21}

\bibitem[\protect\citeauthoryear{{Allison} et~al.,}{{Allison}
  et~al.}{2015}]{Allison:2015a}
{Allison} J.~R.,  et~al., 2015, \mn@doi [MNRAS] {10.1093/mnras/stv1532}, \href
  {http://adsabs.harvard.edu/abs/2015MNRAS.453.1249A} {453, 1249}

\bibitem[\protect\citeauthoryear{{Allison}, {Zwaan}, {Duchesne}  \&
  {Curran}}{{Allison} et~al.}{2016}]{Allison:2016}
{Allison} J.~R.,  {Zwaan} M.~A.,  {Duchesne} S.~W.,   {Curran} S.~J.,  2016,
  \mn@doi [MNRAS] {10.1093/mnras/stw1722}, \href
  {http://adsabs.harvard.edu/abs/2016MNRAS.462.1341A} {462, 1341}

\bibitem[\protect\citeauthoryear{{Allison} et~al.,}{{Allison}
  et~al.}{2017}]{Allison:2017}
{Allison} J.~R.,  et~al., 2017, \mn@doi [MNRAS] {10.1093/mnras/stw2860}, \href
  {http://adsabs.harvard.edu/abs/2017MNRAS.465.4450A} {465, 4450}

\bibitem[\protect\citeauthoryear{{Anderson}, {Johnston}  \&
  {Hunstead}}{{Anderson} et~al.}{2013}]{Anderson:2013}
{Anderson} C.~S.,  {Johnston} H.~M.,   {Hunstead} R.~W.,  2013, \mn@doi [MNRAS]
  {10.1093/mnras/stt406}, \href
  {http://adsabs.harvard.edu/abs/2013MNRAS.431.3269A} {431, 3269}

\bibitem[\protect\citeauthoryear{{Astropy Collaboration} et~al.,}{{Astropy
  Collaboration} et~al.}{2013}]{Astropy:2013}
{Astropy Collaboration} et~al., 2013, \mn@doi [A\&A]
  {10.1051/0004-6361/201322068}, \href
  {http://adsabs.harvard.edu/abs/2013A%26A...558A..33A} {558, A33}

\bibitem[\protect\citeauthoryear{{Bahcall} \& {Ekers}}{{Bahcall} \&
  {Ekers}}{1969}]{Bahcall:1969}
{Bahcall} J.~N.,  {Ekers} R.~D.,  1969, \mn@doi [ApJ] {10.1086/150135}, \href
  {http://adsabs.harvard.edu/abs/1969ApJ...157.1055B} {157, 1055}

\bibitem[\protect\citeauthoryear{{Best} \& {Heckman}}{{Best} \&
  {Heckman}}{2012}]{Best:2012}
{Best} P.~N.,  {Heckman} T.~M.,  2012, \mn@doi [MNRAS]
  {10.1111/j.1365-2966.2012.20414.x}, \href
  {http://adsabs.harvard.edu/abs/2012MNRAS.421.1569B} {421, 1569}

\bibitem[\protect\citeauthoryear{{Bhandari} et~al.,}{{Bhandari}
  et~al.}{2018}]{Bhandari:2018}
{Bhandari} S.,  et~al., 2018, \mn@doi [MNRAS] {10.1093/mnras/sty1157}, \href
  {http://adsabs.harvard.edu/abs/2018MNRAS.478.1784B} {478, 1784}

\bibitem[\protect\citeauthoryear{{Biggs}, {Zwaan}, {Hatziminaoglou},
  {P{\'e}roux}  \& {Liske}}{{Biggs} et~al.}{2016}]{Biggs:2016}
{Biggs} A.~D.,  {Zwaan} M.~A.,  {Hatziminaoglou} E.,  {P{\'e}roux} C.,
  {Liske} J.,  2016, \mn@doi [MNRAS] {10.1093/mnras/stw1786}, \href
  {http://adsabs.harvard.edu/abs/2016MNRAS.462.2819B} {462, 2819}

\bibitem[\protect\citeauthoryear{{Bland-Hawthorn}, {Maloney}, {Stephens},
  {Zovaro}  \& {Popping}}{{Bland-Hawthorn} et~al.}{2017}]{Bland-Hawthorn:2017}
{Bland-Hawthorn} J.,  {Maloney} P.~R.,  {Stephens} A.,  {Zovaro} A.,
  {Popping} A.,  2017, \mn@doi [ApJ] {10.3847/1538-4357/aa8f45}, \href
  {http://adsabs.harvard.edu/abs/2017ApJ...849...51B} {849, 51}

\bibitem[\protect\citeauthoryear{{Blundell} \& {Rawlings}}{{Blundell} \&
  {Rawlings}}{2000}]{Blundell:2000}
{Blundell} K.~M.,  {Rawlings} S.,  2000, \mn@doi [AJ] {10.1086/301254}, \href
  {http://adsabs.harvard.edu/abs/2000AJ....119.1111B} {119, 1111}

\bibitem[\protect\citeauthoryear{{Bolatto}, {Wolfire}  \& {Leroy}}{{Bolatto}
  et~al.}{2013}]{Bolatto:2013}
{Bolatto} A.~D.,  {Wolfire} M.,   {Leroy} A.~K.,  2013, \mn@doi [AR\&A]
  {10.1146/annurev-astro-082812-140944}, \href
  {http://adsabs.harvard.edu/abs/2013ARA%26A..51..207B} {51, 207}

\bibitem[\protect\citeauthoryear{{Bondi}}{{Bondi}}{1952}]{Bondi:1952}
{Bondi} H.,  1952, \mn@doi [MNRAS] {10.1093/mnras/112.2.195}, \href
  {http://adsabs.harvard.edu/abs/1952MNRAS.112..195B} {112, 195}

\bibitem[\protect\citeauthoryear{{Borthakur}, {Tripp}, {Yun}, {Bowen},
  {Meiring}, {York}  \& {Momjian}}{{Borthakur} et~al.}{2011}]{Borthakur:2011}
{Borthakur} S.,  {Tripp} T.~M.,  {Yun} M.~S.,  {Bowen} D.~V.,  {Meiring} J.~D.,
   {York} D.~G.,   {Momjian} E.,  2011, \mn@doi [ApJ]
  {10.1088/0004-637X/727/1/52}, \href
  {http://adsabs.harvard.edu/abs/2011ApJ...727...52B} {727, 52}

\bibitem[\protect\citeauthoryear{{Borthakur}, {Momjian}, {Heckman}, {York},
  {Bowen}, {Yun}  \& {Tripp}}{{Borthakur} et~al.}{2014}]{Borthakur:2014}
{Borthakur} S.,  {Momjian} E.,  {Heckman} T.~M.,  {York} D.~G.,  {Bowen} D.~V.,
   {Yun} M.~S.,   {Tripp} T.~M.,  2014, \mn@doi [ApJ]
  {10.1088/0004-637X/795/1/98}, \href
  {http://adsabs.harvard.edu/abs/2014ApJ...795...98B} {795, 98}

\bibitem[\protect\citeauthoryear{{Bottinelli} et~al.,}{{Bottinelli}
  et~al.}{2009}]{Bottinelli:2009}
{Bottinelli} S.,  et~al., 2009, \mn@doi [ApJL] {10.1088/0004-637X/690/2/L130},
  \href {http://adsabs.harvard.edu/abs/2009ApJ...690L.130B} {690, L130}

\bibitem[\protect\citeauthoryear{{Bowman} et~al.,}{{Bowman}
  et~al.}{2013}]{Bowman:2013}
{Bowman} J.~D.,  et~al., 2013, \mn@doi [PASA] {10.1017/pas.2013.009}, \href
  {http://adsabs.harvard.edu/abs/2013PASA...30...31B} {30, e031}

\bibitem[\protect\citeauthoryear{{Braun}}{{Braun}}{2012}]{Braun:2012}
{Braun} R.,  2012, \mn@doi [ApJ] {10.1088/0004-637X/749/1/87}, \href
  {http://adsabs.harvard.edu/abs/2012ApJ...749...87B} {749, 87}

\bibitem[\protect\citeauthoryear{{Braun}, {Thilker}, {Walterbos}  \&
  {Corbelli}}{{Braun} et~al.}{2009}]{Braun:2009}
{Braun} R.,  {Thilker} D.~A.,  {Walterbos} R.~A.~M.,   {Corbelli} E.,  2009,
  \mn@doi [ApJ] {10.1088/0004-637X/695/2/937}, \href
  {http://adsabs.harvard.edu/abs/2009ApJ...695..937B} {695, 937}

\bibitem[\protect\citeauthoryear{{Burgess} \& {Hunstead}}{{Burgess} \&
  {Hunstead}}{2006}]{Burgess:2006}
{Burgess} A.~M.,  {Hunstead} R.~W.,  2006, \mn@doi [AJ] {10.1086/498679}, \href
  {http://adsabs.harvard.edu/abs/2006AJ....131..114B} {131, 114}

\bibitem[\protect\citeauthoryear{{Burgh}, {France}  \& {McCandliss}}{{Burgh}
  et~al.}{2007}]{Burgh:2007}
{Burgh} E.~B.,  {France} K.,   {McCandliss} S.~R.,  2007, \mn@doi [ApJ]
  {10.1086/511259}, \href {http://adsabs.harvard.edu/abs/2007ApJ...658..446B}
  {658, 446}

\bibitem[\protect\citeauthoryear{{Callingham} et~al.,}{{Callingham}
  et~al.}{2017}]{Callingham:2017}
{Callingham} J.~R.,  et~al., 2017, \mn@doi [ApJ] {10.3847/1538-4357/836/2/174},
  \href {http://adsabs.harvard.edu/abs/2017ApJ...836..174C} {836, 174}

\bibitem[\protect\citeauthoryear{{Campbell-Wilson} \&
  {Hunstead}}{{Campbell-Wilson} \& {Hunstead}}{1994}]{Campbell-Wilson:1994}
{Campbell-Wilson} D.,  {Hunstead} R.~W.,  1994, Proc. of the Astro. Soc. Aust.,
  \href {http://adsabs.harvard.edu/abs/1994PASAu..11...33C} {11, 33}

\bibitem[\protect\citeauthoryear{{Carilli}, {Menten}, {Reid}, {Rupen}  \&
  {Yun}}{{Carilli} et~al.}{1998}]{Carilli:1998}
{Carilli} C.~L.,  {Menten} K.~M.,  {Reid} M.~J.,  {Rupen} M.~P.,   {Yun} M.~S.,
   1998, \mn@doi [ApJ] {10.1086/305191}, \href
  {http://adsabs.harvard.edu/abs/1998ApJ...494..175C} {494, 175}

\bibitem[\protect\citeauthoryear{{Chandra}, {Maheshwari}  \&
  {Sharma}}{{Chandra} et~al.}{1996}]{Chandra:1996}
{Chandra} S.,  {Maheshwari} V.~U.,   {Sharma} A.~K.,  1996, A\&A, \href
  {http://adsabs.harvard.edu/abs/1996A%26AS..117..557C} {117, 557}

\bibitem[\protect\citeauthoryear{{Chippendale} et~al.,}{{Chippendale}
  et~al.}{2015}]{Chippendale:2015}
{Chippendale} A.~P.,  et~al., 2015, in 2015 International Conference on
  Electromagnetics in Advanced Applications (ICEAA). IEEE, pp 541 -- 544
  (\mn@eprint {arXiv} {1509.00544})

\bibitem[\protect\citeauthoryear{{Condon}, {Cotton}, {Greisen}, {Yin},
  {Perley}, {Taylor}  \& {Broderick}}{{Condon} et~al.}{1998}]{Condon:1998}
{Condon} J.~J.,  {Cotton} W.~D.,  {Greisen} E.~W.,  {Yin} Q.~F.,  {Perley}
  R.~A.,  {Taylor} G.~B.,   {Broderick} J.~J.,  1998, \mn@doi [AJ]
  {10.1086/300337}, \href {http://adsabs.harvard.edu/abs/1998AJ....115.1693C}
  {115, 1693}

\bibitem[\protect\citeauthoryear{{Croton} et~al.,}{{Croton}
  et~al.}{2006}]{Croton:2006}
{Croton} D.~J.,  et~al., 2006, \mn@doi [MNRAS]
  {10.1111/j.1365-2966.2005.09675.x}, \href
  {http://adsabs.harvard.edu/abs/2006MNRAS.365...11C} {365, 11}

\bibitem[\protect\citeauthoryear{{Curran} \& {Whiting}}{{Curran} \&
  {Whiting}}{2010}]{Curran:2010}
{Curran} S.~J.,  {Whiting} M.~T.,  2010, \mn@doi [ApJ]
  {10.1088/0004-637X/712/1/303}, \href
  {http://adsabs.harvard.edu/abs/2010ApJ...712..303C} {712, 303}

\bibitem[\protect\citeauthoryear{{Curran}, {Kanekar}  \& {Darling}}{{Curran}
  et~al.}{2004}]{Curran:2004a}
{Curran} S.~J.,  {Kanekar} N.,   {Darling} J.~K.,  2004, \mn@doi [New Astronomy
  Reviews] {10.1016/j.newar.2004.09.004}, \href
  {http://adsabs.harvard.edu/abs/2004NewAR..48.1095C} {48, 1095}

\bibitem[\protect\citeauthoryear{{Curran} et~al.,}{{Curran}
  et~al.}{2011a}]{Curran:2011a}
{Curran} S.~J.,  et~al., 2011a, \mn@doi [MNRAS]
  {10.1111/j.1365-2966.2011.18209.x}, \href
  {http://adsabs.harvard.edu/abs/2011MNRAS.413.1165C} {413, 1165}

\bibitem[\protect\citeauthoryear{{Curran}, {Tanna}, {Koch}, {Berengut}, {Webb},
  {Stark}  \& {Flambaum}}{{Curran} et~al.}{2011b}]{Curran:2011d}
{Curran} S.~J.,  {Tanna} A.,  {Koch} F.~E.,  {Berengut} J.~C.,  {Webb} J.~K.,
  {Stark} A.~A.,   {Flambaum} V.~V.,  2011b, \mn@doi [A\&A]
  {10.1051/0004-6361/201117457}, \href
  {http://adsabs.harvard.edu/abs/2011A%26A...533A..55C} {533, A55}

\bibitem[\protect\citeauthoryear{{Curran}, {Whiting}, {Sadler}  \&
  {Bignell}}{{Curran} et~al.}{2013}]{Curran:2013a}
{Curran} S.~J.,  {Whiting} M.~T.,  {Sadler} E.~M.,   {Bignell} C.,  2013,
  \mn@doi [MNRAS] {10.1093/mnras/sts171}, \href
  {http://adsabs.harvard.edu/abs/2013MNRAS.428.2053C} {428, 2053}

\bibitem[\protect\citeauthoryear{{Darling}}{{Darling}}{2003}]{Darling:2003}
{Darling} J.,  2003, \mn@doi [Physical Review Letters]
  {10.1103/PhysRevLett.91.011301}, \href
  {http://adsabs.harvard.edu/abs/2003PhRvL..91a1301D} {91, 011301}

\bibitem[\protect\citeauthoryear{{Dasyra} \& {Combes}}{{Dasyra} \&
  {Combes}}{2012}]{Dasyra:2012}
{Dasyra} K.~M.,  {Combes} F.,  2012, \mn@doi [A\&A]
  {10.1051/0004-6361/201219229}, \href
  {http://adsabs.harvard.edu/abs/2012A%26A...541L...7D} {541, L7}

\bibitem[\protect\citeauthoryear{{David} et~al.,}{{David}
  et~al.}{2014}]{David:2014}
{David} L.~P.,  et~al., 2014, \mn@doi [ApJ] {10.1088/0004-637X/792/2/94}, \href
  {http://adsabs.harvard.edu/abs/2014ApJ...792...94D} {792, 94}

\bibitem[\protect\citeauthoryear{{Davis} et~al.,}{{Davis}
  et~al.}{2013}]{Davis:2013}
{Davis} T.~A.,  et~al., 2013, \mn@doi [MNRAS] {10.1093/mnras/sts353}, \href
  {http://adsabs.harvard.edu/abs/2013MNRAS.429..534D} {429, 534}

\bibitem[\protect\citeauthoryear{{Davis} et~al.,}{{Davis}
  et~al.}{2015}]{Davis:2015}
{Davis} T.~A.,  et~al., 2015, \mn@doi [MNRAS] {10.1093/mnras/stv597}, \href
  {http://adsabs.harvard.edu/abs/2015MNRAS.449.3503D} {449, 3503}

\bibitem[\protect\citeauthoryear{{Delvecchio} et~al.,}{{Delvecchio}
  et~al.}{2017}]{Delvecchio:2017}
{Delvecchio} I.,  et~al., 2017, \mn@doi [A\&A] {10.1051/0004-6361/201629367},
  \href {http://adsabs.harvard.edu/abs/2017A%26A...602A...3D} {602, A3}

\bibitem[\protect\citeauthoryear{{Dickey}, {Mebold}, {Marx}, {Amy}, {Haynes}
  \& {Wilson}}{{Dickey} et~al.}{1994}]{Dickey:1994}
{Dickey} J.~M.,  {Mebold} U.,  {Marx} M.,  {Amy} S.,  {Haynes} R.~F.,
  {Wilson} W.,  1994, A\&A, \href
  {http://adsabs.harvard.edu/abs/1994A%26A...289..357D} {289, 357}

\bibitem[\protect\citeauthoryear{{Dickey}, {Mebold}, {Stanimirovic}  \&
  {Staveley-Smith}}{{Dickey} et~al.}{2000}]{Dickey:2000}
{Dickey} J.~M.,  {Mebold} U.,  {Stanimirovic} S.,   {Staveley-Smith} L.,  2000,
  \mn@doi [ApJ] {10.1086/308953}, \href
  {http://adsabs.harvard.edu/abs/2000ApJ...536..756D} {536, 756}

\bibitem[\protect\citeauthoryear{{Dutta}, {Gupta}, {Srianand}  \&
  {O'Meara}}{{Dutta} et~al.}{2016}]{Dutta:2016}
{Dutta} R.,  {Gupta} N.,  {Srianand} R.,   {O'Meara} J.~M.,  2016, \mn@doi
  [MNRAS] {10.1093/mnras/stv2980}, \href
  {http://adsabs.harvard.edu/abs/2016MNRAS.456.4209D} {456, 4209}

\bibitem[\protect\citeauthoryear{{Ellingsen}, {Voronkov}, {Breen}  \&
  {Lovell}}{{Ellingsen} et~al.}{2012}]{Ellingsen:2012}
{Ellingsen} S.~P.,  {Voronkov} M.~A.,  {Breen} S.~L.,   {Lovell} J.~E.~J.,
  2012, \mn@doi [ApJL] {10.1088/2041-8205/747/1/L7}, \href
  {http://adsabs.harvard.edu/abs/2012ApJ...747L...7E} {747, L7}

\bibitem[\protect\citeauthoryear{{Ellison}, {Patton}, {Mendel}  \&
  {Scudder}}{{Ellison} et~al.}{2011}]{Ellison:2011}
{Ellison} S.~L.,  {Patton} D.~R.,  {Mendel} J.~T.,   {Scudder} J.~M.,  2011,
  \mn@doi [MNRAS] {10.1111/j.1365-2966.2011.19624.x}, \href
  {http://adsabs.harvard.edu/abs/2011MNRAS.418.2043E} {418, 2043}

\bibitem[\protect\citeauthoryear{{Ellison}, {Mendel}, {Patton}  \&
  {Scudder}}{{Ellison} et~al.}{2013}]{Ellison:2013}
{Ellison} S.~L.,  {Mendel} J.~T.,  {Patton} D.~R.,   {Scudder} J.~M.,  2013,
  \mn@doi [MNRAS] {10.1093/mnras/stt1562}, \href
  {http://adsabs.harvard.edu/abs/2013MNRAS.435.3627E} {435, 3627}

\bibitem[\protect\citeauthoryear{{Fabian}}{{Fabian}}{1999}]{Fabian:1999}
{Fabian} A.~C.,  1999, \mn@doi [MNRAS] {10.1046/j.1365-8711.1999.03017.x},
  \href {http://adsabs.harvard.edu/abs/1999MNRAS.308L..39F} {308, L39}

\bibitem[\protect\citeauthoryear{{Fey}, {Gordon}  \& {Jacobs}}{{Fey}
  et~al.}{2009}]{Fey:2009}
{Fey} A.,  {Gordon} G.,   {Jacobs} C.,  eds, 2009, {The Second Realization of
  the International Celestial Reference Frame by VLBI, IERS Technical Notes
  35}.
Verlad des Bundesamts fur Kartographie und Geodasie, Frankfurt am Main

\bibitem[\protect\citeauthoryear{{Gaensler} \& {Hunstead}}{{Gaensler} \&
  {Hunstead}}{2000}]{Gaensler:2000}
{Gaensler} B.~M.,  {Hunstead} R.~W.,  2000, \mn@doi [PASA] {10.1071/AS00072},
  \href {http://adsabs.harvard.edu/abs/2000PASA...17...72G} {17, 72}

\bibitem[\protect\citeauthoryear{{Gaspari}, {Ruszkowski}  \& {Oh}}{{Gaspari}
  et~al.}{2013}]{Gaspari:2013}
{Gaspari} M.,  {Ruszkowski} M.,   {Oh} S.~P.,  2013, \mn@doi [MNRAS]
  {10.1093/mnras/stt692}, \href
  {http://adsabs.harvard.edu/abs/2013MNRAS.432.3401G} {432, 3401}

\bibitem[\protect\citeauthoryear{{Gibson}, {Taylor}, {Higgs}, {Brunt}  \&
  {Dewdney}}{{Gibson} et~al.}{2005}]{Gibson:2005}
{Gibson} S.~J.,  {Taylor} A.~R.,  {Higgs} L.~A.,  {Brunt} C.~M.,   {Dewdney}
  P.~E.,  2005, \mn@doi [\apj] {10.1086/429870}, \href
  {http://adsabs.harvard.edu/abs/2005ApJ...626..195G} {626, 195}

\bibitem[\protect\citeauthoryear{{Gregory}, {Vavasour}, {Scott}  \&
  {Condon}}{{Gregory} et~al.}{1994}]{Gregory:1994}
{Gregory} P.~C.,  {Vavasour} J.~D.,  {Scott} W.~K.,   {Condon} J.~J.,  1994,
  \mn@doi [ApJS] {10.1086/191862}, \href
  {http://adsabs.harvard.edu/abs/1994ApJS...90..173G} {90, 173}

\bibitem[\protect\citeauthoryear{{Gupta}, {Srianand}, {Petitjean}, {Bergeron},
  {Noterdaeme}  \& {Muzahid}}{{Gupta} et~al.}{2012}]{Gupta:2012}
{Gupta} N.,  {Srianand} R.,  {Petitjean} P.,  {Bergeron} J.,  {Noterdaeme} P.,
   {Muzahid} S.,  2012, \mn@doi [A\&A] {10.1051/0004-6361/201219159}, \href
  {http://adsabs.harvard.edu/abs/2012A%26A...544A..21G} {544, A21}

\bibitem[\protect\citeauthoryear{{Gupta} et~al.,}{{Gupta}
  et~al.}{2016}]{Gupta:2016}
{Gupta} N.,  et~al., 2016, in Proceedings of MeerKAT Science: On the Pathway to
  the SKA. 25-27 May, 2016 Stellenbosch, South Africa (MeerKAT2016). p.~14
  (\mn@eprint {arXiv} {1708.07371})

\bibitem[\protect\citeauthoryear{{Gupta} et~al.,}{{Gupta}
  et~al.}{2018}]{Gupta:2018}
{Gupta} N.,  et~al., 2018, \mn@doi [MNRAS] {10.1093/mnras/sty384}, \href
  {http://adsabs.harvard.edu/abs/2018MNRAS.476.2432G} {476, 2432}

\bibitem[\protect\citeauthoryear{{Hardcastle}, {Evans}  \&
  {Croston}}{{Hardcastle} et~al.}{2007}]{Hardcastle:2007}
{Hardcastle} M.~J.,  {Evans} D.~A.,   {Croston} J.~H.,  2007, \mn@doi [MNRAS]
  {10.1111/j.1365-2966.2007.11572.x}, \href
  {http://adsabs.harvard.edu/abs/2007MNRAS.376.1849H} {376, 1849}

\bibitem[\protect\citeauthoryear{{Hay} \& {O'Sullivan}}{{Hay} \&
  {O'Sullivan}}{2008}]{Hay:2008}
{Hay} S.~G.,  {O'Sullivan} J.~D.,  2008, \mn@doi [Radio Sci.]
  {10.1029/2007RS003798}, 43, 6

\bibitem[\protect\citeauthoryear{{Healey}, {Romani}, {Taylor}, {Sadler},
  {Ricci}, {Murphy}, {Ulvestad}  \& {Winn}}{{Healey}
  et~al.}{2007}]{Healey:2007}
{Healey} S.~E.,  {Romani} R.~W.,  {Taylor} G.~B.,  {Sadler} E.~M.,  {Ricci} R.,
   {Murphy} T.,  {Ulvestad} J.~S.,   {Winn} J.~N.,  2007, \mn@doi [ApJS]
  {10.1086/513742}, \href {http://adsabs.harvard.edu/abs/2007ApJS..171...61H}
  {171, 61}

\bibitem[\protect\citeauthoryear{{Heckman} \& {Best}}{{Heckman} \&
  {Best}}{2014}]{Heckman:2014}
{Heckman} T.~M.,  {Best} P.~N.,  2014, \mn@doi [ARA\&A]
  {10.1146/annurev-astro-081913-035722}, \href
  {http://adsabs.harvard.edu/abs/2014ARA%26A..52..589H} {52, 589}

\bibitem[\protect\citeauthoryear{{Heiles} \& {Troland}}{{Heiles} \&
  {Troland}}{2003}]{Heiles:2003b}
{Heiles} C.,  {Troland} T.~H.,  2003, \mn@doi [ApJ] {10.1086/367828}, \href
  {http://adsabs.harvard.edu/abs/2003ApJ...586.1067H} {586, 1067}

\bibitem[\protect\citeauthoryear{{Helfer}, {Thornley}, {Regan}, {Wong},
  {Sheth}, {Vogel}, {Blitz}  \& {Bock}}{{Helfer} et~al.}{2003}]{Helfer:2003}
{Helfer} T.~T.,  {Thornley} M.~D.,  {Regan} M.~W.,  {Wong} T.,  {Sheth} K.,
  {Vogel} S.~N.,  {Blitz} L.,   {Bock} D.~C.-J.,  2003, \mn@doi [ApJS]
  {10.1086/346076}, \href {http://adsabs.harvard.edu/abs/2003ApJS..145..259H}
  {145, 259}

\bibitem[\protect\citeauthoryear{{Henkel}, {Jethava}, {Kraus}, {Menten},
  {Carilli}, {Grasshoff}, {Lubowich}  \& {Reid}}{{Henkel}
  et~al.}{2005}]{Henkel:2005}
{Henkel} C.,  {Jethava} N.,  {Kraus} A.,  {Menten} K.~M.,  {Carilli} C.~L.,
  {Grasshoff} M.,  {Lubowich} D.,   {Reid} M.~J.,  2005, \mn@doi [\aap]
  {10.1051/0004-6361:20052816}, \href
  {http://adsabs.harvard.edu/abs/2005A%26A...440..893H} {440, 893}

\bibitem[\protect\citeauthoryear{{Heyer} \& {Dame}}{{Heyer} \&
  {Dame}}{2015}]{Heyer:2015}
{Heyer} M.,  {Dame} T.~M.,  2015, \mn@doi [AR\&A]
  {10.1146/annurev-astro-082214-122324}, \href
  {http://adsabs.harvard.edu/abs/2015ARA%26A..53..583H} {53, 583}

\bibitem[\protect\citeauthoryear{{Heyer}, {Krawczyk}, {Duval}  \&
  {Jackson}}{{Heyer} et~al.}{2009}]{Heyer:2009}
{Heyer} M.,  {Krawczyk} C.,  {Duval} J.,   {Jackson} J.~M.,  2009, \mn@doi
  [ApJ] {10.1088/0004-637X/699/2/1092}, \href
  {http://adsabs.harvard.edu/abs/2009ApJ...699.1092H} {699, 1092}

\bibitem[\protect\citeauthoryear{{Holt}, {Tadhunter}, {Morganti}, {Bellamy},
  {Gonz{\'a}lez Delgado}, {Tzioumis}  \& {Inskip}}{{Holt}
  et~al.}{2006}]{Holt:2006}
{Holt} J.,  {Tadhunter} C.,  {Morganti} R.,  {Bellamy} M.,  {Gonz{\'a}lez
  Delgado} R.~M.,  {Tzioumis} A.,   {Inskip} K.~J.,  2006, \mn@doi [MNRAS]
  {10.1111/j.1365-2966.2006.10604.x}, \href
  {http://adsabs.harvard.edu/abs/2006MNRAS.370.1633H} {370, 1633}

\bibitem[\protect\citeauthoryear{{Holt}, {Tadhunter}  \& {Morganti}}{{Holt}
  et~al.}{2008}]{Holt:2008}
{Holt} J.,  {Tadhunter} C.~N.,   {Morganti} R.,  2008, \mn@doi [MNRAS]
  {10.1111/j.1365-2966.2008.13089.x}, \href
  {http://adsabs.harvard.edu/abs/2008MNRAS.387..639H} {387, 639}

\bibitem[\protect\citeauthoryear{{Hopkins} \& {Hernquist}}{{Hopkins} \&
  {Hernquist}}{2006}]{Hopkins:2006}
{Hopkins} P.~F.,  {Hernquist} L.,  2006, \mn@doi [ApJS] {10.1086/505753}, \href
  {http://adsabs.harvard.edu/abs/2006ApJS..166....1H} {166, 1}

\bibitem[\protect\citeauthoryear{{Hopkins}, {Hernquist}, {Cox}  \& {Kere{\v
  s}}}{{Hopkins} et~al.}{2008}]{Hopkins:2008}
{Hopkins} P.~F.,  {Hernquist} L.,  {Cox} T.~J.,   {Kere{\v s}} D.,  2008,
  \mn@doi [ApJS] {10.1086/524362}, \href
  {http://adsabs.harvard.edu/abs/2008ApJS..175..356H} {175, 356}

\bibitem[\protect\citeauthoryear{Hotan et~al.,}{Hotan
  et~al.}{2014}]{Hotan:2014}
Hotan A.~W.,  et~al., 2014, \mn@doi [Publ. Astron. Soc. Aust.]
  {10.1017/pasa.2014.36}, 31, e041

\bibitem[\protect\citeauthoryear{{Hurley-Walker} et~al.,}{{Hurley-Walker}
  et~al.}{2017}]{Hurley-Walker:2017}
{Hurley-Walker} N.,  et~al., 2017, \mn@doi [MNRAS] {10.1093/mnras/stw2337},
  \href {http://adsabs.harvard.edu/abs/2017MNRAS.464.1146H} {464, 1146}

\bibitem[\protect\citeauthoryear{{Intema}, {Jagannathan}, {Mooley}  \&
  {Frail}}{{Intema} et~al.}{2017}]{Intema:2017}
{Intema} H.~T.,  {Jagannathan} P.,  {Mooley} K.~P.,   {Frail} D.~A.,  2017,
  \mn@doi [A\&A] {10.1051/0004-6361/201628536}, \href
  {http://adsabs.harvard.edu/abs/2017A%26A...598A..78I} {598, A78}

\bibitem[\protect\citeauthoryear{{Ishibashi} \& {Fabian}}{{Ishibashi} \&
  {Fabian}}{2012}]{Ishibashi:2012}
{Ishibashi} W.,  {Fabian} A.~C.,  2012, \mn@doi [MNRAS]
  {10.1111/j.1365-2966.2012.22074.x}, \href
  {http://adsabs.harvard.edu/abs/2012MNRAS.427.2998I} {427, 2998}

\bibitem[\protect\citeauthoryear{{Ishwara-Chandra}, {Dwarakanath}  \&
  {Anantharamaiah}}{{Ishwara-Chandra} et~al.}{2003}]{Ishwara-Chandra:2003}
{Ishwara-Chandra} C.~H.,  {Dwarakanath} K.~S.,   {Anantharamaiah} K.~R.,  2003,
  \mn@doi [J. Astrophys. Astron.] {10.1007/BF03012190}, \href
  {http://adsabs.harvard.edu/abs/2003JApA...24...37I} {24, 37}

\bibitem[\protect\citeauthoryear{{Jacobs} et~al.,}{{Jacobs}
  et~al.}{2011}]{Jacobs:2011}
{Jacobs} D.~C.,  et~al., 2011, \mn@doi [ApJ] {10.1088/2041-8205/734/2/L34},
  \href {http://adsabs.harvard.edu/abs/2011ApJ...734L..34J} {734, L34}

\bibitem[\protect\citeauthoryear{{Jahnke} \& {Macci{\`o}}}{{Jahnke} \&
  {Macci{\`o}}}{2011}]{Jahnke:2011}
{Jahnke} K.,  {Macci{\`o}} A.~V.,  2011, \mn@doi [ApJ]
  {10.1088/0004-637X/734/2/92}, \href
  {http://adsabs.harvard.edu/abs/2011ApJ...734...92J} {734, 92}

\bibitem[\protect\citeauthoryear{{Jarvis} et~al.,}{{Jarvis}
  et~al.}{2017}]{Jarvis:2017}
{Jarvis} M.~J.,  et~al., 2017, preprint, \href
  {http://adsabs.harvard.edu/abs/2017arXiv170901901J} {} (\mn@eprint {arXiv}
  {1709.01901})

\bibitem[\protect\citeauthoryear{{Jauncey} et~al.,}{{Jauncey}
  et~al.}{2003}]{Jauncey:2003}
{Jauncey} D.~L.,  et~al., 2003, \mn@doi [Publ. Astron. Soc. Aust.]
  {10.1071/AS03023}, \href {http://adsabs.harvard.edu/abs/2003PASA...20..151J}
  {20, 151}

\bibitem[\protect\citeauthoryear{{Johnston}, {Hunstead}, {Cotter}  \&
  {Sadler}}{{Johnston} et~al.}{2005}]{Johnston:2005}
{Johnston} H.~M.,  {Hunstead} R.~W.,  {Cotter} G.,   {Sadler} E.~M.,  2005,
  \mn@doi [MNRAS] {10.1111/j.1365-2966.2004.08468.x}, \href
  {http://adsabs.harvard.edu/abs/2005MNRAS.356..515J} {356, 515}

\bibitem[\protect\citeauthoryear{{Johnston} et~al.,}{{Johnston}
  et~al.}{2007}]{Johnston:2007}
{Johnston} S.,  et~al., 2007, \mn@doi [PASA] {10.1071/AS07033}, \href
  {http://adsabs.harvard.edu/abs/2007PASA...24..174J} {24, 174}

\bibitem[\protect\citeauthoryear{{Johnston}, {Broderick}, {Cotter}, {Morganti}
  \& {Hunstead}}{{Johnston} et~al.}{2010}]{Johnston:2010}
{Johnston} H.~M.,  {Broderick} J.~W.,  {Cotter} G.,  {Morganti} R.,
  {Hunstead} R.~W.,  2010, \mn@doi [MNRAS] {10.1111/j.1365-2966.2010.16950.x},
  \href {http://adsabs.harvard.edu/abs/2010MNRAS.407..721J} {407, 721}

\bibitem[\protect\citeauthoryear{{Jones}, {Ryu}  \& {Engel}}{{Jones}
  et~al.}{1999}]{Jones:1999}
{Jones} T.~W.,  {Ryu} D.,   {Engel} A.,  1999, \mn@doi [ApJ] {10.1086/306772},
  \href {http://adsabs.harvard.edu/abs/1999ApJ...512..105J} {512, 105}

\bibitem[\protect\citeauthoryear{{Kanekar}}{{Kanekar}}{2011}]{Kanekar:2011b}
{Kanekar} N.,  2011, \mn@doi [ApJL] {10.1088/2041-8205/728/1/L12}, \href
  {http://adsabs.harvard.edu/abs/2011ApJ...728L..12K} {728, L12}

\bibitem[\protect\citeauthoryear{{Kanekar} \& {Briggs}}{{Kanekar} \&
  {Briggs}}{2004}]{Kanekar:2004}
{Kanekar} N.,  {Briggs} F.~H.,  2004, \mn@doi [New Astron. Rev.]
  {10.1016/j.newar.2004.09.030}, \href
  {http://adsabs.harvard.edu/abs/2004NewAR..48.1259K} {48, 1259}

\bibitem[\protect\citeauthoryear{{Kanekar} et~al.,}{{Kanekar}
  et~al.}{2005}]{Kanekar:2005}
{Kanekar} N.,  et~al., 2005, \mn@doi [Physical Review Letters]
  {10.1103/PhysRevLett.95.261301}, \href
  {http://adsabs.harvard.edu/abs/2005PhRvL..95z1301K} {95, 261301}

\bibitem[\protect\citeauthoryear{{Kanekar}, {Langston}, {Stocke}, {Carilli}  \&
  {Menten}}{{Kanekar} et~al.}{2012}]{Kanekar:2012}
{Kanekar} N.,  {Langston} G.~I.,  {Stocke} J.~T.,  {Carilli} C.~L.,   {Menten}
  K.~M.,  2012, \mn@doi [ApJL] {10.1088/2041-8205/746/2/L16}, \href
  {http://adsabs.harvard.edu/abs/2012ApJ...746L..16K} {746, L16}

\bibitem[\protect\citeauthoryear{{Kanekar}, {Gupta}, {Carilli}, {Stocke}  \&
  {Willett}}{{Kanekar} et~al.}{2014}]{Kanekar:2014}
{Kanekar} N.,  {Gupta} A.,  {Carilli} C.~L.,  {Stocke} J.~T.,   {Willett}
  K.~W.,  2014, \mn@doi [ApJ] {10.1088/0004-637X/782/1/56}, \href
  {http://adsabs.harvard.edu/abs/2014ApJ...782...56K} {782, 56}

\bibitem[\protect\citeauthoryear{{Kanekar} et~al.,}{{Kanekar}
  et~al.}{2015}]{Kanekar:2015b}
{Kanekar} N.,  et~al., 2015, \mn@doi [MNRAS] {10.1093/mnrasl/slu206}, \href
  {http://adsabs.harvard.edu/abs/2015MNRAS.448L.104K} {448, L104}

\bibitem[\protect\citeauthoryear{{Kanekar}, {Ghosh}  \& {Chengalur}}{{Kanekar}
  et~al.}{2018}]{Kanekar:2018}
{Kanekar} N.,  {Ghosh} T.,   {Chengalur} J.~N.,  2018, \mn@doi [Physical Review
  Letters] {10.1103/PhysRevLett.120.061302}, \href
  {http://adsabs.harvard.edu/abs/2018PhRvL.120f1302K} {120, 061302}

\bibitem[\protect\citeauthoryear{{Kardashev}}{{Kardashev}}{1962}]{Kardashev:1962}
{Kardashev} N.~S.,  1962, Soviet Ast., \href
  {http://adsabs.harvard.edu/abs/1962SvA.....6..317K} {6, 317}

\bibitem[\protect\citeauthoryear{{Kauffmann} \& {Heckman}}{{Kauffmann} \&
  {Heckman}}{2009}]{Kauffmann:2009}
{Kauffmann} G.,  {Heckman} T.~M.,  2009, \mn@doi [MNRAS]
  {10.1111/j.1365-2966.2009.14960.x}, \href
  {http://adsabs.harvard.edu/abs/2009MNRAS.397..135K} {397, 135}

\bibitem[\protect\citeauthoryear{{Kauffmann} et~al.,}{{Kauffmann}
  et~al.}{2003}]{Kauffmann:2003}
{Kauffmann} G.,  et~al., 2003, \mn@doi [MNRAS]
  {10.1111/j.1365-2966.2003.07154.x}, \href
  {http://adsabs.harvard.edu/abs/2003MNRAS.346.1055K} {346, 1055}

\bibitem[\protect\citeauthoryear{{Kauffmann} et~al.,}{{Kauffmann}
  et~al.}{2007}]{Kauffmann:2007}
{Kauffmann} G.,  et~al., 2007, \mn@doi [ApJS] {10.1086/516647}, \href
  {http://adsabs.harvard.edu/abs/2007ApJS..173..357K} {173, 357}

\bibitem[\protect\citeauthoryear{{Kellermann} \& {Pauliny-Toth}}{{Kellermann}
  \& {Pauliny-Toth}}{1981}]{Kellerman:1981}
{Kellermann} K.~I.,  {Pauliny-Toth} I.~I.~K.,  1981, \mn@doi [ARA\&A]
  {10.1146/annurev.aa.19.090181.002105}, \href
  {http://adsabs.harvard.edu/abs/1981ARA%26A..19..373K} {19, 373}

\bibitem[\protect\citeauthoryear{{Kennicutt}}{{Kennicutt}}{1998}]{Kennicutt:1998}
{Kennicutt} Jr. R.~C.,  1998, \mn@doi [ApJ] {10.1086/305588}, \href
  {http://adsabs.harvard.edu/abs/1998ApJ...498..541K} {498, 541}

\bibitem[\protect\citeauthoryear{{Kewley}, {Geller}  \& {Jansen}}{{Kewley}
  et~al.}{2004}]{Kewley:2004}
{Kewley} L.~J.,  {Geller} M.~J.,   {Jansen} R.~A.,  2004, \mn@doi [AJ]
  {10.1086/382723}, \href {http://adsabs.harvard.edu/abs/2004AJ....127.2002K}
  {127, 2002}

\bibitem[\protect\citeauthoryear{{King}}{{King}}{1994}]{King:1994}
{King} E.,  1994, PhD thesis, Univ. Tasmania, Hobart

\bibitem[\protect\citeauthoryear{{King}}{{King}}{2003}]{King:2003}
{King} A.,  2003, \mn@doi [ApJL] {10.1086/379143}, \href
  {http://adsabs.harvard.edu/abs/2003ApJ...596L..27K} {596, L27}

\bibitem[\protect\citeauthoryear{{Kormendy} \& {Ho}}{{Kormendy} \&
  {Ho}}{2013}]{Kormendy:2013}
{Kormendy} J.,  {Ho} L.~C.,  2013, \mn@doi [AR\&A]
  {10.1146/annurev-astro-082708-101811}, \href
  {http://adsabs.harvard.edu/abs/2013ARA%26A..51..511K} {51, 511}

\bibitem[\protect\citeauthoryear{{Krumholz}, {McKee}  \&
  {Tumlinson}}{{Krumholz} et~al.}{2009}]{Krumholz:2009}
{Krumholz} M.~R.,  {McKee} C.~F.,   {Tumlinson} J.,  2009, \mn@doi [ApJ]
  {10.1088/0004-637X/693/1/216}, \href
  {http://adsabs.harvard.edu/abs/2009ApJ...693..216K} {693, 216}

\bibitem[\protect\citeauthoryear{{LaMassa}, {Heckman}, {Ptak}  \&
  {Urry}}{{LaMassa} et~al.}{2013}]{LaMassa:2013}
{LaMassa} S.~M.,  {Heckman} T.~M.,  {Ptak} A.,   {Urry} C.~M.,  2013, \mn@doi
  [ApJL] {10.1088/2041-8205/765/2/L33}, \href
  {http://adsabs.harvard.edu/abs/2013ApJ...765L..33L} {765, L33}

\bibitem[\protect\citeauthoryear{{Lane}, {Briggs}  \& {Smette}}{{Lane}
  et~al.}{2000}]{Lane:2000}
{Lane} W.~M.,  {Briggs} F.~H.,   {Smette} A.,  2000, \mn@doi [ApJ]
  {10.1086/308578}, \href {http://adsabs.harvard.edu/abs/2000ApJ...532..146L}
  {532, 146}

\bibitem[\protect\citeauthoryear{{Large}, {Mills}, {Little}, {Crawford}  \&
  {Sutton}}{{Large} et~al.}{1981}]{Large:1981}
{Large} M.~I.,  {Mills} B.~Y.,  {Little} A.~G.,  {Crawford} D.~F.,   {Sutton}
  J.~M.,  1981, MNRAS, \href
  {http://adsabs.harvard.edu/abs/1981MNRAS.194..693L} {194, 693}

\bibitem[\protect\citeauthoryear{{Larson}}{{Larson}}{1981}]{Larson:1981}
{Larson} R.~B.,  1981, \mn@doi [MNRAS] {10.1093/mnras/194.4.809}, \href
  {http://adsabs.harvard.edu/abs/1981MNRAS.194..809L} {194, 809}

\bibitem[\protect\citeauthoryear{{Li} \& {Bryan}}{{Li} \&
  {Bryan}}{2014}]{Li:2014a}
{Li} Y.,  {Bryan} G.~L.,  2014, \mn@doi [ApJ] {10.1088/0004-637X/789/2/153},
  \href {http://adsabs.harvard.edu/abs/2014ApJ...789..153L} {789, 153}

\bibitem[\protect\citeauthoryear{{Lim}, {Ao}  \& {Dinh-V-Trung}}{{Lim}
  et~al.}{2008}]{Lim:2008}
{Lim} J.,  {Ao} Y.,   {Dinh-V-Trung} 2008, \mn@doi [ApJ] {10.1086/523664},
  \href {http://adsabs.harvard.edu/abs/2008ApJ...672..252L} {672, 252}

\bibitem[\protect\citeauthoryear{{Liu}, {Song}  \& {Cui}}{{Liu}
  et~al.}{2009}]{Liu:2009}
{Liu} X.,  {Song} H.-G.,   {Cui} L.,  2009, \mn@doi [Astronomische Nachrichten]
  {10.1002/asna.200811142}, \href
  {http://adsabs.harvard.edu/abs/2009AN....330..145L} {330, 145}

\bibitem[\protect\citeauthoryear{{L{\'o}pez-Caniego}, {Gonz{\'a}lez-Nuevo},
  {Herranz}, {Massardi}, {Sanz}, {De Zotti}, {Toffolatti}  \&
  {Arg{\"u}eso}}{{L{\'o}pez-Caniego} et~al.}{2007}]{Lopez-Caniego:2007}
{L{\'o}pez-Caniego} M.,  {Gonz{\'a}lez-Nuevo} J.,  {Herranz} D.,  {Massardi}
  M.,  {Sanz} J.~L.,  {De Zotti} G.,  {Toffolatti} L.,   {Arg{\"u}eso} F.,
  2007, \mn@doi [ApJS] {10.1086/512678}, \href
  {http://cdsads.u-strasbg.fr/abs/2007ApJS..170..108L} {170, 108}

\bibitem[\protect\citeauthoryear{{Maccagni}, {Morganti}, {Oosterloo}  \&
  {Mahony}}{{Maccagni} et~al.}{2014}]{Maccagni:2014}
{Maccagni} F.~M.,  {Morganti} R.,  {Oosterloo} T.~A.,   {Mahony} E.~K.,  2014,
  \mn@doi [A\&A] {10.1051/0004-6361/201424334}, \href
  {http://adsabs.harvard.edu/abs/2014A%26A...571A..67M} {571, A67}

\bibitem[\protect\citeauthoryear{{Maccagni}, {Morganti}, {Oosterloo},
  {Ger{\'e}b}  \& {Maddox}}{{Maccagni} et~al.}{2017}]{Maccagni:2017}
{Maccagni} F.~M.,  {Morganti} R.,  {Oosterloo} T.~A.,  {Ger{\'e}b} K.,
  {Maddox} N.,  2017, \mn@doi [A\&A] {10.1051/0004-6361/201730563}, \href
  {http://adsabs.harvard.edu/abs/2017A%26A...604A..43M} {604, A43}

\bibitem[\protect\citeauthoryear{{Maccagni}, {Morganti}, {Oosterloo}, {Oonk}
  \& {Emonts}}{{Maccagni} et~al.}{2018}]{Maccagni:2018}
{Maccagni} F.~M.,  {Morganti} R.,  {Oosterloo} T.~A.,  {Oonk} J.~B.~R.,
  {Emonts} B.~H.~C.,  2018, \mn@doi [A\&A] {10.1051/0004-6361/201732269}, \href
  {http://adsabs.harvard.edu/abs/2018A%26A...614A..42M} {614, A42}

\bibitem[\protect\citeauthoryear{{Marx-Zimmer}, {Herbstmeier}, {Dickey},
  {Zimmer}, {Staveley-Smith}  \& {Mebold}}{{Marx-Zimmer}
  et~al.}{2000}]{Marx-Zimmer:2000}
{Marx-Zimmer} M.,  {Herbstmeier} U.,  {Dickey} J.~M.,  {Zimmer} F.,
  {Staveley-Smith} L.,   {Mebold} U.,  2000, A\&A, \href
  {http://adsabs.harvard.edu/abs/2000A%26A...354..787M} {354, 787}

\bibitem[\protect\citeauthoryear{{Massardi} et~al.,}{{Massardi}
  et~al.}{2008}]{Massardi:2008}
{Massardi} M.,  et~al., 2008, \mn@doi [MNRAS]
  {10.1111/j.1365-2966.2007.12751.x}, \href
  {http://cdsads.u-strasbg.fr/abs/2008MNRAS.384..775M} {384, 775}

\bibitem[\protect\citeauthoryear{{Mauch}, {Murphy}, {Buttery}, {Curran},
  {Hunstead}, {Piestrzynski}, {Robertson}  \& {Sadler}}{{Mauch}
  et~al.}{2003}]{Mauch:2003}
{Mauch} T.,  {Murphy} T.,  {Buttery} H.~J.,  {Curran} J.,  {Hunstead} R.~W.,
  {Piestrzynski} B.,  {Robertson} J.~G.,   {Sadler} E.~M.,  2003, \mn@doi
  [MNRAS] {10.1046/j.1365-8711.2003.06605.x}, \href
  {http://adsabs.harvard.edu/abs/2003MNRAS.342.1117M} {342, 1117}

\bibitem[\protect\citeauthoryear{{McConnell}, {Sadler}, {Murphy}  \&
  {Ekers}}{{McConnell} et~al.}{2012}]{McConnell:2012}
{McConnell} D.,  {Sadler} E.~M.,  {Murphy} T.,   {Ekers} R.~D.,  2012, \mn@doi
  [MNRAS] {10.1111/j.1365-2966.2012.20726.x}, \href
  {http://cdsads.u-strasbg.fr/abs/2012MNRAS.422.1527M} {422, 1527}

\bibitem[\protect\citeauthoryear{{McConnell} et~al.,}{{McConnell}
  et~al.}{2016}]{McConnell:2016}
{McConnell} D.,  et~al., 2016, \mn@doi [PASA] {10.1017/pasa.2016.37}, \href
  {http://adsabs.harvard.edu/abs/2016PASA...33...42M} {33, e042}

\bibitem[\protect\citeauthoryear{{McMullin}, {Waters}, {Schiebel}, {Young}  \&
  {Golap}}{{McMullin} et~al.}{2007}]{McMullin:2007}
{McMullin} J.~P.,  {Waters} B.,  {Schiebel} D.,  {Young} W.,   {Golap} K.,
  2007, in {Shaw} R.~A.,  {Hill} F.,   {Bell} D.~J.,  eds,  ASP Conf. Ser. Vol.
  376, Astronomical Data Analysis Software and Systems XVI. Astron. Soc. Pac.,
  San Francisco, p.~127

\bibitem[\protect\citeauthoryear{{McNamara} et~al.,}{{McNamara}
  et~al.}{2014}]{McNamara:2014}
{McNamara} B.~R.,  et~al., 2014, \mn@doi [ApJ] {10.1088/0004-637X/785/1/44},
  \href {http://adsabs.harvard.edu/abs/2014ApJ...785...44M} {785, 44}

\bibitem[\protect\citeauthoryear{{Meidt} et~al.,}{{Meidt}
  et~al.}{2014}]{Meidt:2014}
{Meidt} S.~E.,  et~al., 2014, \mn@doi [ApJ] {10.1088/0004-637X/788/2/144},
  \href {http://adsabs.harvard.edu/abs/2014ApJ...788..144M} {788, 144}

\bibitem[\protect\citeauthoryear{{Moffet}}{{Moffet}}{1975}]{Moffet:1975}
{Moffet} A.~T.,  1975, in {Sandage} A.,  {Sandage} M.,   {Kristian} J.,  eds,
  Stars and Stellar Systems Vol. 9, Galaxies and the Universe. University of
  Chicago Press, Chicago, p.~211

\bibitem[\protect\citeauthoryear{{Moore}, {Carilli}  \& {Menten}}{{Moore}
  et~al.}{1999}]{Moore:1999}
{Moore} C.~B.,  {Carilli} C.~L.,   {Menten} K.~M.,  1999, \mn@doi [ApJ]
  {10.1086/311818}, \href {http://adsabs.harvard.edu/abs/1999ApJ...510L..87M}
  {510, L87}

\bibitem[\protect\citeauthoryear{{Moreno}, {Bluck}, {Ellison}, {Patton},
  {Torrey}  \& {Moster}}{{Moreno} et~al.}{2013}]{Moreno:2013}
{Moreno} J.,  {Bluck} A.~F.~L.,  {Ellison} S.~L.,  {Patton} D.~R.,  {Torrey}
  P.,   {Moster} B.~P.,  2013, \mn@doi [MNRAS] {10.1093/mnras/stt1694}, \href
  {http://adsabs.harvard.edu/abs/2013MNRAS.436.1765M} {436, 1765}

\bibitem[\protect\citeauthoryear{{Morganti} \& {Oosterloo}}{{Morganti} \&
  {Oosterloo}}{2018}]{Morganti:2018}
{Morganti} R.,  {Oosterloo} T.,  2018, preprint, \href
  {http://adsabs.harvard.edu/abs/2018arXiv180701475M} {} (\mn@eprint {arXiv}
  {1807.01475})

\bibitem[\protect\citeauthoryear{{Morganti}, {Tadhunter}  \&
  {Oosterloo}}{{Morganti} et~al.}{2005}]{Morganti:2005b}
{Morganti} R.,  {Tadhunter} C.~N.,   {Oosterloo} T.~A.,  2005, \mn@doi [A\&A]
  {10.1051/0004-6361:200500197}, \href
  {http://adsabs.harvard.edu/abs/2005A26A...444L...9M} {444, L9}

\bibitem[\protect\citeauthoryear{{Morganti}, {Sadler}  \& {Curran}}{{Morganti}
  et~al.}{2015}]{Morganti:2015}
{Morganti} R.,  {Sadler} E.~M.,   {Curran} S.,  2015, in {Bourke} T.,  {Braun}
  R.,  {Fender} R.~P.,   {et al.} eds, Advancing Astrophysics with the Square
  Kilometre Array (AASKA14). Proc. Sci., p.~134 (\mn@eprint {arXiv}
  {1501.01091})

\bibitem[\protect\citeauthoryear{{Moss} et~al.,}{{Moss}
  et~al.}{2017}]{Moss:2017}
{Moss} V.~A.,  et~al., 2017, \mn@doi [MNRAS] {10.1093/mnras/stx1679}, \href
  {http://adsabs.harvard.edu/abs/2017MNRAS.471.2952M} {471, 2952}

\bibitem[\protect\citeauthoryear{{Mukherjee}, {Bicknell}, {Wagner},
  {Sutherland}  \& {Silk}}{{Mukherjee} et~al.}{2018}]{Mukherjee:2018}
{Mukherjee} D.,  {Bicknell} G.~V.,  {Wagner} A.~Y.,  {Sutherland} R.~S.,
  {Silk} J.,  2018, \mn@doi [MNRAS] {10.1093/mnras/sty1776}, \href
  {http://adsabs.harvard.edu/abs/2018MNRAS.tmp.1699M} {}

\bibitem[\protect\citeauthoryear{{Muller} et~al.,}{{Muller}
  et~al.}{2013}]{Muller:2013}
{Muller} S.,  et~al., 2013, \mn@doi [A\&A] {10.1051/0004-6361/201220613}, \href
  {http://adsabs.harvard.edu/abs/2013A\%26A...551A.109M} {551, A109}

\bibitem[\protect\citeauthoryear{{Muller} et~al.,}{{Muller}
  et~al.}{2014}]{Muller:2014}
{Muller} S.,  et~al., 2014, \mn@doi [A\&A] {10.1051/0004-6361/201423646}, \href
  {http://adsabs.harvard.edu/abs/2014A\%26A...566A.112M} {566, A112}

\bibitem[\protect\citeauthoryear{{Murgia}}{{Murgia}}{2003}]{Murgia:2003}
{Murgia} M.,  2003, \mn@doi [Publ. Astron. Soc. Aust.] {10.1071/AS02033}, \href
  {http://adsabs.harvard.edu/abs/2003PASA...20...19M} {20, 19}

\bibitem[\protect\citeauthoryear{{Murphy}, {Webb}, {Flambaum}, {Drinkwater},
  {Combes}  \& {Wiklind}}{{Murphy} et~al.}{2001}]{Murphy:2001}
{Murphy} M.~T.,  {Webb} J.~K.,  {Flambaum} V.~V.,  {Drinkwater} M.~J.,
  {Combes} F.,   {Wiklind} T.,  2001, \mn@doi [MNRAS]
  {10.1046/j.1365-8711.2001.04843.x}, \href
  {http://adsabs.harvard.edu/abs/2001MNRAS.327.1244M} {327, 1244}

\bibitem[\protect\citeauthoryear{{Murphy}, {Flambaum}, {Muller}  \&
  {Henkel}}{{Murphy} et~al.}{2008}]{Murphy:2008}
{Murphy} M.~T.,  {Flambaum} V.~V.,  {Muller} S.,   {Henkel} C.,  2008, \mn@doi
  [Science] {10.1126/science.1156352}, \href
  {http://adsabs.harvard.edu/abs/2008Sci...320.1611M} {320, 1611}

\bibitem[\protect\citeauthoryear{{Murray}, {Stanimirovic}, {Goss}, {Heiles},
  {Dickey}, {Babler}  \& {Kim}}{{Murray} et~al.}{2018}]{Murray:2018}
{Murray} C.~E.,  {Stanimirovic} S.,  {Goss} W.~M.,  {Heiles} C.,  {Dickey}
  J.~M.,  {Babler} B.,   {Kim} C.-G.,  2018, preprint, \href
  {http://adsabs.harvard.edu/abs/2018arXiv180606065M} {} (\mn@eprint {arXiv}
  {1806.06065})

\bibitem[\protect\citeauthoryear{{Nulsen} \& {Fabian}}{{Nulsen} \&
  {Fabian}}{2000}]{Nulsen:2000}
{Nulsen} P.~E.~J.,  {Fabian} A.~C.,  2000, \mn@doi [MNRAS]
  {10.1046/j.1365-8711.2000.03038.x}, \href
  {http://adsabs.harvard.edu/abs/2000MNRAS.311..346N} {311, 346}

\bibitem[\protect\citeauthoryear{{O'Dea}}{{O'Dea}}{1998}]{Odea:1998}
{O'Dea} C.~P.,  1998, \mn@doi [PASP] {10.1086/316162}, \href
  {http://adsabs.harvard.edu/abs/1998PASP..110..493O} {110, 493}

\bibitem[\protect\citeauthoryear{{Oosterloo}, {Verheijen}, {van Cappellen},
  {Bakker}, {Heald}  \& {Ivashina}}{{Oosterloo} et~al.}{2009}]{Oosterloo:2009}
{Oosterloo} T.,  {Verheijen} M.~A.~W.,  {van Cappellen} W.,  {Bakker} L.,
  {Heald} G.,   {Ivashina} M.,  2009, in {Torchinsky} S.~A.,  {van Ardenne} A.,
   {van den Brink-Havinga} T.,  {van Es} A.~J.~J.,   {Faulkner} A.~J.,  eds,
  Wide Field Astronomy and Technology for the Square Kilometre Array. Proc.
  Sci., p.~70 (\mn@eprint {arXiv} {0912.0093})

\bibitem[\protect\citeauthoryear{{Oosterloo}, {Raymond Oonk}, {Morganti},
  {Combes}, {Dasyra}, {Salom{\'e}}, {Vlahakis}  \& {Tadhunter}}{{Oosterloo}
  et~al.}{2017}]{Oosterloo:2017}
{Oosterloo} T.,  {Raymond Oonk} J.~B.,  {Morganti} R.,  {Combes} F.,  {Dasyra}
  K.,  {Salom{\'e}} P.,  {Vlahakis} N.,   {Tadhunter} C.,  2017, \mn@doi [A\&A]
  {10.1051/0004-6361/201731781}, \href
  {http://adsabs.harvard.edu/abs/2017A%26A...608A..38O} {608, A38}

\bibitem[\protect\citeauthoryear{{Orienti}, {Dallacasa}  \&
  {Stanghellini}}{{Orienti} et~al.}{2010}]{Orienti:2010}
{Orienti} M.,  {Dallacasa} D.,   {Stanghellini} C.,  2010, \mn@doi [MNRAS]
  {10.1111/j.1365-2966.2010.17179.x}, \href
  {http://adsabs.harvard.edu/abs/2010MNRAS.408.1075O} {408, 1075}

\bibitem[\protect\citeauthoryear{{Ostorero} et~al.,}{{Ostorero}
  et~al.}{2010}]{Ostorero:2010}
{Ostorero} L.,  et~al., 2010, \mn@doi [ApJ] {10.1088/0004-637X/715/2/1071},
  \href {http://adsabs.harvard.edu/abs/2010ApJ...715.1071O} {715, 1071}

\bibitem[\protect\citeauthoryear{{Papadopoulos}, {Kovacs}, {Evans}  \&
  {Barthel}}{{Papadopoulos} et~al.}{2008}]{Papadopoulos:2008}
{Papadopoulos} P.~P.,  {Kovacs} A.,  {Evans} A.~S.,   {Barthel} P.,  2008,
  \mn@doi [A\&A] {10.1051/0004-6361:200810513}, \href
  {http://adsabs.harvard.edu/abs/2008A%26A...491..483P} {491, 483}

\bibitem[\protect\citeauthoryear{{Planck Collaboration VII}}{{Planck
  Collaboration VII}}{2011}]{Planck:2011}
{Planck Collaboration VII} 2011, \mn@doi [A\&A] {10.1051/0004-6361/201116474},
  \href {http://adsabs.harvard.edu/abs/2011A%26A...536A...7P} {536, A7}

\bibitem[\protect\citeauthoryear{{Purll}, {Lobb}, {Barnes}, {Talbot}, {Rolt},
  {Robertson}, {Closs}  \& {te Plate}}{{Purll} et~al.}{2017}]{Purll:2017}
{Purll} D.~J.,  {Lobb} D.~R.,  {Barnes} A.~R.,  {Talbot} R.~G.,  {Rolt} S.,
  {Robertson} D.~J.,  {Closs} M.~F.,   {te Plate} M.,  2017, in Society of
  Photo-Optical Instrumentation Engineers (SPIE) Conference Series. p. 105650J,
  \mn@doi{10.1117/12.2309221}

\bibitem[\protect\citeauthoryear{{Ramos Almeida}, {Tadhunter}, {Inskip},
  {Morganti}, {Holt}  \& {Dicken}}{{Ramos Almeida}
  et~al.}{2011}]{RamosAlmeida:2011}
{Ramos Almeida} C.,  {Tadhunter} C.~N.,  {Inskip} K.~J.,  {Morganti} R.,
  {Holt} J.,   {Dicken} D.,  2011, \mn@doi [MNRAS]
  {10.1111/j.1365-2966.2010.17542.x}, \href
  {http://adsabs.harvard.edu/abs/2011MNRAS.410.1550R} {410, 1550}

\bibitem[\protect\citeauthoryear{{Ramos Almeida} et~al.,}{{Ramos Almeida}
  et~al.}{2012}]{RamosAlmeida:2012}
{Ramos Almeida} C.,  et~al., 2012, \mn@doi [MNRAS]
  {10.1111/j.1365-2966.2011.19731.x}, \href
  {http://adsabs.harvard.edu/abs/2012MNRAS.419..687R} {419, 687}

\bibitem[\protect\citeauthoryear{{Reynolds}}{{Reynolds}}{1994}]{Reynolds:1994}
{Reynolds} J.,  1994, AT Technical Document AT/39.3/040

\bibitem[\protect\citeauthoryear{{Ricci}, {Prandoni}, {Gruppioni}, {Sault}  \&
  {de Zotti}}{{Ricci} et~al.}{2006}]{Ricci:2006}
{Ricci} R.,  {Prandoni} I.,  {Gruppioni} C.,  {Sault} R.~J.,   {de Zotti} G.,
  2006, \mn@doi [A\&A] {10.1051/0004-6361:20053797}, \href
  {http://cdsads.u-strasbg.fr/abs/2006A%26A...445..465R} {445, 465}

\bibitem[\protect\citeauthoryear{{Russell}, {McNamara}, {Edge}, {Hogan}, {Main}
   \& {Vantyghem}}{{Russell} et~al.}{2013}]{Russell:2013}
{Russell} H.~R.,  {McNamara} B.~R.,  {Edge} A.~C.,  {Hogan} M.~T.,  {Main}
  R.~A.,   {Vantyghem} A.~N.,  2013, \mn@doi [MNRAS] {10.1093/mnras/stt490},
  \href {http://adsabs.harvard.edu/abs/2013MNRAS.432..530R} {432, 530}

\bibitem[\protect\citeauthoryear{{Russell} et~al.,}{{Russell}
  et~al.}{2016}]{Russell:2016}
{Russell} H.~R.,  et~al., 2016, \mn@doi [\mnras] {10.1093/mnras/stw409}, \href
  {http://adsabs.harvard.edu/abs/2016MNRAS.458.3134R} {458, 3134}

\bibitem[\protect\citeauthoryear{{Salom{\'e}} et~al.,}{{Salom{\'e}}
  et~al.}{2006}]{Salome:2006}
{Salom{\'e}} P.,  et~al., 2006, \mn@doi [A\&A] {10.1051/0004-6361:20054745},
  \href {http://adsabs.harvard.edu/abs/2006A%26A...454..437S} {454, 437}

\bibitem[\protect\citeauthoryear{{Sanders}, {Soifer}, {Elias}, {Madore},
  {Matthews}, {Neugebauer}  \& {Scoville}}{{Sanders}
  et~al.}{1988}]{Sanders:1988}
{Sanders} D.~B.,  {Soifer} B.~T.,  {Elias} J.~H.,  {Madore} B.~F.,  {Matthews}
  K.,  {Neugebauer} G.,   {Scoville} N.~Z.,  1988, \mn@doi [ApJ]
  {10.1086/165983}, \href {http://adsabs.harvard.edu/abs/1988ApJ...325...74S}
  {325, 74}

\bibitem[\protect\citeauthoryear{{Sault}, {Teuben}  \& {Wright}}{{Sault}
  et~al.}{1995}]{Sault:1995}
{Sault} R.~J.,  {Teuben} P.~J.,   {Wright} M.~C.~H.,  1995, in {Shaw} R.~A.,
  {Payne} H.~E.,   {Hayes} J.~J.~E.,  eds,  ASP Conf. Ser. Vol. 77,
  Astronomical Data Analysis Software and Systems IV. Astron. Soc. Pac., San
  Francisco, p.~433 (\mn@eprint {} {arXiv:astro-ph/0612759})

\bibitem[\protect\citeauthoryear{{Schawinski}, {Thomas}, {Sarzi}, {Maraston},
  {Kaviraj}, {Joo}, {Yi}  \& {Silk}}{{Schawinski}
  et~al.}{2007}]{Schawinski:2007}
{Schawinski} K.,  {Thomas} D.,  {Sarzi} M.,  {Maraston} C.,  {Kaviraj} S.,
  {Joo} S.-J.,  {Yi} S.~K.,   {Silk} J.,  2007, \mn@doi [MNRAS]
  {10.1111/j.1365-2966.2007.12487.x}, \href
  {http://adsabs.harvard.edu/abs/2007MNRAS.382.1415S} {382, 1415}

\bibitem[\protect\citeauthoryear{{Schlegel}, {Finkbeiner}  \&
  {Davis}}{{Schlegel} et~al.}{1998}]{Schlegel:1998}
{Schlegel} D.~J.,  {Finkbeiner} D.~P.,   {Davis} M.,  1998, \mn@doi [ApJ]
  {10.1086/305772}, \href {http://adsabs.harvard.edu/abs/1998ApJ...500..525S}
  {500, 525}

\bibitem[\protect\citeauthoryear{{Serra} et~al.,}{{Serra}
  et~al.}{2012}]{Serra:2012}
{Serra} P.,  et~al., 2012, \mn@doi [MNRAS] {10.1111/j.1365-2966.2012.20219.x},
  \href {http://adsabs.harvard.edu/abs/2012MNRAS.422.1835S} {422, 1835}

\bibitem[\protect\citeauthoryear{{Shabala}, {Deller}, {Kaviraj}, {Middelberg},
  {Turner}, {Ting}, {Allison}  \& {Davis}}{{Shabala}
  et~al.}{2017}]{Shabala:2017}
{Shabala} S.~S.,  {Deller} A.,  {Kaviraj} S.,  {Middelberg} E.,  {Turner}
  R.~J.,  {Ting} Y.~S.,  {Allison} J.~R.,   {Davis} T.~A.,  2017, \mn@doi
  [MNRAS] {10.1093/mnras/stw2536}, \href
  {http://adsabs.harvard.edu/abs/2017MNRAS.464.4706S} {464, 4706}

\bibitem[\protect\citeauthoryear{{Shankar}, {Weinberg}  \&
  {Miralda-Escud{\'e}}}{{Shankar} et~al.}{2009}]{Shankar:2009}
{Shankar} F.,  {Weinberg} D.~H.,   {Miralda-Escud{\'e}} J.,  2009, \mn@doi
  [ApJ] {10.1088/0004-637X/690/1/20}, \href
  {http://adsabs.harvard.edu/abs/2009ApJ...690...20S} {690, 20}

\bibitem[\protect\citeauthoryear{{Siemiginowska}, {Sobolewska}, {Migliori},
  {Guainazzi}, {Hardcastle}, {Ostorero}  \& {Stawarz}}{{Siemiginowska}
  et~al.}{2016}]{Siemiginowska:2016}
{Siemiginowska} A.,  {Sobolewska} M.,  {Migliori} G.,  {Guainazzi} M.,
  {Hardcastle} M.,  {Ostorero} L.,   {Stawarz} {\L}.,  2016, \mn@doi [ApJ]
  {10.3847/0004-637X/823/1/57}, \href
  {http://adsabs.harvard.edu/abs/2016ApJ...823...57S} {823, 57}

\bibitem[\protect\citeauthoryear{{Silk}}{{Silk}}{2013}]{Silk:2013}
{Silk} J.,  2013, \mn@doi [ApJ] {10.1088/0004-637X/772/2/112}, \href
  {http://adsabs.harvard.edu/abs/2013ApJ...772..112S} {772, 112}

\bibitem[\protect\citeauthoryear{{Silk} \& {Rees}}{{Silk} \&
  {Rees}}{1998}]{Silk:1998}
{Silk} J.,  {Rees} M.~J.,  1998, A\&A, \href
  {http://adsabs.harvard.edu/abs/1998A%26A...331L...1S} {331, L1}

\bibitem[\protect\citeauthoryear{{Simpson}, {Clements}, {Rawlings}  \&
  {Ward}}{{Simpson} et~al.}{1993}]{Simpson:1993}
{Simpson} C.,  {Clements} D.~L.,  {Rawlings} S.,   {Ward} M.,  1993, MNRAS,
  \href {http://adsabs.harvard.edu/abs/1993MNRAS.262..889S} {262, 889}

\bibitem[\protect\citeauthoryear{{Skrutskie} et~al.,}{{Skrutskie}
  et~al.}{2006}]{Skrutskie:2006}
{Skrutskie} M.~F.,  et~al., 2006, \mn@doi [AJ] {10.1086/498708}, \href
  {http://adsabs.harvard.edu/abs/2006AJ....131.1163S} {131, 1163}

\bibitem[\protect\citeauthoryear{{Snellen}, {Schilizzi}, {de Bruyn}, {Miley},
  {Rengelink}, {Roettgering}  \& {Bremer}}{{Snellen}
  et~al.}{1998}]{Snellen:1998}
{Snellen} I.~A.~G.,  {Schilizzi} R.~T.,  {de Bruyn} A.~G.,  {Miley} G.~K.,
  {Rengelink} R.~B.,  {Roettgering} H.~J.,   {Bremer} M.~N.,  1998, \mn@doi
  [A\&AS] {10.1051/aas:1998281}, \href
  {http://adsabs.harvard.edu/abs/1998A%26AS..131..435S} {131, 435}

\bibitem[\protect\citeauthoryear{{Sofia}, {Lauroesch}, {Meyer}  \&
  {Cartledge}}{{Sofia} et~al.}{2004}]{Sofia:2004}
{Sofia} U.~J.,  {Lauroesch} J.~T.,  {Meyer} D.~M.,   {Cartledge} S.~I.~B.,
  2004, \mn@doi [ApJ] {10.1086/382592}, \href
  {http://adsabs.harvard.edu/abs/2004ApJ...605..272S} {605, 272}

\bibitem[\protect\citeauthoryear{{Solomon}, {Rivolo}, {Barrett}  \&
  {Yahil}}{{Solomon} et~al.}{1987}]{Solomon:1987}
{Solomon} P.~M.,  {Rivolo} A.~R.,  {Barrett} J.,   {Yahil} A.,  1987, \mn@doi
  [ApJ] {10.1086/165493}, \href
  {http://adsabs.harvard.edu/abs/1987ApJ...319..730S} {319, 730}

\bibitem[\protect\citeauthoryear{{Srianand}, {Gupta}, {Rahmani}, {Momjian},
  {Petitjean}  \& {Noterdaeme}}{{Srianand} et~al.}{2013}]{Srianand:2013}
{Srianand} R.,  {Gupta} N.,  {Rahmani} H.,  {Momjian} E.,  {Petitjean} P.,
  {Noterdaeme} P.,  2013, \mn@doi [MNRAS] {10.1093/mnras/sts190}, \href
  {http://adsabs.harvard.edu/abs/2013MNRAS.428.2198S} {428, 2198}

\bibitem[\protect\citeauthoryear{{Stanghellini}, {Bondi}, {Dallacasa}, {O'Dea},
  {Baum}, {Fanti}  \& {Fanti}}{{Stanghellini} et~al.}{1997}]{Stanghellini:1997}
{Stanghellini} C.,  {Bondi} M.,  {Dallacasa} D.,  {O'Dea} C.~P.,  {Baum} S.~A.,
   {Fanti} R.,   {Fanti} C.,  1997, A\&A, \href
  {http://adsabs.harvard.edu/abs/1997A%26A...318..376S} {318, 376}

\bibitem[\protect\citeauthoryear{{Tengstrand}, {Guainazzi}, {Siemiginowska},
  {Fonseca Bonilla}, {Labiano}, {Worrall}, {Grandi}  \&
  {Piconcelli}}{{Tengstrand} et~al.}{2009}]{Tengstrand:2009}
{Tengstrand} O.,  {Guainazzi} M.,  {Siemiginowska} A.,  {Fonseca Bonilla} N.,
  {Labiano} A.,  {Worrall} D.~M.,  {Grandi} P.,   {Piconcelli} E.,  2009,
  \mn@doi [A\&A] {10.1051/0004-6361/200811284}, \href
  {http://adsabs.harvard.edu/abs/2009A%26A...501...89T} {501, 89}

\bibitem[\protect\citeauthoryear{{Tingay}, {Jauncey}, {King}, {Tzioumis},
  {Lovell}  \& {Edwards}}{{Tingay} et~al.}{2003}]{Tingay:2003}
{Tingay} S.~J.,  {Jauncey} D.~L.,  {King} E.~A.,  {Tzioumis} A.~K.,  {Lovell}
  J.~E.~J.,   {Edwards} P.~G.,  2003, PASJ, \href
  {http://adsabs.harvard.edu/abs/2003PASJ...55..351T} {55, 351}

\bibitem[\protect\citeauthoryear{{Tingay} et~al.,}{{Tingay}
  et~al.}{2013}]{Tingay:2013}
{Tingay} S.~J.,  et~al., 2013, \mn@doi [PASA] {10.1017/pasa.2012.007}, \href
  {http://adsabs.harvard.edu/abs/2013PASA...30....7T} {30, e007}

\bibitem[\protect\citeauthoryear{{Tingay} et~al.,}{{Tingay}
  et~al.}{2015}]{Tingay:2015}
{Tingay} S.~J.,  et~al., 2015, \mn@doi [\aj] {10.1088/0004-6256/149/2/74},
  \href {http://adsabs.harvard.edu/abs/2015AJ....149...74T} {149, 74}

\bibitem[\protect\citeauthoryear{{Treister}, {Schawinski}, {Urry}  \&
  {Simmons}}{{Treister} et~al.}{2012}]{Treister:2012}
{Treister} E.,  {Schawinski} K.,  {Urry} C.~M.,   {Simmons} B.~D.,  2012,
  \mn@doi [ApJL] {10.1088/2041-8205/758/2/L39}, \href
  {http://adsabs.harvard.edu/abs/2012ApJ...758L..39T} {758, L39}

\bibitem[\protect\citeauthoryear{{Tremblay} et~al.,}{{Tremblay}
  et~al.}{2016}]{Tremblay:2016}
{Tremblay} G.~R.,  et~al., 2016, \mn@doi [Nature] {10.1038/nature17969}, \href
  {http://adsabs.harvard.edu/abs/2016Natur.534..218T} {534, 218}

\bibitem[\protect\citeauthoryear{{Tribble}}{{Tribble}}{1993}]{Tribble:1993}
{Tribble} P.~C.,  1993, \mn@doi [MNRAS] {10.1093/mnras/261.1.57}, \href
  {http://adsabs.harvard.edu/abs/1993MNRAS.261...57T} {261, 57}

\bibitem[\protect\citeauthoryear{Tuthill, Hampson, Bunton, Brown, Neuhold,
  Bateman, de Souza  \& Joseph}{Tuthill et~al.}{2012}]{Tuthill:2012}
Tuthill J.,  Hampson G.,  Bunton J.,  Brown A.,  Neuhold S.,  Bateman T.,  de
  Souza L.,   Joseph J.,  2012, in Int. Conf. on Electromagnetics in Advanced
  Applications (ICEAA). IEEE, pp 1067--1070,
  \mn@doi{10.1109/ICEAA.2012.6328788}

\bibitem[\protect\citeauthoryear{{Ueda}, {Akiyama}, {Ohta}  \& {Miyaji}}{{Ueda}
  et~al.}{2003}]{Ueda:2003}
{Ueda} Y.,  {Akiyama} M.,  {Ohta} K.,   {Miyaji} T.,  2003, \mn@doi [ApJ]
  {10.1086/378940}, \href {http://adsabs.harvard.edu/abs/2003ApJ...598..886U}
  {598, 886}

\bibitem[\protect\citeauthoryear{{Umemura}}{{Umemura}}{2001}]{Umemura:2001}
{Umemura} M.,  2001, \mn@doi [ApJL] {10.1086/324063}, \href
  {http://adsabs.harvard.edu/abs/2001ApJ...560L..29U} {560, L29}

\bibitem[\protect\citeauthoryear{{Uson}, {Bagri}  \& {Cornwell}}{{Uson}
  et~al.}{1991}]{Uson:1991}
{Uson} J.~M.,  {Bagri} D.~S.,   {Cornwell} T.~J.,  1991, \mn@doi [Phys. Rev.
  Lett.] {10.1103/PhysRevLett.67.3328}, \href
  {http://adsabs.harvard.edu/abs/1991PhRvL..67.3328U} {67, 3328}

\bibitem[\protect\citeauthoryear{{Van Kempen} et~al.,}{{Van Kempen}
  et~al.}{2014}]{VanKempen:2014}
{Van Kempen} T.~A.,  et~al., 2014, ALMA Memo 599

\bibitem[\protect\citeauthoryear{{Vermeulen} et~al.,}{{Vermeulen}
  et~al.}{2003}]{Vermeulen:2003}
{Vermeulen} R.~C.,  et~al., 2003, \mn@doi [A\&A] {10.1051/0004-6361:20030468},
  \href {http://adsabs.harvard.edu/abs/2003A26A...404..861V} {404, 861}

\bibitem[\protect\citeauthoryear{{Wagner}, {Bicknell}  \& {Umemura}}{{Wagner}
  et~al.}{2012}]{Wagner:2012}
{Wagner} A.~Y.,  {Bicknell} G.~V.,   {Umemura} M.,  2012, \mn@doi [ApJ]
  {10.1088/0004-637X/757/2/136}, \href
  {http://adsabs.harvard.edu/abs/2012ApJ...757..136W} {757, 136}

\bibitem[\protect\citeauthoryear{{Wall}, {Shimmins}  \& {Bolton}}{{Wall}
  et~al.}{1975}]{Wall:1975}
{Wall} J.~V.,  {Shimmins} A.~J.,   {Bolton} J.~G.,  1975, Aust. J. Phys.
  Astrophys. Suppl., \href {http://adsabs.harvard.edu/abs/1975AuJPA..34...55W}
  {34, 55}

\bibitem[\protect\citeauthoryear{{Wiklind} \& {Combes}}{{Wiklind} \&
  {Combes}}{1994}]{Wiklind:1994}
{Wiklind} T.,  {Combes} F.,  1994, A\&A, \href
  {http://adsabs.harvard.edu/abs/1994A%26A...286L...9W} {286}

\bibitem[\protect\citeauthoryear{{Wiklind} \& {Combes}}{{Wiklind} \&
  {Combes}}{1995}]{Wiklind:1995}
{Wiklind} T.,  {Combes} F.,  1995, A\&A, \href
  {https://ui.adsabs.harvard.edu/#abs/1995A&A...299..382W} {299, 382}

\bibitem[\protect\citeauthoryear{{Wiklind} \& {Combes}}{{Wiklind} \&
  {Combes}}{1996a}]{Wiklind:1996b}
{Wiklind} T.,  {Combes} F.,  1996a, A\&A, \href
  {http://adsabs.harvard.edu/abs/1996A%26A...315...86W} {315, 86}

\bibitem[\protect\citeauthoryear{{Wiklind} \& {Combes}}{{Wiklind} \&
  {Combes}}{1996b}]{Wiklind:1996a}
{Wiklind} T.,  {Combes} F.,  1996b, \mn@doi [Nature] {10.1038/379139a0}, \href
  {http://adsabs.harvard.edu/abs/1996Natur.379..139W} {379, 139}

\bibitem[\protect\citeauthoryear{{Wiklind} \& {Combes}}{{Wiklind} \&
  {Combes}}{1997}]{Wiklind:1997}
{Wiklind} T.,  {Combes} F.,  1997, A\&A, \href
  {http://adsabs.harvard.edu/abs/1997A%26A...328...48W} {328, 48}

\bibitem[\protect\citeauthoryear{{Wiklind} \& {Combes}}{{Wiklind} \&
  {Combes}}{1998}]{Wiklind:1998}
{Wiklind} T.,  {Combes} F.,  1998, \mn@doi [ApJ] {10.1086/305701}, \href
  {http://adsabs.harvard.edu/abs/1998ApJ...500..129W} {500, 129}

\bibitem[\protect\citeauthoryear{{Wiklind}, {Combes}  \& {Kanekar}}{{Wiklind}
  et~al.}{2018}]{Wiklind:2018}
{Wiklind} T.,  {Combes} F.,   {Kanekar} N.,  2018, preprint, \href
  {http://adsabs.harvard.edu/abs/2018arXiv180405377W} {} (\mn@eprint {arXiv}
  {1804.05377})

\bibitem[\protect\citeauthoryear{{Wild}, {Heckman}  \& {Charlot}}{{Wild}
  et~al.}{2010}]{Wild:2010}
{Wild} V.,  {Heckman} T.,   {Charlot} S.,  2010, \mn@doi [\mnras]
  {10.1111/j.1365-2966.2010.16536.x}, \href
  {http://adsabs.harvard.edu/abs/2010MNRAS.405..933W} {405, 933}

\bibitem[\protect\citeauthoryear{{Wolfe}, {Gawiser}  \& {Prochaska}}{{Wolfe}
  et~al.}{2005}]{Wolfe:2005}
{Wolfe} A.~M.,  {Gawiser} E.,   {Prochaska} J.~X.,  2005, \mn@doi [ARA\&A]
  {10.1146/annurev.astro.42.053102.133950}, \href
  {http://adsabs.harvard.edu/abs/2005ARA%26A..43..861W} {43, 861}

\bibitem[\protect\citeauthoryear{{Wolfire}, {McKee}, {Hollenbach}  \&
  {Tielens}}{{Wolfire} et~al.}{2003}]{Wolfire:2003}
{Wolfire} M.~G.,  {McKee} C.~F.,  {Hollenbach} D.,   {Tielens} A.~G.~G.~M.,
  2003, \mn@doi [ApJ] {10.1086/368016}, \href
  {http://adsabs.harvard.edu/abs/2003ApJ...587..278W} {587, 278}

\bibitem[\protect\citeauthoryear{{Wolfire}, {Hollenbach}  \& {McKee}}{{Wolfire}
  et~al.}{2010}]{wolfire:2010}
{Wolfire} M.~G.,  {Hollenbach} D.,   {McKee} C.~F.,  2010, \mn@doi [ApJ]
  {10.1088/0004-637X/716/2/1191}, \href
  {http://adsabs.harvard.edu/abs/2010ApJ...716.1191W} {716, 1191}

\bibitem[\protect\citeauthoryear{{Wright}, {Griffith}, {Burke}  \&
  {Ekers}}{{Wright} et~al.}{1994}]{Wright:1994}
{Wright} A.~E.,  {Griffith} M.~R.,  {Burke} B.~F.,   {Ekers} R.~D.,  1994,
  \mn@doi [ApJS] {10.1086/191939}, \href
  {http://cdsads.u-strasbg.fr/abs/1994ApJS...91..111W} {91, 111}

\bibitem[\protect\citeauthoryear{{Wright} et~al.,}{{Wright}
  et~al.}{2010}]{Wright:2010}
{Wright} E.~L.,  et~al., 2010, \mn@doi [AJ] {10.1088/0004-6256/140/6/1868},
  \href {http://adsabs.harvard.edu/abs/2010AJ....140.1868W} {140, 1868}

\bibitem[\protect\citeauthoryear{{Yan}, {Stocke}, {Darling}, {Momjian},
  {Sharma}  \& {Kanekar}}{{Yan} et~al.}{2016}]{Yan:2016}
{Yan} T.,  {Stocke} J.~T.,  {Darling} J.,  {Momjian} E.,  {Sharma} S.,
  {Kanekar} N.,  2016, \mn@doi [AJ] {10.3847/0004-6256/151/3/74}, \href
  {http://adsabs.harvard.edu/abs/2016AJ....151...74Y} {151, 74}

\bibitem[\protect\citeauthoryear{{Young} et~al.,}{{Young}
  et~al.}{2011}]{Young:2011}
{Young} L.~M.,  et~al., 2011, \mn@doi [MNRAS]
  {10.1111/j.1365-2966.2011.18561.x}, \href
  {http://adsabs.harvard.edu/abs/2011MNRAS.414..940Y} {414, 940}

\bibitem[\protect\citeauthoryear{{de Vries}, {Barthel}  \& {O'Dea}}{{de Vries}
  et~al.}{1997}]{deVries:1997}
{de Vries} W.~H.,  {Barthel} P.~D.,   {O'Dea} C.~P.,  1997, A\&A, \href
  {http://adsabs.harvard.edu/abs/1997A%26A...321..105D} {321, 105}

\bibitem[\protect\citeauthoryear{{de Vries}, {Snellen}, {Schilizzi}, {Mack}  \&
  {Kaiser}}{{de Vries} et~al.}{2009}]{deVries:2009}
{de Vries} N.,  {Snellen} I.~A.~G.,  {Schilizzi} R.~T.,  {Mack} K.-H.,
  {Kaiser} C.~R.,  2009, \mn@doi [A\&A] {10.1051/0004-6361/200811145}, \href
  {http://adsabs.harvard.edu/abs/2009A%26A...498..641D} {498, 641}

\bibitem[\protect\citeauthoryear{{van de Voort} et~al.,}{{van de Voort}
  et~al.}{2018}]{VanDeVoort:2018}
{van de Voort} F.,  et~al., 2018, \mn@doi [MNRAS] {10.1093/mnras/sty228}, \href
  {http://adsabs.harvard.edu/abs/2018MNRAS.476..122V} {476, 122}

\makeatother
\end{thebibliography}

\bsp	
\label{lastpage}
\end{document}